\documentclass[twocolumn,english,aps,prx,twocolum,superscriptaddress,bibnotes,amsmath,amssymb,floatfix]{revtex4-2}

\usepackage[colorlinks=true,citecolor=blue,linkcolor=blue]{hyperref}
\usepackage{changes}
\usepackage{lineno}
\usepackage{stmaryrd}
\usepackage{bm}
\usepackage{graphicx}
\usepackage{epsfig}
\usepackage{color}
\usepackage{amssymb}
\usepackage{amsmath}
\usepackage{array}
\usepackage{braket}
\usepackage{soul}
\usepackage{cleveref}
\usepackage[caption=false]{subfig}
\usepackage[english]{babel}
\usepackage{letltxmacro}
\usepackage{siunitx}
\usepackage{dsfont}
\usepackage{verbatim}
\usepackage{lipsum}  
\usepackage{booktabs}
\usepackage{pdfpages}
\usepackage{pgffor}

\makeatletter
\AtBeginDocument{\let\LS@rot\@undefined}
\makeatother

\def\ii{{\rm i}}  \def\ee{{\rm e}}
\def\rb{{\bf r}}  \def\Rb{{\bf R}}    \def\vb{{\bf v}}
    \def\zz{\hat{\bf z}}    \def\rr{\hat{\bf r}}

\def\me{m_{\rm e}}  \def\kB{{k_{\rm B}}}
\def\Eb{{\bf E}}  \def\Bb{{\bf B}}  
  \def\jb{{\bf j}}
  
    \def\Fb{ {\bf F}} \def\zb{{\bf z}} \def\nb{{\bf n}}

\newcommand{\MPINAT}{Max Planck Institute for Multidisciplinary Sciences, D-37077 G\"{o}ttingen, Germany}
\newcommand{\UGOE}{4th Physical Institute - Solids and Nanostructures, Georg-August-Universit\"{a}t G\"{o}ttingen, D-37077 G\"{o}ttingen, Germany}

\begin{document}


\title{Coulomb Crystallization in Free Electron Beams}
\author{Valerio Di Giulio}
\email{valerio.digiulio@mpinat.mpg.de}
\affiliation{\MPINAT}
\affiliation{\UGOE}

\author{Rudolf Haindl}
\altaffiliation{Present address: Photonics Laboratory, ETH Z\"urich, CH-8093 Z\"urich, Switzerland.}
\affiliation{\MPINAT}
\affiliation{\UGOE}

\author{Armin Feist}
\affiliation{\MPINAT}
\affiliation{\UGOE}

\author{Claus Ropers}
\email{claus.ropers@mpinat.mpg.de}
\affiliation{\MPINAT}
\affiliation{\UGOE}


\date{\today}

\begin{abstract}
Classical Coulomb crystals and quantum Wigner crystals typically arise when repulsive interactions order like-charged particles in spatially confined environments. Here, we show that transient crystallization can also emerge in freely propagating electron pulses without external trapping. Experimentally, we use a transmission electron microscope to measure number-resolved energy correlations in photoemitted pulses containing up to nine electrons, revealing strong correlations at small particle numbers that progressively disappear in the pulse interior as the number increases. To explain these observations, we combine many-body simulations with a one-dimensional statistical-mechanical model, showing that the leading and trailing electrons generate a self-induced longitudinal potential that transiently confines the inner particles, enabling the formation of an ordered chain during propagation. The resulting crystallization phase diagram reveals how many-body ordering dynamics can be understood in terms of evolving macroscopic beam properties, namely its geometry and the ratio of the mean Coulomb interaction energy to the longitudinal thermal energy. Extending this framework, we theoretically demonstrate that weak transverse guiding through electrostatic or optical ponderomotive forces can sustain crystallization at larger particle numbers and predict spontaneous longitudinal periodic bunching in continuous electron beams. Our results identify conditions under which stochastic Coulomb interactions can be harnessed to generate self-structured, strongly correlated electron beams in both pulsed and continuous regimes.
\end{abstract}

\maketitle
\clearpage

\section{Introduction}

Long-range interactions can govern the behavior of many-body systems, promoting the spatial stabilization of their components and giving rise to ordered phases in the limit of strong coupling. For like-charged particles, the equilibrium configuration takes the form of a crystalline structure when the average Coulomb energy $E_{\rm Coul}$, arising from repulsive inter-particle interactions, dominates over the average kinetic energy $E_{\rm kin}$ \cite{MBD25_arxiv}. In classical systems, where the kinetic energy is set by thermal motion, this condition can be achieved by increasing the particle density or by lowering the temperature. Due to the repulsive nature of the interaction, Coulomb-induced crystallization generally requires the presence of a confining potential. This has so far been realized through the external confinement of laser-cooled ions in Paul traps \cite{PHK19} or through alternating static electric and magnetic multipoles in storage rings \cite{ST92,BKW92,SSH01_2,SH04}, giving rise to so-called Coulomb crystals \cite{H1973_2,S96_2,MBD25_arxiv}. In quantum systems, in which the Fermi energy sets the relevant kinetic-energy scale, decreasing the density can instead drive a quantum transition toward spatial ordering \cite{KBL22}. This regime is typically reached in cryogenically cooled solid-state systems, where confinement is naturally provided by the atomic potential landscape, allowing conduction electrons to crystallize. Such ordered structures are known as Wigner crystals \cite{W1934} and have been observed in carbon nanotubes \cite{DB08,SHP19_2}, in twisted transition-metal dichalcogenides \cite{LXR24}, and twisted bilayer graphene \cite{SSE25}. Beyond equilibrium crystalline phases, Coulomb-induced correlations have been observed in electron pairs exhibiting interaction-driven antibunching and partitioning  when traveling in cryogenic semiconductor nanostructures, including ballistic edge-channel beam splitters \cite{FPR23,UFP23} and surface-acoustic-wave circuits \cite{WER23}. Moving few-electron droplets in a GaAs heterostructure have further revealed signatures of a correlated Coulomb liquid \cite{SPSW25}.    

Electron microscopes inherently rely on beams of equally charged particles that are focused onto a specimen, with atomic-resolution images and spectroscopic information reconstructed from the properties of the transmitted and scattered electrons \cite{WC09,E11,paper149}. Precise electron microscopy and spectroscopy therefore require electron beams (e-beams) with tightly controlled distributions in energy, space, and time, such that scattering-induced changes can be distinguished from the intrinsic spread of the incident beam. In this context, Coulomb repulsion has traditionally been regarded as detrimental because it gives rise to the Boersch \cite{B1954_2,K1979,KNA16} and L\"{o}ffler effects \cite{L1969,J1988_2} and, more generally, increases the occupied phase-space volume while reducing beam brightness \cite{J1988_2,O09}. More recently, however, Coulomb-driven electron-electron interactions in free space have attracted interest as a means of forming well-defined correlations in the spatial, temporal, and energy degrees of freedom of few-electron pulses photoemitted from Schottky field emitters. Early evidence for such interaction-induced correlations was found in the form of arrival-time anticorrelations and sub-Poissonian number statistics, observed for ultrashort photoelectron pulses from nanotips \cite{KBG21}, spin-polarized beams from semiconductor photocathodes \cite{KYN21}, and continuous field-emission beams probed with sub-picosecond resolution \cite{BRG24}. Their formation has since been demonstrated in transmission electron microscopes \cite{HFD23,KNZ24} and from free-standing nanotips \cite{MHH23, HLW24}, and they have been characterized via delayed near-field excitation in the time \cite{KNZ24,SWW25}, energy \cite{KNZ24,TNR26}, and energy-time domains \cite{HDF25}. Until now, their formation has been limited to $N\leq 4$ electrons per pulse, while the conditions required to extend these correlations to larger electron numbers remain not fully understood, thus hindering further advances toward the synthesis of strongly correlated e-beams at higher $N$.

Overcoming this limitation would open the possibility of harnessing these correlations at high currents. Combined with spatial and energy filtering, this could enable the generation of sub-shot-noise e-beams \cite{GND12,FHR25} with deterministic $N$ and controlled phase-space properties \cite{FHR25}. Furthermore, Coulomb crystallized e-beams spontaneously bunch in time, similarly to charged particles in plasma wakefields \cite{PGM1990,RBD15,AAA18,AAA19}. Such bunching may in turn give rise to coherent superradiant emission \cite{GIF19}, akin to that generated in free-electron lasers \cite{SSY1}. Indeed, as demonstrated for ions \cite{SSH01_2}, collective interactions can lead to the formation of short pulses without external compression fields. Applied to pulsed and continuous electron emission, this approach could improve control over particle-number and current fluctuations, benefiting electron-beam lithography \cite{KSJ04}, microscopy, and spectroscopy through enhanced dose stability, precision, and sensitivity \cite{MCT09,HLL17,SIS18}.

\begin{figure*}
    \centering
\includegraphics[width=1\linewidth]{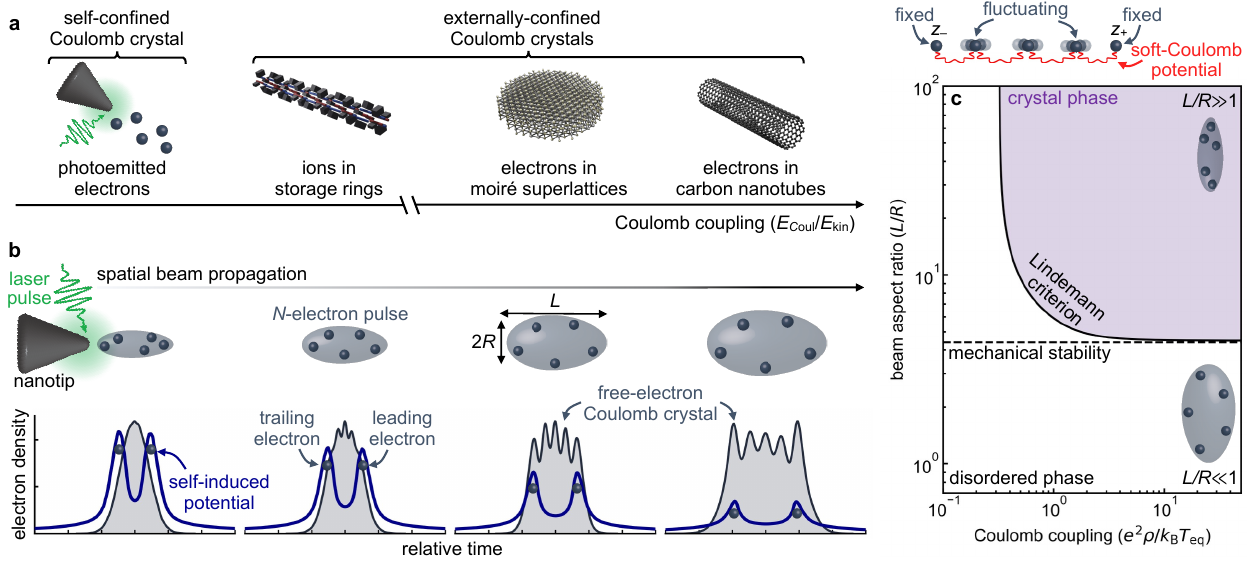}
  \caption{{\bf Formation of self-trapped Coulomb crystals.} {\bf (a)} Overview of selected systems in which Coulomb crystals have been observed, shown as a function of the Coulomb coupling, i.e., the ratio between Coulomb energy $E_{\rm Coul}$ and kinetic energy $E_{\rm kin}$. These include externally trapped classical systems, such as ions in storage rings \cite{ST92,BKW92,SSH01_2,SH04}, and quantum Coulomb crystals (Wigner crystals), such as electrons in carbon nanotubes \cite{DB08,SHP19} and moiré superlattices in two-dimensional heterostructures \cite{LXR24}. As demonstrated in this work, $N$-electron beams ($N$e-beams) photoemitted from a metallic nanotip in an electron microscope can also crystallize under their self-induced potential. {\bf (b)} During propagation, a photoemitted $N$e-beam with longitudinal extension $L$ and lateral size $R$ evolves from a large aspect ratio ${\rm AR}=L/R$, where leading and trailing electrons generate a strong self-trapping potential (blue lines), to a smaller aspect ratio due to beam divergence and Coulomb expansion. The interplay between Coulomb repulsion and self-induced trapping orders the electrons within the pulse, producing a Coulomb crystal visible in the electron density (dark blue). {\bf (c)} Phase diagram obtained from the one-dimensional softened-Coulomb model for a 5-electron pulse with fixed leading and trailing electrons and fluctuating inner electrons, plotted as a function of aspect ratio AR and Coulomb coupling $E_{\rm Coul}/E_{\rm kin}=e^2\rho/\kB T_{\rm eq}$, with $\rho=(N-1)/L$. The solid-disordered transition line is determined from the Lindemann criterion (see Eq.~(\ref{Lindemann})), while the dashed black line indicates the minimum aspect ratio $L/R\approx 4.5$ required for mechanical stability of the chain.}
    \label{Fig1}
\end{figure*}

In this work, we demonstrate that electron pulses photoemitted from a metallic nanotip can form free-space Coulomb crystals, confined not by an external trap but by a potential generated by the pulse itself. We show that, in one-dimensional beams with high pulse current, the leading and trailing electrons form a self-induced potential that confines the inner particles in free space. Such an emergence of order stands in stark contrast to previously studied Coulomb crystals in storage rings, moiré superlattices, and carbon nanotubes (see Fig.~\ref{Fig1}a), where order formation fundamentally relies on external trapping forces. By combining Monte Carlo simulations with experimental measurements of $N$-electron beams ($N$e-beams) with $4$--$9$ particles per pulse in a transmission electron microscope, we determine the phase diagram of a one-dimensional chain model and demonstrate that mechanically and thermally stable crystalline configurations can form in high-aspect-ratio $N$e-beams. A transition from spatial order to disorder is observed with increasing $N$ due to a decrease in the time spent in such a one-dimensional configuration for the electrons to order through intra-beam scattering.

We then numerically show that such loss of order can be overcome by introducing a weak transverse confinement between the extractor and linear accelerator that preserves a high pulse aspect ratio, thus sustaining correlations at larger $N$. Our simulations predict that even a single refocusing event enhances the final electron-electron correlations relative to the case with no trapping, while a longer millimeter-scale trapping stage can produce crystalline pulses with larger particle numbers, reaching up to $N=10$ electrons per pulse under the same experimental conditions.

Finally, using molecular dynamics simulations, we extend the same principle to transversely confined continuous beams, i.e., to the limit $N\rightarrow\infty$. We show that this equilibrium can exhibit strong longitudinal bunching and Coulomb ordering, with universal properties controlled by a normalized linear electron density and the Coulomb coupling. By linking these quantities to the input-beam properties and trapping parameters through energy conservation arguments, we theoretically establish the conditions for realizing continuous Coulomb-ordered free-electron beams in electron microscopes.

\section{Crystallization of free $N$-electron pulses}

\begin{figure*}
    \centering
\includegraphics[width=1\linewidth]{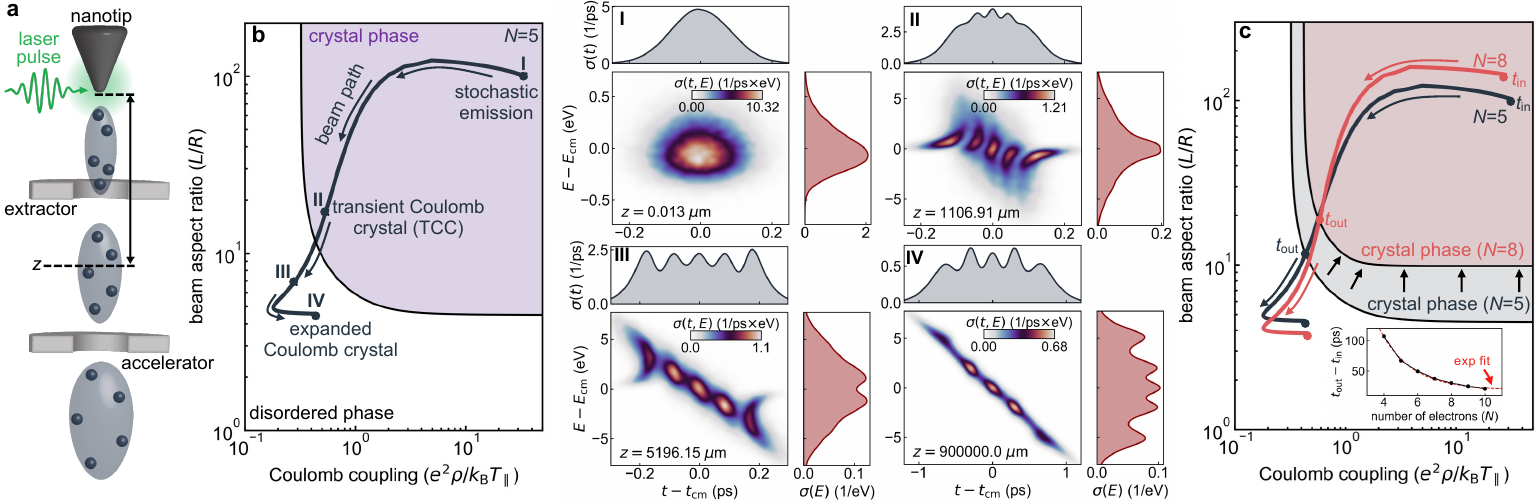}
    \caption{{\bf Coulomb crystal formation in a pulsed electron beam.} {\bf (a)} A laser pulse generates an $N$-electron pulse near a nanotip. The pulse evolves under the influence of accelerating fields provided by the extractor, held at 2 kV, and the accelerator, at 200 kV, located at distances of $500$ $\mu$m and $0.9$ m from the tip, respectively. {\bf (b)} Evolution of the properties of an $N$e-beam with $N=5$ during propagation along the column against the phase diagram that determines the conditions for crystallization. We plot the aspect ratio ${\rm AR}=L/R$ and the longitudinal Coulomb coupling parameter $\Gamma_\parallel=e^2 \rho/\kB T_\parallel$, computed using the longitudinal beam temperature $\kB T_\parallel=\me \langle (v_z-v_{\rm cm})^2\rangle_{z}$ and the average linear density $\rho$ (see main text). {\bf (I-IV)} Longitudinal phase-space density at different $z$-planes corresponding to the specific path coordinates marked in panel (b) (roman numerals). Time and energy marginals are shown in the top and right panels, respectively. {\bf (c)} Evolution of the e-beam properties within the phase diagram for $N=5$ and $8$ electrons per pulse. The crystallization region shrinks for larger particle numbers, yielding an exponential decrease in the time the beam path spends within it, as shown by the fit (dashed red line) in the inset. The photoemission parameters used in all panels are reported in Appendix~\ref{appendixA}.}
    \label{Fig2}
\end{figure*}

In an ultrafast transmission electron microscope, pulsed beams are generated by photoemission from a metallic nanotip \cite{FBR17} (see Fig.~\ref{Fig1}b) and subsequently accelerated by an extractor electrode \cite{HFD23,HDF25}. During propagation along the microscope column, the beam configuration evolves under the combined effects of the accelerating fields, its initial divergence, dispersion, and Coulomb-driven expansion, which modify properties such as the beam aspect ratio. Close to the tip, the finite duration of photoemission $\sigma_{\rm L}$ gives the electrons relative emission delays, and the strong extraction field then produces an initial axial stretching that renders the pulse effectively one dimensional. Its approximate geometry can be determined by the initial transverse beam size $\sigma_\perp$, the semi-angle spread $\sigma_\alpha$, and the extractor potential $U_{\rm ext}$. For instance, at the extractor plane, located a distance $d_{\rm ext}$ from the tip, the longitudinal and transverse dimensions of a non-interacting beam are $L\sim \sigma_{\rm L}(2eU_{\rm ext}/\me)^{1/2}$ and $R\sim (\sigma_\perp^2+\sigma_\alpha^2d_{\rm ext}^2)^{1/2}/2$, respectively. For the experimental values $\sigma_{\rm L}=200$ fs, $U_{\rm ext}=2$ kV, $\sigma_\alpha=1$ mrad, $\sigma_\perp=100$ nm, and $d_{\rm ext}=500$ $\mu$m, these expressions yield an aspect ratio ${\rm AR}=L/R\sim20$ and a linear density $\rho=(N-1)/L\sim1$ $\mu$m${}^{-1}$.

The combination of a large aspect ratio and high linear density provides the physical conditions for the emergence of a spatially ordered electron phase without external transverse confinement. In the quasi-one-dimensional regime, the leading and trailing electrons generate a well-defined potential that acts as an effective boundary (see the blue curves in Fig.~\ref{Fig1}b), transiently confining the inner electrons during propagation and enabling the same Coulomb-driven ordering mechanism underlying crystallization in trapped and solid-state systems (Fig.~\ref{Fig1}a). This self-induced confinement therefore provides the finite time window required for a freely propagating electron pulse to develop crystalline order. As the beam expands and its aspect ratio decreases, the confining potential gradually weakens and eventually disappears. Nevertheless, if its lifetime exceeds the characteristic ordering time, the electrons can rearrange into a spatially ordered configuration, producing a longitudinal density profile with one distinct peak per electron, a hallmark of a free-electron Coulomb crystal (see the time-dependent densities in Fig.~\ref{Fig1}b).

In electron microscopy, this high-aspect-ratio limit is commonly referred to as the pencil-beam regime, whereas the opposite limit, characterized by ${\rm AR}\sim1$, is known as the Holtsmark regime \cite{J1988_2}. Both regimes are conventionally described in terms of the statistical broadening of the energy and angular distributions \cite{CVH10}, under the assumption that the beam retains a single-peaked phase-space density. This description, however, breaks down when strong electron-electron correlations generate multi-lobed distributions, as occurs in the present work. In what follows, we therefore adopt a framework based on the melting behavior of a strongly correlated electron chain, combined with a full numerical treatment of the many-body dynamics.
  
\subsection{One-dimensional equilibrium model}

Because the Coulomb forces between particles in a quasi-one-dimensional arrangement depend primarily on their axial separations and only weakly on their exact transverse positions, the formation of longitudinal order in $N$e-beams can be described using a simple one-dimensional model. In this model, the electrons interact along a single line through a softened Coulomb potential within a non-relativistic framework $\varphi_{\rm soft}(z_i,z_{i'}) = -e/\sqrt{(z_i - z_{i'})^2 + R^2}$, which we can adopt throughout this work because most of the interaction occurs at kinetic energies below 10 keV. 

For a finite number of electrons, a purely repulsive Coulomb interaction cannot form an equilibrium state in the absence of an external restoring potential, which is not present in the setup considered here. Nevertheless, an alternative form of stability can arise by assuming that the required trapping potential takes the form $U_{\rm self\text{-}trap}(z)=-e\sum_\pm \varphi_{\rm soft}(z,\pm L/2)$ and is generated self-consistently by the two outermost charges, which are treated as fixed (see Fig.~\ref{Fig1}c). This assumption is justified if the ordering of the inner electrons occurs on a shorter time scale than that associated with the evolution of the global pulse geometry. In this regime, the expansion can be treated quasistatically, with the two outermost electrons regarded as effectively fixed while the remaining $N-2$ particles relax toward the equilibrium configuration determined by the instantaneous values of $\rho$ and $R$. Using Monte Carlo simulations (see below), we validate this approximation by comparing the ordering time $\tau_{\rm ord}=1/\omega_{\rm p}$, with $\omega_{\rm p}=(4\pi e^2\rho^3/\me)^{1/2}$, to the characteristic longitudinal and transverse evolution times, $\tau_\parallel=|\rho/\partial_t\rho|$ and $\tau_\perp=R/\partial_t R$, respectively. While in the high-aspect-ratio regime the transverse evolution plays only a minor role, we find that, during the high-density stage in which ordering develops, $\tau_{\rm ord}\ll\tau_\parallel$ (see Fig.~\ref{FigA1}h).

In the presence of the self-trapping potential, crystalline order can be quantified through the mean-squared displacement $\langle\delta z^2\rangle$ \cite{D1993} of the particles from their equilibrium positions. We evaluate these fluctuations within the harmonic approximation, following an approach analogous to that used to calculate phonon modes in solids \cite{AM1976}. We first determine the equilibrium positions ${z_i^0}$ of the $N-2$ inner electrons by minimizing the potential energy $U_{\rm soft}=\sum_{i=2}^{N-1}U_{\text{self-trap}}(z_i)-e\sum_{i<i'}\varphi_{\rm soft}(z_i,z_{i'})$, and then expand it to second order around equilibrium as $U_{\rm soft}\approx\delta\zb^{\rm T}D\delta\zb/2$, where $D$ is the corresponding force-constant matrix. For an average kinetic energy characterized by an equilibrium temperature $T_{\rm eq}$, the positional fluctuations are related to $D$ through $\langle\delta z^2\rangle=\kB T_{\rm eq}{\rm Tr}\{D^{-1}\}/(N-2)$ (see Appendix~\ref{appendixB}).

Mechanical stability is a necessary condition for spatial ordering, whereas thermal fluctuations progressively weaken interparticle correlations. In practice, the absence of mechanical stability leads to imaginary displacements, while thermal motion produces displacements that remain compatible with an ordered chain only below a critical value. This threshold is conventionally determined using the Lindemann melting criterion \cite{L1910}, stating that if the average displacement is larger than a fraction of the average inter-particle distance $\rho^{-1}$, the crystal transitions to a gas or liquid phase. The condition leads to the the transition-line equation 
\begin{align}
\langle \delta z^2\rangle\rho^2=\alpha_{\rm L}^2. \label{Lindemann}
\end{align} 

The curve defined by Eq.~(\ref{Lindemann}) depends on only two parameters, the $\rm AR$ and the longitudinal Coulomb coupling $\Gamma_{\rm eq}=e^2 \rho /\kB T_{\rm eq}$, which quantifies the ratio between Coulomb and kinetic energy in the one-dimensional chain model. In Fig.~\ref{Fig1}c, we plot the melting line for $N=5$ by setting $\alpha_{\rm L}=1$ over the full range of aspect ratios for which the chain is mechanically stable (${\rm AR}\geq 4.5$). As expected, spatial ordering is favored in electron pulses with high aspect ratios, whereas lower aspect ratios lead to disordered configurations. Stability with respect to stochastic thermal fluctuations is instead captured by the Coulomb coupling which should exceed 0.3 to obtain ordered structures. We emphasize that, because of the long-range nature of the Coulomb interaction, the resulting phase diagram depends strongly on the particle number $N$, making the formation of stable chains increasingly difficult at large particle numbers (see Fig.~\ref{FigA3}c).  

\subsection{Formation of a transient crystal state in a transmission electron microscope}

\begin{figure*}
    \centering
\includegraphics[width=1\linewidth]{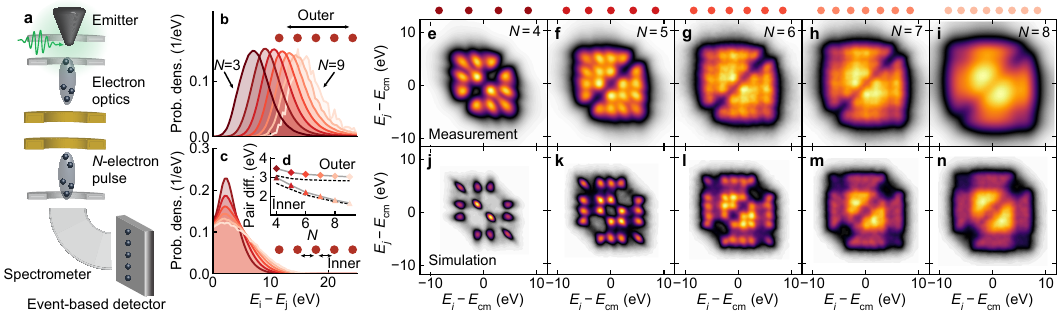}

    \caption{{\bf Experimental observation of TCC by recorded energy-energy correlations in a transmission electron microscope.} {\bf (a)} Experimental setup. $N$-electron pulses are generated by near-threshold femtosecond photoemission from a Schottky field emitter, accelerated to \SI{200}{keV} in the transmission electron microscope, energy-dispersed and detected by an event-driven camera, yielding $N$ and $\{E_1, \dots, E_N\}$ for each pulse. {\bf (b,c)} Experimental histograms of the energy difference between the two outermost electrons (b) and all pairs of inner electrons (c) of the energy-sorted pulse, for $N=3$--$9$ (dark to light). The outermost pair remains well separated, with a separation growing monotonically with $N$, whereas the energy differences among the inner electrons show a correlation gap at small values that progressively fills in with increasing $N$. {\bf (d)} Mean energy difference of the outermost (diamonds) and innermost (triangles) adjacent electron pairs versus $N$. Dashed lines: uncorrelated reference from event mixing, i.e., $N$-electron events assembled from electrons of $N$ different measured events. The Coulomb-induced excess over this reference persists for the outermost pair but vanishes for the innermost pair with increasing $N$. {\bf (e--i)} Measured axial energy correlations for $N=4$ (e), $5$ (f), $6$ (g), $7$ (h), and $8$ (i) electrons, displaying a transition from a crystalline to non-crystalline beam. {\bf (j--n)} Monte Carlo simulations reproducing the experimental conditions in (e--i). Maps (e--n) show the joint probability density, normalized to its maximum (linear scale, white to yellow).}
    
    \label{Fig3}
\end{figure*}

The time-dependent statistical properties of the photoemitted ensemble of $N$ classical electrons are commonly described by the phase-space distribution $\sigma(\rb_1,\vb_1,\dots,\rb_N,\vb_N,t)$, which gives the probability of finding the $i$-th particle at position $\rb_i$ with velocity $\vb_i$ at a given time.  To account for the strong correlations induced by inter-electron scattering, we numerically solve the Liouville equation using a Monte Carlo method, in which the initial particle velocities and positions are sampled from a prescribed distribution and subsequently evolved in time (see Appendix~\ref{appendixA} for more details). In particular, we simulate a photoemission geometry where $N$ electrons are emitted from the nanotip and rapidly accelerated by an extractor held at kilovolt potentials before being further accelerated to kinetic energies of several hundred keV by the linear accelerator, as schematically illustrated in Fig.~\ref{Fig2}a.

We use these simulations to characterize the confinement experienced by the inner electrons due to the particles occupying the leading and trailing positions of the pulse. A key result is that, despite the random nature of the photoemission process and the resulting shot-to-shot variations in the microscopic configurations, under the studied conditions their ensemble average converges to a well-defined and reproducible axial self-confinement. Remarkably, this averaged potential is accurately described by the one-dimensional softened-Coulomb form constructed from the instantaneous linear density $\rho$ and aspect ratio ${\rm AR}$ extracted from the Monte Carlo simulations (see Fig.~\ref{FigA2}). The underlying many-particle dynamics can therefore be represented as emerging from a deterministic potential parametrized by the macroscopic beam properties. In this regard, together with the longitudinal temperature $\kB T_\parallel=\me\langle(v_z-v_{\rm cm})^2\rangle_z$ (averaged over electrons at a fixed longitudinal plane $z$), these parameters establish a direct mapping between the evolving $N$e-beam and the corresponding phase diagram.

In Fig.~\ref{Fig2}b we trace, for pulses containing 5 electrons, the path followed by the beam within the phase diagram of Fig.~\ref{Fig1}c. As expected from the rapid initial temporal stretching of the beam, the near-emission configurations yield an average ${\rm AR}\sim 100$ and relatively strong Coulomb coupling due to the large pulse current, $I=e \rho v_{\rm cm}\sim 3.5$~$\mu{\rm A}$, which gives a high initial linear density $\rho\sim 1$~$\mu{\rm m}^{-1}$ and a moderate initial thermal energy of $\kB T_\parallel\sim 0.1$~ meV (see Fig.~\ref{FigA1}). This combination of parameters is clearly located in the crystallization region of the chain phase diagram (see Fig.~\ref{Fig2}b), thus promoting the formation of an ordered state. However, at this stage the $N$ electrons still occupy an uncorrelated energy-time phase space, as depicted in Fig.~\ref{Fig2}I. After emission, the beam undergoes several structural transformations: Coulomb scattering leads to a temperature increase and to a three-dimensional beam expansion, thereby reducing the pulse current and the aspect ratio, the latter being further decreased by the initial beam divergence (see Fig.~\ref{FigA1}c-g). In terms of the path in the phase diagram, after spending a certain time ordering themselves in the crystal-allowed region, the electrons cross the stability line and remain in the disordered phase until the end of the acceleration stage. Interestingly, close to the crossing of the phase-transition line (Fig.~\ref{Fig2}II), the electrons already appear to form a well-structured time density, where each electron is represented by its own peak, in stark contrast with their single-peaked unstructured energy distribution, which closely resembles an uncorrelated equilibrium distribution. Because of its close correspondence to the spatial densities typical of one-dimensional electron crystals \cite{DB08,SHP19}, together with the uncorrelated energy distribution, we define this short-lived state of $N$ free electrons as a {\it transient Coulomb crystal} (TCC). 

Once the TCC is formed, the $N$e-beam undergoes a quasi-deterministic expansion characterized by the appearance of multiple separated energy peaks and a better definition of the temporal peaks (Fig.~\ref{Fig2}III). The expansion process, which also cools down the beam and thus yields an increase in $\Gamma_\parallel$ (see Fig.~\ref{Fig2}b), is finally completed at very large times, when the number of energy peaks becomes equal to the number of electrons in the pulse and their distribution no longer evolves in time (Fig.~\ref{Fig2}IV).

A completely different behavior would be observed if the TCC state was not reached during the evolution, as displayed by the longitudinal phase-space evolution for higher-$N$ pulses (see Fig.~S1e-h of the SI \cite{SI} for the $N=8$ case). Indeed, even  maintaining the same initial emission conditions, increasing $N$ leads to several consequences that are detrimental to the formation of spatio-temporal ordering. In this regard, the larger number of charges composing the $N$e-beam leads to a faster three-dimensional expansion, as well as to a modification of the order-disorder phase diagram. Mechanical and thermal stability are weakened for higher $N$ and fixed $\rm AR$, as a consequence of the stronger inner electron repulsion, which can more easily overcome the barrier imposed by $U_{\rm self\text{-}trap}$ (see Fig.~\ref{FigA3}c). In Fig.~\ref{Fig2}c, we show how the area occupied by the crystal region for $N=8$ shrinks in comparison with that occupied for $N=5$. All these effects reduce the time that the $N$e-beam spends inside the crystallization region, $t_{\rm out}-t_{\rm in}$, while ordering itself, displaying an exponential decay as a function of the number of electrons in the pulse (see inset in Fig.~\ref{Fig2}c).

Since the ordering time remains at $\tau_{\rm ord}\sim 20$~ps over the entire initial part of the $N$e-beam dynamics (see Fig.~\ref{FigA1}h), we expect, under the simulated conditions, crystallization for pulses with $N\leq 6$ electrons, whereas selecting higher-$N$ electron pulses will lead to a loss of long-range correlations.

\subsection{Experimental observation of TCC in a transmission electron microscope}

To test these predictions, we performed number-resolved spectroscopy of few-electron pulses in the Göttingen Ultrafast Transmission Electron Microscope (UTEM), a platform based on a modified JEOL JEM-F200. Building on our previous studies~\cite{HFD23,HDF25}, we access $N$-electron pulses with substantially higher electron numbers. The pulses are generated via near-threshold photoemission from a W(100)/ZrO Schottky emitter, accelerated to \SI{2}{keV} and \SI{7}{keV} by an extraction and focusing electrode assembly, and brought to a final beam energy of $E_0 = \SI{200}{keV}$ for propagation through the microscope column (see Fig.~\ref{Fig3}a). At the column exit, the electrons are energy-dispersed by an imaging energy filter and recorded by a Timepix3-based event-driven camera with near-unity detection efficiency. The nanosecond-scale temporal resolution of the detector unambiguously correlates each transmitted electron to its laser pulse, giving direct access to the number of electrons ($N$) per pulse and its full set of kinetic energies $\{E_1,\dots,E_N\}$ for subsequent spectral analysis. The emitted electron number follows Poisson statistics and scales with the excitation laser power~\cite{HFD23,MHH23}. Unlike previous experiments, we intentionally operate at substantially higher mean electron numbers, up to $\langle N\rangle \approx 1.0$, accessing events with $N=4$--$9$.

A first indication of the internal pulse structure comes from the pairwise energy differences within each event. In Fig.~\ref{Fig3}b\&c, we compare the probability densities of the energy difference between the two outermost electrons of the energy-sorted pulse (b) and between all pairs of inner electrons (c), for $N=3$--$9$. The two classes behave in strikingly different ways. The outermost pair exhibits a well-defined, single-peaked distribution centered at a large energy separation that grows monotonically with $N$, reflecting the Coulomb expansion of the leading and trailing electrons. The inner electrons, in contrast, show a correlation gap that progressively fills with increasing $N$. To quantify this, we turn to the mean spacing of adjacent electrons at the edges and in the center of the sorted pulse (Fig.~\ref{Fig3}d). Part of its $N$-dependence is purely statistical: as more electrons share the same spectrum, adjacent energies of the sorted pulse move closer together even without any interaction. We therefore compare the mean spacing of the outermost and innermost adjacent pairs with an uncorrelated reference obtained by event mixing, i.e., from $N$-electron events assembled from electrons of $N$ different measured events (Fig.~\ref{Fig3}d, dashed lines). The Coulomb-induced excess over this reference behaves very differently for the two classes. At the pulse edges it persists for all $N$, mirroring the special dynamical role of the two outermost charges, which, per the self-trapping mechanism introduced above, generate the confining potential experienced by the inner particles while remaining only weakly affected by fluctuations in the interior of the pulse. For the innermost pair, the excess gradually vanishes with increasing $N$, signaling that the inner electrons progressively lose their ordering, consistent with the predicted disappearance of the crystalline phase at higher particle numbers, while the self-trapping boundary electrons remain well separated throughout.

This picture is confirmed by the complete two-particle spectral statistics. For each electron number, we generate the energy-energy correlation maps in Fig.~\ref{Fig3}e-i by binning pairwise energy differences within a pulse into a two-dimensional histogram, with each electron's energy expressed relative to the pulse's center-of-mass energy. The maps reveal a pronounced dependence on electron number. For $N=4$ and $5$ (Fig.~\ref{Fig3}e\&f), they break up into a regular lattice of well-separated maxima. This discretized pattern is the spectral fingerprint of the expansion of an initially space-ordered configuration (cf.\ Fig.~\ref{Fig2}III, IV) and is therefore consistent with the transient formation of a Coulomb crystal during early propagation. With increasing electron number (Fig.~\ref{Fig3}g-i), the discrete structure progressively washes out from the interior outward: the inner-electron maxima broaden and merge, while the outermost features persist, in agreement with the pair-resolved analysis of Fig.~\ref{Fig3}b-d. For $N=8$, the map approaches the smooth distribution expected for a stochastically expanding, disordered ensemble.

These observations are quantitatively reproduced by Monte Carlo simulations carried out under the experimental emission and acceleration conditions (Fig.~\ref{Fig3}j-n). The agreement between measured and simulated maps, including both the persistence of the outer-pair anticorrelations and the gradual loss of inner-electron discretization with increasing $N$, supports the interpretation that the final-state energy ordering is inherited from a transient, spatially ordered configuration formed during initial propagation, rather than emerging only during the final acceleration stage.

\section{Crystallization of trapped $N$-electron pulses}

\begin{figure*}
    \centering
\includegraphics[width=1.0\linewidth]{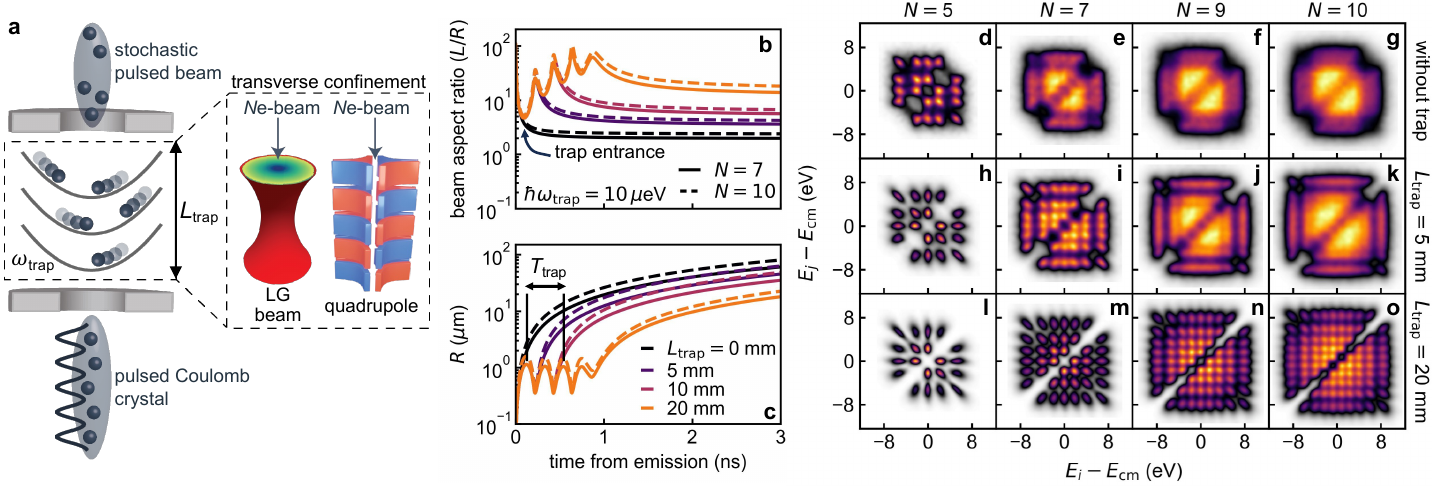}
    \caption{{\bf Trapping of $N$-electron pulses}  {\bf (a)} A photoemitted electron pulse composed of $N$ electrons is accelerated by the extractor and then injected into a harmonic transverse trap with trapping frequency $\hbar\omega_{\rm trap}$, realized either by sending the beam through a Laguerre-Gauss LG$_{01}$ (see inset) mode via ponderomotive forces or by a series of static quadrupoles with alternating polarity. After the trapping stage of length $L_{\rm trap}$, a pulsed Coulomb crystal beam is formed thanks to the one-dimensional confinement provided by the trap, and is then brought to $E_k=200$ keV by the accelerator. {\bf (b,c)} Beam aspect ratio (b) and average transverse electron-electron distance (c) as a function of time elapsed since the photoemission of the pulse, for $N=7,10$ electrons and for different trap lengths. ({\bf d-o}) Energy-energy correlations after the final acceleration stage for $N=5,7,9,10$ (columns) electrons per pulse, with no trap (uppermost row) and with $L_{\rm trap}=5$ and $20$ mm (middle and bottom rows). A trapping frequency of $\hbar \omega_{\rm trap}=10$ $\mu$eV, achievable with state-of-the-art laser systems and miniaturized electronics (see Appendix~\ref{appendixC}), is used in all panels. Initial source parameters are kept the same as in Fig.~\ref{Fig2}.}
    \label{Fig4}
\end{figure*}

Since the formation of spatially ordered structures is promoted by preserving a large beam aspect ratio over extended propagation times, a natural strategy to mitigate the loss of long-range order at high $N$ is to axially constrain the electron trajectories, i.e., to refocus the electrons. Furthermore, this must be done in such a way that Coulomb collisions do not lead to excessive heating of the beam, thereby pushing it toward small values of $\Gamma_\parallel$. Lateral trapping of charged particles can be achieved in several ways, either by resorting to static electrodes or to electrodynamic potentials. In this work, we propose two practical designs to achieve lateral confinement, namely by pondomerotive forces shaped in a Laguerre-Gauss mode and a series of static quadrupoles with alternating polarity. Both mechanisms are described in more detail in Appendix~\ref{appendixC}.

\subsection{Electron-electron correlations of trapped few-electron pulses}

Because, in both trapping mechanisms, the effective force can be approximated as harmonic, we include this force during an acceleration-free stage of length $L_{\rm trap}$ placed between the end of the extractor and the linear accelerator (see Fig.~\ref{Fig4}a), while maintaining the same configuration as in Fig.~\ref{Fig2}. We investigate the dependence of the energy correlations on the trapping length $L_{\rm trap}$ and the number of particles per pulse at a fixed, relatively low trapping frequency of $\hbar\omega_{\rm trap}=10$~$\mu$eV (see Tab.~\ref{Tab1} for a proposed set of experimentally feasible parameters that yield such value).

The effect of the trapping stage can be followed by examining the beam dimensions as a function of the time elapsed since the emission of the last electron. Immediately after the extractor, the beam is refocused every half period ($\sim 207$~ps), until it exits the trapping stage after a time $L_{\rm trap}/v_{\rm cm}$ ($\sim 0.76$ ns for $L_{\rm trap}=20$ mm), before being accelerated again to $200$~keV (see Fig.~\ref{Fig4}b,c). Under this sequence of focusing and refocusing events, Coulomb repulsion induces an axial expansion of the beam, bringing the aspect ratio back to very high values at the expense of a rapid decrease in the linear density, as shown in Fig.~\ref{Fig4}b. For instance at $N=7$, this rapid expansion, together with moderate beam heating, brings the Coulomb coupling to $0.03 \lesssim\Gamma_\parallel \lesssim 0.4$ thus right below the boundary defining the crystal region of the phase diagram placed at $\Gamma_{\rm eq}\approx 0.4$ (see Fig.~\ref{FigA3}c) even if in a high-aspect-ratio regime. Nevertheless, this does not imply a sudden loss of correlations: in the vicinity of the crystallization boundary, the system still retains a strongly correlated liquid character. This can be appreciated by considering the Lindemann condition in Eq.~(\ref{Lindemann}) which, for high aspect ratio and allowing fluctuations up to three times the average lattice spacing, i.e., $\alpha_{\rm L}=3$, places the corresponding order-disorder line at $\Gamma_{\rm eq}=0.04$ ($N=7$). As a result, the electrons complete the propagation with very strong energy-energy and time-time correlations. Similar arguments apply to the other explored electron numbers. 

We notice that even a single refocusing of the beam ($L_{\rm trap}=5$~mm) enhances electron-electron correlations in the final accelerated beam with respect to the trap-free case. Indeed, comparing panels d--g with h--k in Fig.~\ref{Fig4} reveals a marked difference, especially for small numbers of electrons. Remarkably, in this configuration, the cross-over promotes ordering within the $N$e-beam rather than solely producing the detrimental statistical broadening typically associated with cross-overs in electron microscopy~\cite{WN00}. By contrast, a longer trapping stage is required for pulses containing a larger number of particles. As shown in Fig.~\ref{Fig4}l--o, trapping over $L_{\rm trap}=20$~mm at $2$~keV is sufficient to produce crystalline pulses containing up to $N=10$ electrons per pulse.

\section{Continuous beam crystallization}

The preceding results demonstrate that transverse guiding can extend the timescale over which Coulomb ordering develops in few-electron pulses and preserve strong correlations at higher particle numbers. However, the finite longitudinal extent of these pulses ultimately limits the persistence of their periodic structures, as Coulomb repulsion drives longitudinal expansion. This limitation motivates the investigation of ordering in continuous electron beams, where the number of electrons approaches infinity and longitudinal expansion is suppressed. In this regime, Coulomb interactions may give rise to a stationary, strongly correlated states exhibiting spontaneous temporal bunching, opening a potential route to generating structured electron beams at high average currents.

Here, we investigate this regime for transversely guided electron beams in electron microscopes. Related systems have been studied in the context of ion Coulomb crystals in storage rings \cite{ST92,SSH01_2,BKW92}, where beam shearing can disrupt spatial ordering \cite{ST97}. In contrast, electrons in electron microscopes predominantly propagate  along straight trajectories. We therefore focus on the equilibrium spatial properties of trapped continuous electron beams and identify the physical conditions under which Coulomb-induced ordering forms.

\subsection{Equilibrium properties of a trapped continuous Coulomb-bunched beam}

We consider an initially uncorrelated beam entering the trapping stage and subsequently thermalizing within the confining potential to form an equilibrium state (see Fig.~\ref{Fig5}). After a sufficiently long thermalization time \cite{H92_2,CD93}, the phase-space distribution $\sigma(t)$ approaches the Boltzmann distribution $\sigma_{\rm eq}\propto \ee^{-\mathcal{H}_{\rm trap}/\kB T_{\rm eq}}$ (see Eq.~\ref{htrap}).

Because of the characteristic spatial scaling of the Coulomb interaction, a one-dimensional lattice
cannot sustain true crystalline order at any finite temperature. As a result the Lindemann-criterion approach employed above for finite $N$ is not suitable for characterizing particle thermal fluctuations (see SI \cite{SI}). We therefore employ standard molecular-dynamics methods to exactly sample $\sigma_{\rm eq}$ (see Appendix~\ref{appendixD}). Scaled beam properties are universally governed by two dimensionless parameters: the normalized linear density $\lambda=\rho a_{\rm ws}$ and the equilibrium Coulomb coupling $\Gamma_{\rm eq}=e^2/a_{\rm ws}\kB T_{\rm eq}$. Here, $a_{\rm ws}=(3e^2/2\me\omega_{\rm trap}^2)^{1/3}$ is the Wigner-Seitz radius, which is set directly by the transverse trapping frequency.

At fixed trapping frequency, beam ordering can be achieved either by decreasing the average distance between electrons or through cooling. By increasing the linear density from $\lambda=0.1$ to $0.7$, the beam transitions from a gas phase, showing a spatially flat longitudinal phase-space density, first to a liquid and then to a strongly ordered structure that resembles a solid but still shows a decay in the density-density correlation function $g(z)$ (see Fig.~\ref{Fig5}b-d and the corresponding insets for a scan along the $\Gamma_{\rm eq}=23$ line). For values above the critical value $\lambda_{\rm c}\approx 0.71$, the internal Coulomb repulsion pushes the electrons away from the minimum of the trapping potential, yielding a structural transition from a one-dimensional string to higher-order structures such as zig-zag and helical configurations \cite{HS90}. This beam conformation can be directly visualized in the transverse density, which shows a single peak in the string phase (Fig.~\ref{Fig5}e,f) and a donut-shaped profile in the higher-order phases (Fig.~\ref{Fig5}g).

As the beam transitions further into the crystalline state, the traveling electrons occupy increasingly narrow temporal regions, whose duration is limited by fluctuations determined by the beam temperature, linear density, and trapping frequency. For this reason, the beam duty cycle $\delta$, i.e., the ratio between the time duration of one pulse and the crystal period (see sketch in Fig.~\ref{Fig5}h), provides a convenient measure of crystallization and of the degree of longitudinal compression. We define $\delta=I\Delta t/e$, where $\Delta t$ is the effective temporal duration of a single-electron pulse obtained from a generalized definition of the longitudinal beam emittance (see Eq.~(\ref{Deltat})). A duty cycle of $50$\% provides a useful threshold separating the disordered and crystalline regimes (see Fig.~\ref{Fig5}h). Interestingly, even in the limit of high Coulomb coupling, $\Gamma_{\rm eq}\gg 1$, and large normalized density $\lambda$, the duty cycle approaches the non-zero value of $\delta\sim11$\% due to interaction-induced fluctuations.

\subsection{Superradiant light emission from continuous Coulomb crystals}

\begin{figure*}
    \centering
\includegraphics[width=1\linewidth]{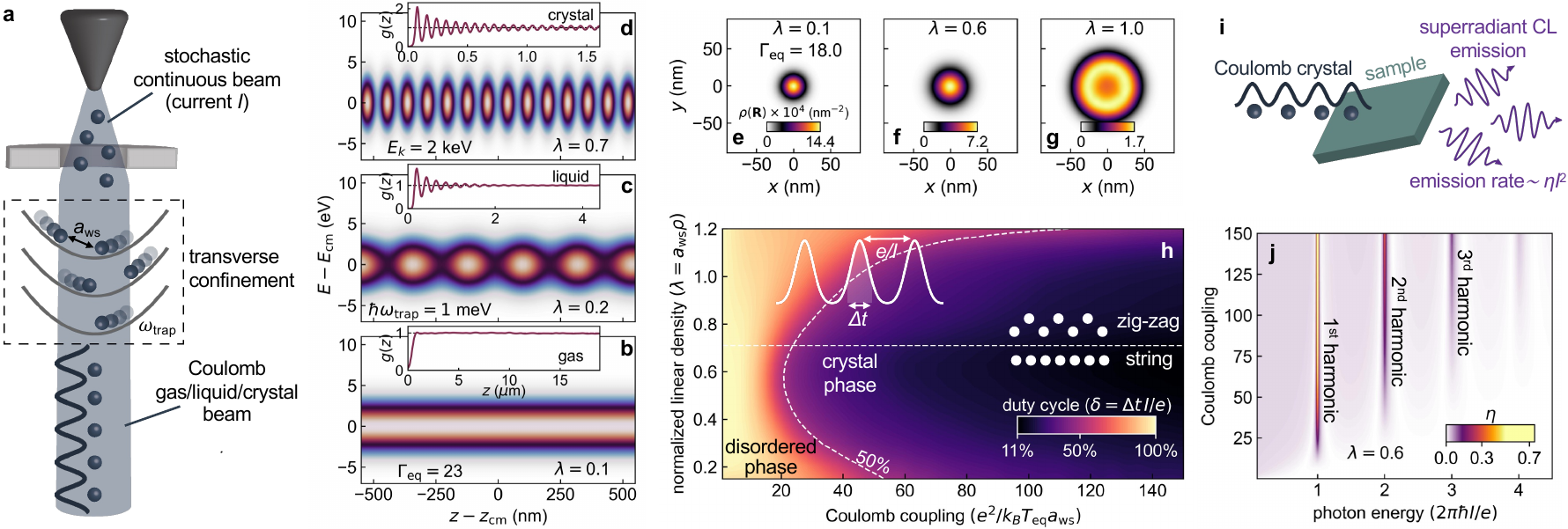}
    \caption{{\bf Crystallization of a continuous electron beam.} {\bf (a)} A continuous source produces an electron current $I$, which is accelerated by the extractor and injected into a transverse trap with frequency $\hbar\omega_{\rm trap}$, similarly to Fig.~\ref{Fig4}a, but with a trap length sufficient for thermalization. The thermalized beam exits in a gas, liquid, or crystal phase depending on the Coulomb coupling $\Gamma_{\rm eq}$, with the Wigner-Seitz radius $a_{\rm ws}$, and on the normalized linear density $\lambda= a_{\rm ws} I/ev_{\rm cm}$. {\bf (b-d)} Longitudinal phase-space density at the trap exit for $\lambda = 0.1$ (b), $0.2$ (c), and $0.7$ (d), at fixed $\Gamma_{\rm eq} = 23$. As $\lambda$ decreases, the beam undergoes crystal-liquid and liquid-gas transitions, as shown by the density-density correlation functions in the insets. Here, $\hbar \omega_{\rm trap} = 1$~meV and $E_{\rm cm}=2$~keV. {\bf (e-g)} Beam density cross-section as a function of $\lambda$. For higher-order structural phases relative to the string configuration, the beam evolves from a single-peak distribution to a donut-shaped profile. We set $\hbar \omega_{\rm trap}$ as in b-d.  {\bf (h)} Universal plot (see main text) of the duty cycle of the Coulomb-bunched beam as a function of $\Gamma_{\rm eq}$ and $\lambda$. The 50\% contour line highlights the transition between disordered and ordered phases, while the horizontal dashed line at $\lambda_{\rm c}\approx 0.71$ marks the string-to-zig-zag structural transition. {\bf (i)} Coulomb crystal beam impinging on a sample and yielding cathodoluminescence (CL) with a superradiant photon-emission rate scaling as $\sim \eta I^2$. {\bf (j)} Superradiant CL enhancement factor $\eta$ for a Coulomb beam as a function of $\Gamma_{\rm eq}$ at $\lambda=0.6$.}
    \label{Fig5}
\end{figure*}

Longitudinal electron bunching plays a central role in the generation of coherent radiation by free-electron beams. In relativistic free-electron lasers, microbunching can develop self-consistently through the collective interaction between electrons and the radiation field, as in self-amplified spontaneous emission, or be externally induced through coherent seeding schemes \cite{SSY1}. At lower electron energies, laser-induced bunching enables coherent cathodoluminescence (CL) through coupling to nanostructures \cite{CH17,KZ17}, coherent control over the excitation of quantum systems \cite{HDF25}, laser-electron pump-probe experiments \cite{paper373}, and superradiant scaling of light emission \cite{SH1969,FFK1971,ABJ05,paper371,DHR25}. However, in these low-energy approaches, temporal bunching does not imply correlations between individual electrons, which remain uncorrelated. In contrast, the longitudinal ordering considered here emerges spontaneously from bare Coulomb interactions between electrons, without externally imposed temporal modulation. We therefore compute the probability associated with superradiant CL emission from such self-organized beams, both as a potential route to coherent light emission and as a direct optical probe of the degree of beam crystallization.

For the strongly correlated electrons described by the equilibrium distribution $\sigma_{\rm eq}$, the angle-integrated CL emission rate accumulated over an illumination time grows linearly with time, with a proportionality factor $I^2(d\Gamma_{\rm CL}/d\omega)\eta(\omega)/e^2$ that scales quadratically with the electron-beam current. Here, $d\Gamma_{\rm CL}/d\omega$ denotes the single-electron CL emission probability per unit frequency \cite{paper149}, as derived in Appendix~\ref{appendixG}. The quadratic current dependence is a signature of superradiance, whose magnitude is governed by the enhancement factor
\begin{align}
\eta(\omega)=\frac{2\rho}{N}\int_{0}^\infty \!\!\! dz\, g(z)\cos{(\omega z/v)}.\label{superrad}
\end{align}
Because the bunching is determined solely by Coulomb repulsion, the temporal spacing between electrons is directly set by the beam current. We therefore expect coherent emission from a crystalline beam to appear at frequencies $2\pi m I/e$, with $m$ a positive integer. By taking the Fourier transform of the density-density correlation function (insets in Fig.~\ref{Fig5}b-d), we compute Eq.~(\ref{superrad}) as a function of frequency and Coulomb coupling (see Fig.~\ref{Fig5}j). Interestingly, this quantity clearly identifies the transition from a disordered to a crystalline structure through the appearance of multiple orders of coherent emission as the Coulomb coupling increases. In this regard, the enhancement factor shows a strong coherent build-up of radiation already for $\Gamma_{\rm eq}\gtrsim 50$, reaching values $\eta \gtrsim 0.5$ at the first harmonic and considerably smaller values for the second and the third (Fig.~\ref{Fig5}j).

\subsection{Requirements on the input beam to achieve continuous beam crystallization}
While the preceding analysis establishes the beam properties attainable under certain equilibrium conditions, assessing whether such states can be reached requires relating them to the characteristics of the beam entering the trap: its transverse size $\sigma_\perp$, divergence or convergence semi-angle $\sigma_\alpha$, and energy spread $\sigma_E$. Since the beam current is conserved, the equilibrium temperature $T_{\rm eq}$ is the only remaining quantity to be determined from these initial conditions. To this end, we impose conservation of the average energy per particle between the incoming beam and the equilibrium state characterized by $\sigma_{\rm eq}$. Assuming that the Coulomb energy of the incoming beam equals that of the zero-temperature equilibrium configuration, corresponding to perfect axial ordering, we derive an upper bound on the equilibrium coupling parameter $\Gamma_{\rm eq}$ and determine the trapping frequency that maximizes it. This procedure yields the optimal frequency
$\omega^{\rm opt}_{\rm trap}=(\sigma_\alpha/\sigma_\perp)(\sigma_E^2/8E_{\rm cm}\me\sigma_\alpha^2+E_{\rm cm}/\me)^{1/2}$,
together with the corresponding bound on the equilibrium Coulomb coupling (see Appendix~\ref{appendixF} for a detailed derivation)
\begin{align}
\Gamma_{\rm eq}\lesssim 1.16 \left[\frac{\sigma_E^2 \sigma_\perp}{4 E_{\rm cm} e^2}+\frac{2E_{\rm cm} \sigma_\alpha^2 \sigma_\perp}{e^2}\right]^{-2/3}.\label{gamma_eq}
\end{align}
Interestingly, according to $\omega_{\rm opt}$, beams with larger $\sigma_\perp$ require weaker transverse confinement, because their larger initial potential energy is converted into thermal energy as the system approaches equilibrium.

Eq.~(\ref{gamma_eq}) predicts a considerable dependence on longitudinal and transverse beam temperatures, i.e., on the center-of-mass energy, energy spread, and convergence angle. In particular, low-energy beams are favored for more monochromatic beams, whereas higher energies should be chosen for beams with large energy spread. For instance, we obtain $\Gamma_{\rm eq}\lesssim 61$ with $\hbar \omega_{\rm trap}\sim 34$~$\mu$eV for a beam with $\sigma_E=0.1$~eV, focused into the trap with $\sigma_\perp=50$~nm and convergence angle $\sigma_\alpha=0.5$~mrad at $E_{\rm cm}=100$~eV (see Fig.~\ref{FigA4}f-i). This would form a beam with a good level of crystallization, corresponding to a duty cycle of $\delta \sim 25$\% if formed at a current $I\sim 1$~$\mu$A (see Fig.~\ref{Fig5}h). High transverse beam temperatures strongly undermines crystallization. Indeed, a beam with equal divergence and lateral size but accelerated at $2$~keV and with current $I\sim 10$ $\mu$A would not crystallize, since $\Gamma_{\rm eq}\lesssim 11$ with even a higher required trapping frequency of $\sim 120$~$\mu$eV (see Fig.~\ref{FigA4}f-i).

These results highlight that continuous-beam crystallization is not determined by the electron current alone, but requires a suitable balance between the initial beam properties and transverse guiding. In particular, high linear density must be combined with sufficiently low longitudinal and transverse energy spreads to prevent thermal fluctuations from suppressing Coulomb-induced ordering. The resulting conditions arising from  Eq.~(\ref{gamma_eq}) connect the equilibrium phase diagram in Fig.~\ref{Fig5}h to experimentally accessible beam parameters, providing a practical guideline for the generation of crystallized continuous e-beams.

\section{Conclusions}
Ordered forms of matter are ubiquitous in nature and range from solid-state systems, where atoms attract each other, to charged-particle systems, where mutually repelling electrons or ions can form Coulomb crystals when confined by an external field. In this work, we have shown that the paradigm connecting Coulomb crystallization with external confinement breaks down in photoemitted few-particle electron pulses. Under suitable low-dimensional conditions, the ensemble-averaged Coulomb potential created by the two outermost electrons gives rise to a well-defined effective potential that acts as a dynamical boundary for the inner electrons. As a result, the beam can self-trap for a sufficiently long time for the electrons to form an ordered state (Fig.~\ref{Fig1}). Remarkably, since this ordering develops on a faster timescale than beam expansion, the conditions for crystallization at finite electron number $N$ can be predicted through a phase diagram defined solely by macroscopic dynamical beam parameters. During propagation, this mechanism drives the dynamical evolution of the beam into what we call a {\it transient Coulomb crystal} (TCC), in which the electrons are strongly correlated in time but remain uncorrelated in energy (Fig.~\ref{Fig2}). 

While this behavior was demonstrated through particle simulations, we also provided experimental evidence of the final product of the expanded TCC by recording energy maps of $N=4$--$8$ electron pulses in a transmission electron microscope. In the experiment, we observed signatures of TCC formation for pulses containing 4, 5, and 6 electrons (Fig.~\ref{Fig3}), whereas higher-$N$ pulses remained uncorrelated because the stronger intrabeam Coulomb expansion, together with the increased mechanical and thermal instability at larger particle number, prevents the beam from ordering within the time spent in order-favored region of the phase diagram.

We then explored how to sustain the pulse self-trapping potential for longer times by introducing a $\mu$eV transverse trap during beam propagation in the microscope. Remarkably, we found that only a few refocusing events, corresponding to a $20$~mm-long trap, are sufficient to synthesize electron-beam Coulomb crystals with tens of particles (Fig.~\ref{Fig4}). We provided values to implement two viable guiding designs: using free-space ponderomotive forces with a intensity profile displaying a minimum in the center, or stacking a series of quadrupoles with alternating polarity along the direction of electron motion. The former could be implemented using commercially available spatial-light modulators \cite{MAI08,FOD21,FDM16} and by exploiting recent advances in light-induced free-electron shaping \cite{paper397,MWS22}. The latter could benefit from rapid progress in micro- and nanofabricated charged-particle optics \cite{HFK11,HKK22,SBS26}, which can provide electric fields of order $10^7$~V/m over distances of tens of microns. 

As a further step, we explored the possibility of creating Coulomb crystals from a continuous electron source (Fig.~\ref{Fig5}). Because fluctuations increase with the number of particles, continuous crystallization imposes severe constraints on the input beam entering the trap, requiring simultaneously high current, small transverse size, low divergence, and narrow energy spread. Nevertheless, recent progress in electron-source technology suggests that these requirements are not so out of reach. Cold-field emitters and laser-driven nanotips already provide high brightness, high coherence, and sub-eV energy spreads \cite{SS09,FBR17,HCW18_2}, while nanostructured LaB$_6$ field emitters have demonstrated ultrabright monochromatic emission and stable microampere-level operation \cite{ZTY16,GTH25}. In parallel, ultracold electron sources based on near-threshold photoionization of laser-cooled atoms offer a complementary route toward extremely low transverse temperatures and low emittance, although so far mainly in pulsed operation \cite{MSJ13,FRN19}. In this regard, a remaining fundamental challenge is the heating induced by stochastic intrabeam scattering at high current \cite{KMW05}. In crystalline ion beams, this problem can be counteracted by laser cooling \cite{SH04}; however, an analogous strategy cannot be directly transferred to e-beams because free electrons lack internal transitions. This highlights the need for applying to e-beams cooling or phase-space control methods capable of reducing the three-dimensional effective electron temperature without sacrificing a significant fraction of the beam current, such as stochastic \cite{JLR22}, cavity-assisted sideband \cite{MCC07} and Doppler cooling \cite{VCB01}.

Beyond the fundamental interest associated with the formation of this new type of Coulomb crystal, such strong correlations could be harnessed to address several challenges in electron microscopy. Marked temporal and energy correlations between electrons could be exploited to extract electron bunches with pre-selected properties. For instance, electron-current fluctuations could be strongly suppressed by applying filters that select a precise energy distribution corresponding to a definite electron number \cite{HFD23,FHR25}. Time-ordered electron distributions, tunable either through the degree of lateral confinement or by placing the sample at different axial planes, could also drive superradiant THz sources based on CL emission with tunable quantum statistics \cite{DHR25}, and enable new forms of electron microscopy based on timed multi-electron pulses.

More generally, the combination of Coulomb repulsion with external static and time-dependent fields creates a rich many-body dynamics that can open new routes not only for controlling free-electron beams, but also for studying strongly correlated systems in free space. This setting offers direct access to diagnostic tools while allowing the surrounding electromagnetic environment to be modified with a high degree of flexibility. This idea could be further extended by tailoring the electron-electron interaction through nearby dielectric or nanophotonic structures, opening the way to a new research direction devoted to strongly correlated beams with tunable interparticle interaction.

\section*{Acknowledgements}

We acknowledge the continued support from the Göttingen UTEM team. We thank Martin Koz\'{a}k for insightful discussions. The experiments were conducted at the Göttingen UTEM Lab, funded by the Deutsche Forschungsgemeinschaft (German Research Foundation) through Project-ID 432680300 (SFB 1456, project C01 to C.R.), the Gottfried Wilhelm Leibniz program (RO 3936/4-1 to C.R.) and the European Union’s Horizon 2020 research and innovation programme under grant agreement no. 101017720 (FET-Proactive EBEAM).

\section*{Author contributions}
C.R. directed the study. V.D.G. prepared the figures, with contributions from R.H.. V.D.G. developed the theoretical framework and conducted all particle and molecular dynamics simulations. R.H. conducted the experiments and data analysis.  All authors discussed the results and interpretation. V.D.G., R.H., and C.R. wrote the manuscript.
\section*{Competing Interests}
The authors declare no competing interests.


\appendix 

\section*{Appendix}
\label{appendix}

\setcounter{equation}{0}
\renewcommand{\theequation}{A\arabic{equation}}
\setcounter{figure}{0}
\renewcommand{\thefigure}{A\arabic{figure}}
\renewcommand{\thesubsection}{\Alph{subsection}}

\subsection{Particle simulations of photoemitted $N$-electron pulses}
\label{appendixA}

\begin{figure*}
\centering
\includegraphics[width=1\linewidth]{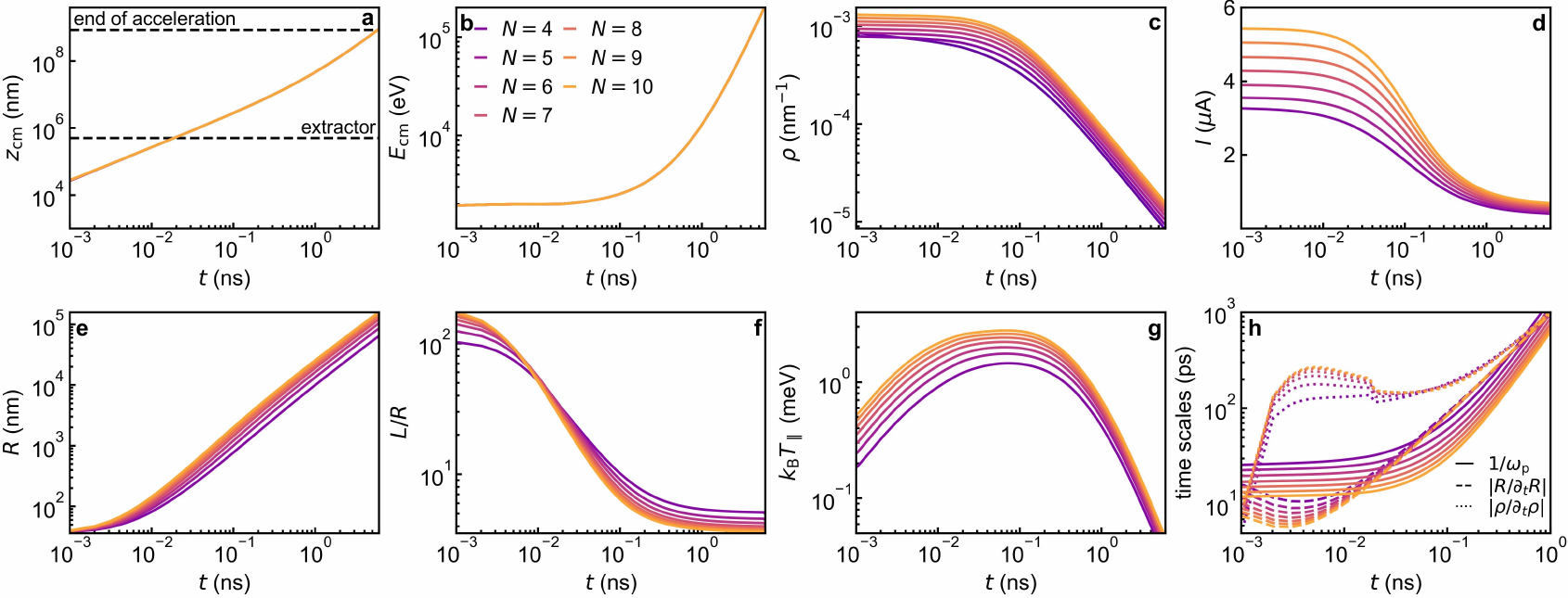}
\caption{\label{FigA1}{\bf Evolution of the $N$e-beam properties in the absence of lateral trapping.} Time evolution of the beam properties as a function of the time elapsed since pulse photoemission, defined from the ejection of the last electron, for different numbers of electrons per pulse (see legend in panel {\bf (b)}). We show the center-of-mass axial coordinate {\bf (a)} and energy {\bf (b)}, the linear density {\bf (c)}, the pulse current $I=e\rho v_{\rm cm}$ {\bf (d)}, the transverse beam size {\bf (e)}, the pulse aspect ratio {\bf (f)}, and the longitudinal temperature $\kB T_\parallel$ (see main text) {\bf (g)}. Panel {\bf (h)} compares the characteristic time scales governing the beam evolution: the ordering time $1/\omega_{\rm p}$, the transverse expansion time $|R/\partial_t R|$, and the longitudinal expansion time $|\rho/\partial_t\rho|$.}
\end{figure*}

We compute the evolution of the phase-space density $\sigma(\rb_1,\vb_1,\dots,\rb_N,\vb_N,t)$ related to the $N$ photoemitted electrons using Monte Carlo trajectory simulations. For each realization, the initial positions and velocities are sampled from the photoemission distribution (see below), and the classical
equations of motion are integrated,
\begin{align}
\me\ddot{\rb}_i=\Fb_{\rm ext}(\rb_i)+\Fb_{\rm int}(\rb_i),
\end{align}
where $\rb_i=(\Rb_i,z_i)$ and $\Fb_{\rm int}(\rb_i)
=-e\sum_{i'\neq i}\nabla_{\rb_i}
\varphi_{\rm Coul}(\rb_i,\rb_{i'})$ with $
\varphi_{\rm Coul}(\rb_i,\rb_{i'})
=-e/|\rb_i-\rb_{i'}|$. 
Ensemble averages over the resulting trajectories yield the time-dependent
phase-space density and the beam observables discussed in the main text.

Photoemission from the nanotip is modeled following Refs.~\cite{HFD23,HDF25}. Electrons are emitted uniformly from a flat circular surface of radius $100$ nm. Their emission times and initial kinetic energies are sampled from Gaussian distributions. The emission-time distribution has zero mean and standard deviation $\sigma_{\rm L}\approx {\rm FWHM}/2.35$, with ${\rm FWHM}=220$ fs, while the kinetic-energy distribution has mean $E_0=0.1$ eV and standard deviation $\sigma_{0,E}=0.35$ eV. The initial emission angle is restricted to $\theta_{\rm max}=2$ mrad, as determined by comparison with experimental results.

The simulated propagation consists of an extraction stage, an optional transverse guide of length $L_{\rm trap}$, and final acceleration to $200$ keV. The external force is modeled piecewise in these three regions. In the
extraction region ($0<z<d_{\rm ext}$), with $d_{\rm ext}=500~\mu$m and $U_{\rm ext}=2$ kV, we use the sphere--plane
approximation \cite{BZD12} and retain its axial component $\Fb_{\rm ext}^{(i)}(\rb)
=eU_{\rm ext}R_{\rm tip}\rb/r^3\approx
eU_{\rm ext}R_{\rm tip}{\bf z}/r^3 $ with $R_{\rm tip}=1$ $\mu$m. Within the transverse guide
($d_{\rm ext}<z<d_{\rm ext}+L_{\rm trap}$), the force is harmonic, $\Fb_{\rm ext}^{(ii)}(\rb)
=-\me\omega_{\rm trap}^2\Rb$, where $\hbar\omega_{\rm trap}=10~\mu$eV
(see Appendix~\ref{appendixC}). Finally, in the acceleration region
($d_{\rm ext}+L_{\rm trap}<z<d_{\rm acc}$), with $d_{\rm acc}=85$ cm and
$U_{\rm acc}=200$ kV, the electric field is approximated as uniform, $\Fb_{\rm ext}^{(iii)}(\rb)
=eU_{\rm acc}\zz/[d_{\rm acc}-(d_{\rm ext}+L_{\rm trap})]\,$.
Figure~\ref{FigA1} shows the simulated evolution of the beam properties in
the absence of transverse confinement ($L_{\rm trap}=0$).

\subsection{Fluctuations of a finite electron chain:
softened-Coulomb model}
\label{appendixB}

\begin{figure*}
\centering
\includegraphics[width=1\linewidth]{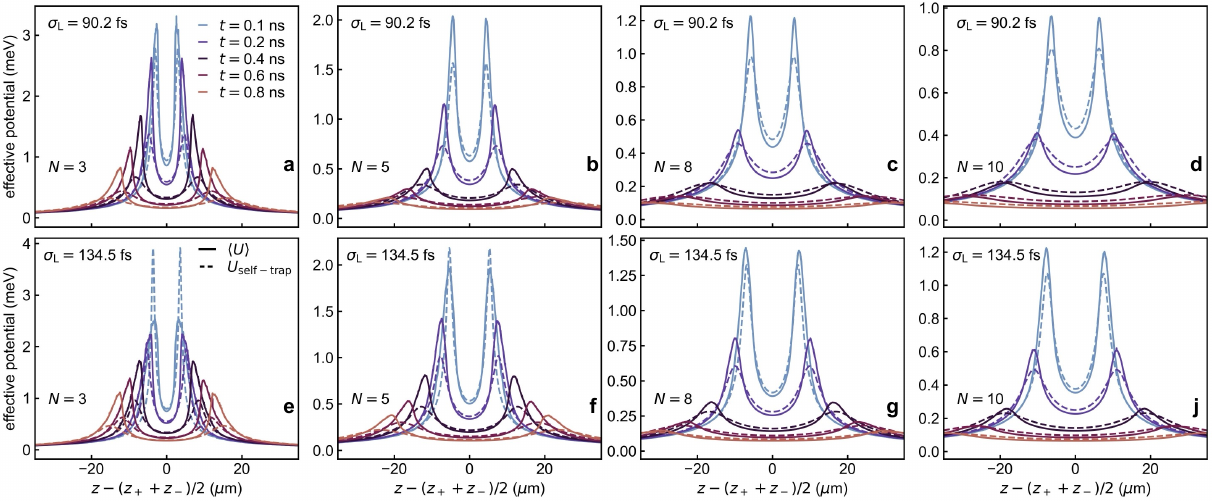}
\caption{\label{FigA2}{\bf Effective one-dimensional softened potential.} Comparison between the average potential $\langle U \rangle$ experienced by the inner electrons of the pulse due to the leading and trailing electrons and the effective self-trapping potential $U_{\rm self\text{-}trap}$, for pulse durations $\sigma_{\rm L}=90.2$ fs {\bf (a–d)} and $\sigma_{\rm L}=134.5$ fs {\bf (e–j)}. Each column corresponds to a different number of electrons in the pulse, $N=3,5,8,10$. The average potential $\langle U \rangle$ is extracted directly from the time-dependent particle simulations according to Eq.~(\ref{eq:simulation-potential}), whereas $U_{\rm self\text{-}trap}$ is evaluated analytically using the parameters $\rho$ and $R$ obtained from the same simulations (see Appendix~\ref{appendixB}).}
\end{figure*}

To describe the stability of a freely expanding $N$e-beam, we
consider an effective one-dimensional model in which the electrons occupy
a segment of length $L$, with linear density $\rho=(N-1)/L$, and interact
through the softened Coulomb potential $\varphi_{\rm soft}(z_i,z_{i'})
=-e/\sqrt{(z_i-z_{i'})^2+R^2}$, where $R$ is the characteristic transverse electron-electron separation
extracted from the particle simulations.

In this model, the leading and trailing electrons are fixed at
$z_\pm=\pm L/2$, whereas the remaining $N-2$ electrons are allowed to
relax and fluctuate around their equilibrium positions.
The two boundary electrons generate an effective self-trapping potential,
$U_{\rm self\text{-}trap}(z)=e^2(1/r_++1/r_-)$, with
$r_\pm=\sqrt{(z-z_\pm)^2+R^2}$.
The effective Hamiltonian of the inner electrons can then be written as
$\mathcal{H}_{\rm soft}=\sum_{i=2}^{N-1}p_{i,z}^2/2\me+U_{\rm soft}$,
where
\begin{align}
U_{\rm soft}=
\sum_{i=2}^{N-1}U_{\rm self\text{-}trap}(z_i)
-\frac{e}{2}\sum_{\substack{i,i'=2\\i\neq i'}}^{N-1}
\varphi_{\rm soft}(z_i,z_{i'}).
\label{eq:soft-hamiltonian}
\end{align}
We determine the equilibrium positions $\{z_i^0\}$ by minimizing
the potential energy in Eq.~(\ref{eq:soft-hamiltonian}).
Small fluctuations around this configuration are governed by the
force-constant matrix
\begin{align}
D_{ii'}=&\delta_{ii'}\sum_{k\neq i}V_{ik}
-(1-\delta_{ii'})V_{ii'},
\label{eq:force-constant-matrix}
\end{align}
with
$V_{ii'}=
e^2[2(z_i^0-z_{i'}^0)^2-R^2]/
[(z_i^0-z_{i'}^0)^2+R^2]^{5/2}$.
Here, $i,i'=2,\ldots,N-1$, whereas the sum over $k$ also includes
the two fixed boundary electrons.

\begin{figure*}
\centering
\includegraphics[width=0.9\linewidth]{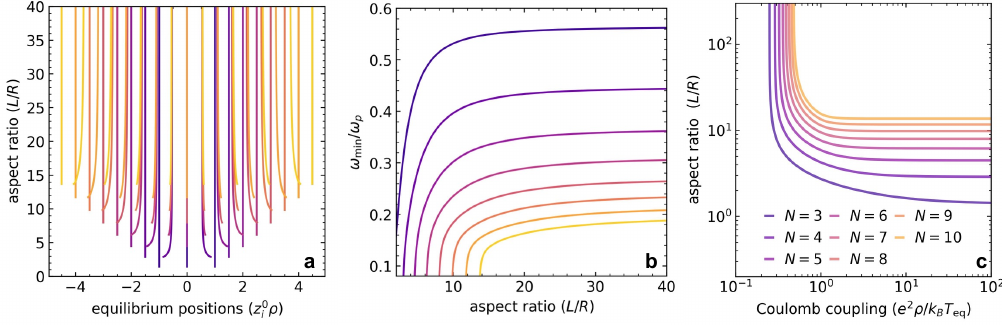}
\caption{\label{FigA3}{\bf Equilibrium positions and normal frequencies of a $N$-electron Coulomb chain.} {\bf (a)} Equilibrium positions $z_i^0$ normalized to the linear density $\rho=(N-1)/L$, as a function of the aspect ratio $L/R$ for chains with different number of particles (see legen in panel c). Each curve terminates at the minimum aspect ratio compatible with mechanical stability. For instance, $L/R\approx 4.5$ for $N=5$, $\approx 6.2$ for $N=6$, $\approx 8.0$ for $N=7$, and $\approx 9.8$ for $N=8$. {\bf (b)} Minimum normal-mode frequency as a function of the aspect ratio and particle number, normalized to the plasma frequency $\omega_{\rm p}=\sqrt{4\pi e^2 \rho^3/\me}$. {\bf (c)} Order-disorder transition lines computed from the Lindemann criterion (see Eq.~(\ref{Lindemann})) for a chain containing from $N=3$ to 10 particles and for $\alpha_{\rm L}=1$.}
\end{figure*}

Expanding the potential energy to second order around the equilibrium
positions yields the harmonic approximation
$U_{\rm soft}\approx
\sum_{i,i'=2}^{N-1}\delta z_i D_{ii'}\delta z_{i'}/2$,
up to an additive constant, where $\delta z_i=z_i-z_i^0$.
Since $D$ is real and symmetric, it can be diagonalized in terms
of orthonormal normal modes $\delta\zb^{(n)}$ satisfying
\begin{align}
D\,\delta\zb^{(n)}
=\me\omega_n^2\delta\zb^{(n)}.
\end{align}
Mechanical stability requires all eigenvalues of $D$ to be positive,
or equivalently, all normal-mode frequencies to be real.
The equilibrium positions and lowest normal-mode frequencies are
shown in Fig.~\ref{FigA3}a,b.

To analyze the thermal stability of the chain, we express the
displacement vector as a linear combination of normal modes,
$\delta\zb=\sum_n\alpha_n\delta\zb^{(n)}$.
In this basis, the harmonic potential energy becomes
$U_{\rm soft}\approx\sum_n\me\omega_n^2\alpha_n^2/2$.
Averaging over the corresponding Boltzmann distribution then gives
$\langle\alpha_n^2\rangle=\kB T_{\rm eq}/\me\omega_n^2$.
Consequently, the mean-square displacement of the $i$-th inner
electron is
\begin{align}
\langle\delta z_i^2\rangle
=\kB T_{\rm eq}
\sum_n\frac{\big[\delta z_i^{(n)}\big]^2}{\me\omega_n^2}
=\kB T_{\rm eq}\big[D^{-1}\big]_{ii},
\end{align}
where the last equality follows from the orthonormality of the
eigenvectors of $D$.
Averaging over the $N-2$ inner electrons yields
$\langle\delta z^2\rangle
=\kB T_{\rm eq}\,{\rm Tr}\!\left\{D^{-1}\right\}/(N-2)$.
We determine the order-disorder boundary using the Lindemann criterion
$\langle\delta z^2\rangle\rho^2=\alpha_{\rm L}^2$,
with $\alpha_{\rm L}=1$ for the phase diagrams in Figs.~\ref{Fig1},\ref{Fig2}, and \ref{FigA3}.
The resulting melting lines for different $N$ are shown in
Fig.~\ref{FigA3}c.

The model just described relies on two main assumptions:
{\it (i)} the validity of the softened Coulomb potential as an
effective self-trapping mechanism; and
{\it (ii)} the separation of the time scales associated with the
ordering of the inner electrons and the transverse and longitudinal
expansion of the $N$e-pulse.

To validate assumption {\it (i)}, we compare
$U_{\rm self\text{-}trap}$ with the potential energy experienced by
the inner electrons in the trajectory simulations.
Denoting the leading and trailing electrons by $i_\pm$, with
coordinates $\rb_\pm=(\Rb_\pm,z_\pm)$, the simulation-based potential
is given by
\begin{align}
\langle U(z) \rangle_{\sigma(t)} &
=\frac{e^2}{N-2}\label{eq:simulation-potential}\\
\times & \sum_{i\neq i_-,i_+}\sum_{s=\pm}
\left\langle
\frac{1}
{\sqrt{(z-z_s)^2+|\Rb_i-\Rb_s|^2}}
\right\rangle_{\sigma(t)}.\nonumber
\end{align}
At each propagation time, the parameters of the effective model
are determined from the simulated pulse length, through
$\rho=(N-1)/\langle z_+-z_-\rangle_{\sigma(t)}$,
and the corresponding mean transverse electron-electron separation $R=\sum_{i<i'}\langle |\Rb_i-\Rb_{i'}|\rangle_{\sigma(t)}/N(N-1)$.
As shown in Fig.~\ref{FigA2}, the potential in
Eq.~(\ref{eq:simulation-potential}) is accurately reproduced by
$U_{\rm self\text{-}trap}(z)$.
This agreement establishes a direct connection between the stochastic
many-particle dynamics and the effective description in terms of
macroscopic beam parameters.

Assumption {\it (ii)} is justified by comparing the characteristic
ordering time $\tau_{\rm ord}=1/\omega_{\rm p}$, where
$\omega_{\rm p}=(4\pi e^2\rho^3/\me)^{1/2}$,
with the longitudinal and transverse expansion times,
$\tau_\parallel=|\rho/\partial_t\rho|$ and
$\tau_\perp=|R/\partial_tR|$, respectively.
As shown in Fig.~\ref{FigA1}h,
$\tau_{\rm ord}\ll\tau_\parallel$ during the high-density stage
in which ordering develops.
Although $\tau_{\rm ord}$ can become comparable to $\tau_\perp$,
the large beam aspect ratio makes the effective interaction only
weakly sensitive to variations in $R$.

\subsection{Lateral trapping designs}
\label{appendixC}
We consider two experimentally feasible strategies for realizing the weak transverse harmonic confinement used in the particle simulations shown in Fig.~\ref{Fig4}: an alternating sequence of electrostatic quadrupoles and ponderomotive guiding by a structured optical field (see Fig.~\ref{Fig4}a). Both schemes generate an approximately harmonic potential in the transverse plane. In particular, alternating static quadrupoles produce an effective Paul-trap potential~\cite{P90,G1995} in the beam frame, whereas the dark central core of a Laguerre-Gaussian LG$_{p\ell}$ optical mode~\cite{ABS92,YLW22} provides confinement through the ponderomotive force.

For electrostatic quadrupolar guiding, the spatially periodic field is experienced as a temporal modulation by electrons traveling with mean velocity $v_{\rm cm}$. This situation is analogous to the Paul-trap pseudopotential and has been implemented for electron guiding~\cite{ZSH21}. In this configuration, the electron beam samples the modulation at $\omega_0=2\pi v_{\rm cm}/d$, with stability parameter $q=eV/\me r^2\omega_0^2$, where $V$ denotes the applied quadrupole-voltage scale, $r$ the effective distance from the beam axis to the electrodes, and $d$ the pitch of the alternating sequence. The resulting secular trapping frequency is (see SI~\cite{SI})
\begin{align}
\omega_{\rm trap}=\frac{|q|\omega_0}{\sqrt{8}}.
\end{align}

Ponderomotive guiding exploits the cycle-averaged potential $U_{\rm pond}=e^2|\vec{\mathcal{E}}(\rb)|^2/\me\omega^2$ associated with a free-space electric field $\Eb(\rb,t)=2{\rm Re}\{\vec{\mathcal{E}}(\rb)\ee^{-\ii \omega t}\}$, which drives electrons toward minima of the optical intensity. We use a Gaussian laser pulse of duration $\sigma_t$, prepared in the LG$_{01}$ mode and with peak electric-field-envelope amplitude $E_{10}$. The beam is focused by a lens with numerical aperture $\mathrm{NA}$ and has a Rayleigh range $z_{\rm R}=2c/\omega\,\mathrm{NA}^2$. In this configuration, the harmonic trapping frequency generated near the on-axis intensity minimum at the focus is (see SI~\cite{SI})
\begin{align}
\omega_{\rm trap}=\sqrt{\frac{2}{\pi}}\,
\frac{eE_{10}\,\mathrm{NA}}{\me c}.
\end{align}

Table~\ref{Tab1} reports representative quadrupolar and ponderomotive parameter sets yielding $\hbar\omega_{\rm trap}=10~\mu$eV. For the quadrupolar implementation, the parameters are chosen in the conservative range $0.1\leq |q|\leq 0.4$, which provides appreciable confinement while limiting the micromotion. In the ponderomotive case, to obtain confinement that remains nearly constant over the trapping length required in the particle simulations of Fig.~\ref{Fig4}, we choose a Rayleigh range exceeding $L_{\rm trap}$.

\begin{table*}[t]
\centering
\label{tab:guiding_parameters}
\begin{tabular}{c c c c c c @{\hspace{1.5em}} c c c c c c}
\toprule
\multicolumn{1}{c}{Trapping frequency}
& \multicolumn{5}{c}{Ponderomotive guiding}
& \multicolumn{6}{c}{Electrostatic quadrupolar guiding} \\
\cmidrule(r){1-1}
\cmidrule(lr){2-6}
\cmidrule(l){7-12}

$\hbar\omega_{\rm trap}$ ($\mu$eV)
& NA
& $w_0$ ($\mu$m)
& $E_{10}$ (V/m)
& $\hbar\omega$ (eV)
& $U_p$ (J)
& $V$ (V)
& $r$ ($\mu$m)
& $d$ (mm)
& $\hbar\omega_0$ ($\mu$eV)
& $E_k$ (keV)
& $q$ \\
\cmidrule(r){1-1}
\cmidrule(lr){2-6}
\cmidrule(l){7-12}

10 & 0.0041 & 82 & $7.9\times10^{9}$ & 1.17 & 2.0
& 262 & 100 & 0.77 & 141 & 2 & 0.2 \\

10 & 0.0036 & 70 & $9.0\times10^{9}$ & 1.55 & 1.9
& 175 & 100 & 1.16 & 94.3 & 2 & 0.3 \\

10 & 0.0029 & 58 & $1.1\times10^{10}$ & 2.33 & 2.0
& 33 & 50 & 1.55 & 70.7 & 2 & 0.4 \\

\bottomrule
\end{tabular}
\caption{{\bf Ponderomotive and quadropolar trapping.}  Different choices of parameters yielding a harmonic trapping frequency $\hbar\omega_{\rm trap}=10~\mu{\rm eV}$. We report parameters for ponderomotive guiding with a Laguerre-Gaussian laser pulse of electric-field standard deviation $\sigma_t$ with $p=0$ and $\ell=1$, for which $\omega_{\rm trap}=\sqrt{2/\pi}eE_{10}{\rm NA}/m_e c$, pulse energy $U_p=cw_0^2E_{10}^2\sigma_t/2\sqrt{\pi}$, and waist $w_0=2c/\omega {\rm NA}$. The optical parameters are chosen such that the Rayleigh length is $\sim 20~{\rm mm}$ and for a pulse duration of $\sigma_t=500$ ps. For electrostatic quadrupolar guiding, the parameters are chosen for 2 keV electrons and lie within the stability region $0.1\leq q\leq 0.4$; the trapping frequency is computed as $\omega_{\rm trap}=q\omega_0/\sqrt{8}$, with $\omega_0=2\pi v_{\rm cm}/d$, where $v_{\rm cm}$ is the electron velocity and $d$ is the quadrupole pitch.}
\label{Tab1}
\end{table*}

\subsection{Simulations of equilibrium properties of a trapped continuous electron beam}
\label{appendixD}
To characterize a continuous electron beam beyond the harmonic approximation (see SI \cite{SI}), we perform Metropolis simulations \cite{BST1966,PSP04}. Specifically, we consider a beam composed by $N=50$ electrons evolving in a squared box of size $L$ with periodic boundary conditions and subjected to a transverse harmonic potential with trapping frequency $\omega_{\rm trap}$. The Hamiltonian governing the stable configuration of the system is given by 
\begin{align}
\mathcal{H}_{\rm trap}=\frac{1}{2}\sum_{i=1}^N\bigg[\frac{p_i^2}{\me}+\me & \omega_{\rm trap}R_i^2  \bigg]-\frac{e}{2}\sum_{i\neq i'} \varphi_{\rm Coul}(\rb_i,\rb_{i'})\label{htrap}
\end{align}
To accurately account for the long-range Coulomb interaction while minimizing finite-size and surface effects, we employ the Ewald summation method for charges embedded in a uniform positive background, corresponding to the jellium model \cite{BST1966,H1973_2,HS90}. Within this framework, the Coulomb pair-interaction energy $-e\varphi_{\rm Coul}$ is replaced in Eq.~(\ref{htrap}) by the Ewald potential energy 
\begin{align}
\!\!\! \!\! U_{\rm Ewald}(\rb_{i}, \rb_{i'})&= e^2\sum_{\nb} \bigg\{ \frac{{\rm erfc}\big(\sqrt{\pi}|\rb_{i}-\rb_{i'}-\nb L |/L\big)}{|\rb_i-\rb_{i'}-\nb L|} \nonumber \\
+&(1-\delta_{\nb,0})\frac{\ee^{-\pi n^2 }}{\pi n^2 L}\cos\bigg[\frac{2\pi}{L} (\rb_i-\rb_{i'}) \cdot \nb\bigg]\bigg\}, \nonumber
\end{align}
where the summation runs over the lattice vectors $\nb$ of a squared lattice, and ${\rm erfc}(x)=2\int_x^\infty dt\, \ee^{-t^2}$ denotes the complementary error function \cite{AS1972}.

We implement the Metropolis algorithm by constructing a Markov chain in which, at each step, one of the $N$ electrons is selected at random and subjected to a trial displacement. This procedure samples the Boltzmann distribution $\sigma_{\rm eq}=\ee^{-\mathcal{H}_{\rm trap}/\kB T_{\rm eq}}/Z$. To visualize and characterize the Coulomb-induced spatial ordering, we calculate the density $\rho(z)=\sum_{i=1}^N\big \langle \delta(z-z_i)\big \rangle/N$ and the density-density correlation function $g(z-z')\equiv  \sum_{i\neq i'}^N\langle \delta(z-z_i)\delta(z'-z_{i'})\rangle/\rho^2=L\sum_{i\neq i'}^N\langle \delta[z-z'-(z_{i}-z_{i'})]\rangle/N^2$ along the periodic direction \cite{C1974_2}. 

An important consequence of the structure of $\mathcal{H}_{\rm trap}$ is that any quantity of the form $\langle f(\rb_1 \rho ,\dots,\rb_N\rho)\rangle$ depends only on $\Gamma_{\rm eq}$ and $\lambda$, rather than separately on $\omega_{\rm trap}$. Consequently, the average energy per particle normalized to
\begin{align}
\frac{\langle \mathcal{H}_{\rm trap}\rangle}{N E_{\rm trap}}= \frac{3}{2 \Gamma_{\rm eq}}+\frac{3}{4\lambda^2 N}&\sum_{i=1}^N\langle R_i^2 \rho^2 \rangle
\\&+\frac{\lambda}{N}\sum_{i<i'}\langle U_{\rm Ewald}(\rb_i \rho,\rb_{i'} \rho)\rangle ,\nonumber
\end{align}
where $E_{\rm trap}= (2e^4 \me \omega_{\rm trap}^2/3)^{1/3}=e^2/a_{\rm ws}$, is a function only of the two parameters $\lambda$ and $\Gamma_{\rm eq}$. We exploit this scaling property to determine both the duty cycle in Appendix~\ref{appendixE} and the final equilibrium temperature in Appendix \ref{appendixF} for arbitrary trapping frequencies while performing simulations at only a single value of $\omega_{\rm trap}$.

\subsection{Calculation of longitudinal emittance and duty cycle}
\label{appendixE}

The root-mean-square (rms) emittance is the conventional quantity used to
characterize the phase-space area occupied by an electron beam. For a
one-particle phase-space distribution $\sigma(x_s,\beta_s)$, where
$s=x,y,z$ and $\beta_s=v_s/c$, it is defined as $\epsilon_s^{\rm rms}
=\sqrt{
\left\langle \delta x_s^2\right\rangle
\left\langle \delta\beta_s^2\right\rangle
-\left\langle \delta x_s\delta\beta_s\right\rangle^2},$
where $\delta x_s=x_s-\langle x_s\rangle$ and
$\delta\beta_s=\beta_s-\langle\beta_s\rangle$. For a single Gaussian
distribution, $\epsilon_s^{\rm rms}$ provides a direct and robust measure of
the occupied phase-space area.

In the strongly correlated regime considered here, however, the longitudinal
phase-space distribution can develop several separated lobes. In this case,
the rms emittance need not faithfully represent the physically occupied area. We therefore define a generalized
emittance from the union of the contours $\gamma_i$ satisfying $\sigma(x_s,\beta_s)
=\max_{x_s,\beta_s}
\sigma(x_s,\beta_s)/\ee$.

The generalized emittance is the enclosed area divided by $2\pi$,
\begin{align}
\epsilon_s=\frac{1}{2\pi}
\sum_i\int_{\gamma_i}dx_s\,d\beta_s.
\label{eq:generalized-emittance}
\end{align}
This definition is independent of the detailed distribution shape and reduces
to $\epsilon_s^{\rm rms}$ for a single Gaussian lobe. It is therefore suited
to the multi-lobed distributions obtained in the equilibrium simulations of
Appendix~\ref{appendixD}.

At equilibrium, the single-particle distribution assumes the separable form $\sigma_s(x_s,\beta_s)
=g(x_s)\ee^{-\beta_s^2/2\sigma_{\beta_s}^2}/\sqrt{2\pi\sigma_{\beta_s}^2}$ which tranforms
Eq.~(\ref{eq:generalized-emittance}) into $
\epsilon_s=\sqrt{2}\sigma_{\beta_s}
\sum_i\int_{\Delta x_{s,i}}dx_s
\sqrt{
\log\left[\ee\, g(x_s)/\max_{x_s}g(x_s)
\right]}/\pi,$
where $\Delta x_{s,i}$ denotes the intervals over which
$g(x_s)>\max_{x_s}g(x_s)/\ee$. For a Gaussian spatial profile,
$g(x_s)\propto\exp\{-x_s^2/2\sigma_{x_s}^2\}$, this expression reduces to $\epsilon_s
=\sqrt{2}\sigma_{\beta_s}
\int_{-\sqrt{2}\sigma_{x_s}}^{\sqrt{2}\sigma_{x_s}}
dx_s\sqrt{1-x_s^2/2\sigma_{x_s}^2}/\pi=\epsilon_s^{\rm rms}$. Since the length occupied by the electrons in the velocity part of the phase space is given by $2\sqrt{2}\sigma_{\beta_s}$, the spatial segment is given by $\pi \epsilon_s /\sqrt{2}\sigma_{\beta_s}$.

For the longitudinal direction, $x_s=z$, we use the same construction to define the duty cycle as $\delta=I \Delta t/e$, i.e., the average time occupied by an electron peak $\Delta t$ respect to the bunching period $I/e$, with
\begin{align}
\Delta t=\frac{1}{ N v_{\rm cm}}\sum_i\int_{\Delta z_i}dz
\sqrt{
\log\left[
\frac{\ee\,\sigma_{\rm eq}(z)}
{\max_z\sigma_{\rm eq}(z)}
\right]}.
\label{Deltat}
\end{align}
Here, $\sigma_{\rm eq}(z)$ is the equilibrium single-particle longitudinal density and $\Delta z_i$ is the
interval associated with the $i$-th density peak. In the string regime, where
the distribution consists of identical peaks with linear density $\rho$, the duty cycle becomes $\delta=\rho\int_{\Delta z}dz \sqrt{\log\left[\ee\,\sigma_{\rm eq}(z)/\max_z\sigma_{\rm eq}(z)\right]}$. If each peak is approximately Gaussian with spatial standard deviation
$\sigma_z$, the duty cycle reduces to
$\delta=\pi\rho\sigma_z/\sqrt{2}
=\pi I\sigma_t/\sqrt{2}e$,
where $\sigma_t=\sigma_z/v_{\rm cm}$.

\subsection{Conditions for continuous electron beam crystallization: equilibrium temperature through average conservation of energy}
\label{appendixF}

\begin{figure*}
    \centering
\includegraphics[width=1\linewidth]{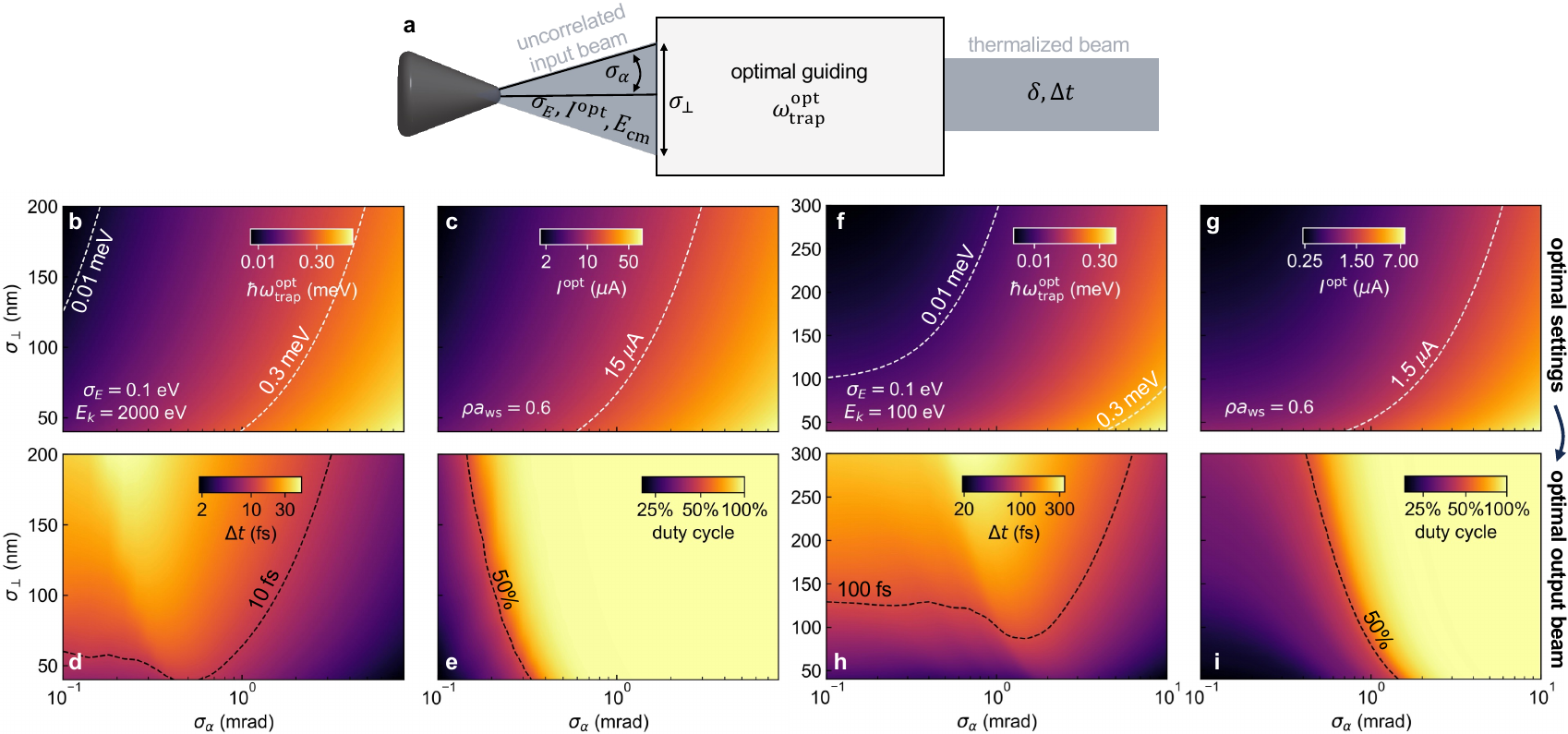}
    \caption{{\bf Relation between optimal input and equilibrium beam properties} {\bf (a)} An input continuous electron beam with current $I^{\rm opt}$, transverse size $\sigma_\perp$, divergence half-angle $\sigma_\alpha$, and energy spread $\sigma_E$ enters a trap with an optimal frequency $\omega^{\rm opt}_{\rm trap}$ at a kinetic energy $E_{\rm cm}$. After thermalization, the beam exits the trap in a state characterized by a duty cycle $\delta$ and a temporal spread $\Delta t$. {\bf (b-e)} Optimal trapping frequency $\omega^{\rm opt}_{\rm trap}$ (b), input current $I^{\rm opt}=0.6\,e/a_{\rm ws} v_{\rm cm}$ (c), and output pulse temporal duration $\Delta t$ (d) and duty cycle $\delta$ (e) as a function of $\sigma_\alpha$ and $\sigma_\perp$ for $E_{\rm cm}=2$ keV and $\sigma_E=0.1$ eV. {\bf (f-i)} Same as in panels b-e but for $E_{\rm cm}=100$ eV. In all panels $\lambda=0.6$ to reach the minimum duty cycle with highest temperatures (see Fig.~\ref{Fig5}h).}
    \label{FigA4}
\end{figure*}

In this section, we determine the beam characteristics required at the entrance
of the trap to reach a given equilibrium phase, assuming that the trapping
structure is sufficiently long for the beam to thermalize \cite{H92_2,CD93}.
Because the Hamiltonian is time independent, the average energy is conserved,
i.e., $\partial_t \langle \mathcal{H}\rangle_{\sigma(t)}\equiv
\partial_t \langle \mathcal{H}\rangle_{t}=0$. We denote by
$E_{\rm in}=\langle \mathcal{H}_{\rm trap}\rangle_{t_{\rm in}}$ the mean
energy of the beam when it is fully contained within the trap at time
$t_{\rm in}$, and by $E_{\rm eq}=\langle \mathcal{H}_{\rm trap}\rangle_{t_{\rm eq}}$
its energy after equilibration. Imposing $E_{\rm in}=E_{\rm eq}$ and dividing
by the number of particles yields
\begin{align}
\frac{E_{\rm eq}(T_{\rm eq},\rho)}{N}&=\frac{\me}{2}\langle (\vb-\zz v_{\rm cm})^2 \rangle_{t_{\rm in}} \label{energy_cons} \\
+ \frac{1}{2} & \me \omega_{\rm trap}^2 \langle R^2 \rangle_{t_{\rm in}}+\frac{1}{2N}\sum_{i \neq i'}^N\langle U_{\rm Ewald} (\rb_i,\rb_{i'})\rangle_{t_{\rm in}}. \nonumber
\end{align}
Since the current $I=v_{\rm cm}\rho$ and the center-of-mass velocity
$v_{\rm cm}$ are conserved during propagation through the acceleration-free
trapping region, the only remaining unknown is $T_{\rm eq}$, which can be
obtained by inverting Eq.~(\ref{energy_cons}). Together with $\rho$ and
$\omega_{\rm trap}$, this quantity determines the equilibrium phase reached
by the beam for a given input configuration. In principle, all terms on the
right-hand side of Eq.~(\ref{energy_cons}) should be obtained from
nonequilibrium particle simulations, whereas $E_{\rm eq}$ is calculated using
the Metropolis algorithm described in Appendix~\ref{appendixD}. Figure~S2a in the
SI \cite{SI} shows the equilibrium energy per particle
$E_{\rm eq}/N E_{\rm trap}$ over the relevant parameter range.

Because $E_{\rm eq}/N E_{\rm trap}$ depends monotonically on
$\Gamma_{\rm eq}$, Eq.~(\ref{energy_cons}) can be inverted to determine the
equilibrium Coulomb coupling. The equilibrium energy per particle is
accurately described by a quadratic function of $1/\Gamma_{\rm eq}$ with
$\lambda$-dependent coefficients (see the scaling properties discussed in
Appendix~\ref{appendixD}),
$E_{\rm eq}/N E_{\rm trap}=e_1(\lambda)+e_2(\lambda)/\Gamma_{\rm eq}+e_3(\lambda)/\Gamma_{\rm eq}^2$.
The only positive root is
\begin{align}
\Gamma_{\rm eq}&=\frac{e_2(\lambda)}{2|E_{\rm in}/N E_{\rm trap}-e_1(\lambda)|}\label{gammaeq}\\
+&\sqrt{\frac{e_2^2(\lambda)}{4|E_{\rm in}/N E_{\rm trap}-e_1(\lambda)|^2}+\frac{e_3(\lambda)}{|E_{\rm in}/N E_{\rm trap}-e_1(\lambda)|}},\nonumber 
\end{align}
where $E_{\rm in}/N E_{\rm trap}$ must exceed the zero-temperature equilibrium
energy $e_1(\lambda)$, corresponding to
$\Gamma_{\rm eq}\rightarrow\infty$, and where all coefficients remain positive
over the range of $\lambda$ considered here.

To estimate $\Gamma_{\rm eq}$ without nonequilibrium simulations, we
approximate the input Coulomb-energy contribution in $E_{\rm in}$ by the
average potential energy of the equilibrated beam at zero temperature and
fixed $\lambda$. This amounts to setting
$E_{\rm in}/N=\me \langle (\vb-\zz v_{\rm cm})^2 \rangle_{t_{\rm in}}/2+\me \omega_{\rm trap}^2 \langle R^2 \rangle_{t_{\rm in}} /2+e_1(\lambda)E_{\rm trap}$
in Eq.~(\ref{gammaeq}). The remaining energy contributions can be expressed
in terms of experimentally relevant beam parameters through
$\langle (v_z-v_{\rm cm})^2\rangle_{t_{\rm in}}=\sigma_E^2/\me^2v_{\rm cm}^2$,
$\langle R^2\rangle_{t_{\rm in}}=2\sigma_\perp^2$, and
$\langle v_x^2\rangle_{t_{\rm in}}=\langle v_y^2\rangle_{t_{\rm in}}=v_{\rm cm}^2\sigma_{\rm \alpha}^2$,
where $\sigma_\alpha$ and $\sigma_\perp$ denote the input angular and
transverse beam spreads, respectively. We then obtain
\begin{align}
\frac{\Gamma_{\rm eq} E_{\rm res}}{E_{\rm trap}}= \frac{e_2(\lambda)}{2}+\sqrt{\frac{e^2_2(\lambda)}{4}+ \frac{E_{\rm res} e_3(\lambda)}{E_{\rm trap}}}, \label{gamma_eq}
\end{align}
with $E_{\rm res}=\sigma_E^2/4E_{\rm cm}+2 E_{\rm cm}\sigma_{\rm \alpha}^2+\me \omega_{\rm trap}^2\sigma_\perp^2$
the residual beam energy and $E_{\rm cm}=\me v^2_{\rm cm}/2$ the
center-of-mass energy.

Numerical simulations show that the quadratic coefficient $e_3(\lambda)$ is
negligible compared with $e_2(\lambda)$ over the entire range of $\lambda$
explored in this work (see Fig.~S2 in the SI \cite{SI}). Equation
(\ref{gamma_eq}) can therefore be approximated as
$\Gamma_{\rm eq}\approx e_2(\lambda)E_{\rm trap}/E_{\rm res}$, which reaches a
maximum at the optimal trapping frequency
$\omega_{\rm trap}^{\rm opt}=\sqrt{\sigma_E^2/8 E_{\rm cm}\me \sigma_\perp^2+E_{\rm cm }\sigma_{\alpha}^2/\me \sigma_\perp^2}$.
This frequency yields the maximum Coulomb coupling
$\Gamma^{\rm max}_{\rm eq}\approx [e_2(\lambda)/2.163]\left[\sigma_E^2 \sigma_\perp/4 E_{\rm cm} e^2+2E_{\rm cm} \sigma_\alpha^2 \sigma_\perp/ e^2\right]^{-2/3}$.
Because $\Gamma_{\rm eq}^{\rm max}$ is obtained by assigning the minimum
Coulomb energy to the input beam at fixed $\lambda$, any other input beam with the equal properties ($\sigma_E$, $\sigma_\alpha$, $\sigma_\perp$, $I$, $E_{\rm cm}$) but with a different particle spatial distribution  
satisfies $\Gamma_{\rm eq}\leq\Gamma_{\rm eq}^{\rm max}$.

In Fig.~(\ref{FigA4}), we show the effect of the input-beam properties on the
equilibrated bunch duration $\Delta t$ and duty cycle $\delta$ for 2 keV
(Fig.~\ref{FigA4}b--e) and 100 eV (Fig.~\ref{FigA4}f--i) electrons. In all
cases, we use $\lambda=0.6$, which approximately provides the smallest duty cycle at small
values of $\Gamma_{\rm eq}$ (see Fig.~\ref{Fig5}h) and yields an optimum current $I^{\rm opt}$.

\subsection{CL emission from a strongly-correlated  continuous e-beam}
\label{appendixG}

We now derive the CL emission probability generated by the interaction of $N$ correlated electrons with a sample capable of supporting coherent CL emission \cite{paper149}. We begin with the angle- and frequency-resolved, time-integrated far-field photon-emission probability produced by the fluctuating current density $\jb(\rb,t)=-ev_{\rm cm}\zz\sum_{i=1}^N\delta(z_i+v_{\rm cm}t-z)\delta(\Rb-\Rb_i)$ associated with $N$ electrons propagating with mean velocity $\vb=v_{\rm cm}\zz$. This probability is given by \cite{paper373}
\begin{align}
\frac{d\Gamma_{\rm CL}^{(N)}}{d\Omega_{\rr}d\omega}=\lim_{kr\rightarrow\infty} & \frac{r^2}{4\pi^2\hbar k}\label{cl}\\
&\times {\rm Re}\big\{\big\langle \Eb(\rb,\omega)\times \Bb^*(\rb,\omega)\big\rangle\big\}\cdot \rr,\nonumber
\end{align}
where $k=\omega/c$, $\jb(\rb,\omega)=\int_{-\infty}^\infty dt\,\ee^{\ii \omega t}\jb(\rb,t)$, and the electromagnetic fields are expressed in terms of the Green tensor $G(\rb,\rb',\omega)$ as $\Eb(\rb,\omega)=-4\pi \ii \omega \int d^3\rb ' G(\rb,\rb',\omega)\jb(\rb',\omega)$ and $\Bb(\rb,\omega)=-4\pi c\nabla \times \int d^3\rb' G(\rb,\rb',\omega)\jb(\rb',\omega)$ \cite{paper149}. For a dielectric environment with a linear response described by a spatially and frequency-dependent permittivity $\epsilon(\rb,\omega)$, the Green tensor is implicitly defined by the equation $\nabla_\rb \times \nabla_\rb \times G(\rb,\rb',\omega)-k^2\epsilon(\rb,\omega)G(\rb,\rb')=-\delta(\rb-\rb')/c^2$. The average in Eq.~(\ref{cl}) is taken over the Boltzmann distribution associated with the Hamiltonian $\mathcal{H}_{\rm trap}$.

Because the interaction between electrons and optical excitations depends only weakly on $v_{\rm cm}$ over electron-volt variations in kinetic energy, we neglect fluctuations of this quantity within the beam. Moreover, the characteristic transverse size of a beam in the string phase, $\sqrt{\kB T_{\rm eq} /\me \omega_{\rm trap}^2}\sim 46$ nm ($\kB T_{\rm eq}=1$ meV and $\hbar \omega_{\rm trap}=1$ meV), can be reduced to few nanometers using electron optics. We therefore assume that the CL emission varies negligibly across the transverse extent of the crystalline beam and set $\Rb_i\approx \Rb_0$. Under these assumptions, Eq.~(\ref{cl}) can be recast as
\begin{align}
\frac{d\Gamma_{\rm CL}^{(N)}}{d\Omega_{\rr}d\omega} = \frac{d\Gamma_{\rm CL}}{d\Omega_{\rr}d\omega}\Big[N+\sum_{i\neq i'}^N \Big\langle \ee^{-\ii \omega (z_i-z_{i'})/v_{\rm cm}}\Big\rangle \Big],\label{cl2}
\end{align}
where $d\Gamma_{\rm CL}/d\Omega_{\rr}d\omega$ denotes the single-electron CL emission probability density. It is defined by Eq.~(\ref{cl}), evaluated using the current density $\jb(\rb,\omega)=-e\zz\,\delta(\Rb-\Rb_0)\,\ee^{\ii \omega z/v_{\rm cm}}$.

For a continuous beam, the second term in Eq.~(\ref{cl2}) can be expressed in terms of the density--density correlation function as 
\begin{align}
\sum_{i\neq i'}^N \big\langle \ee^{-\ii \omega (z_i-z_{i'})/v_{\rm cm}}\big\rangle=N^2 \eta(\omega),\nonumber
\end{align} where $\eta(\omega)=2\rho\int_0^\infty dz\, g(z)\,\cos( \omega z/v_{\rm cm})/N$ is the superradiant enhancement factor introduced in the main text. Importantly, in the limit of $N\rightarrow \infty$ at fixed $\rho$, the enhancement factor remains finite and in particular it is bounded by $|\eta(\omega)|\leq 1$. Finally, integrating Eq.~(\ref{cl}) over all solid angles and dividing by the measurement time $\tau_{\rm meas}$ yields the total photon-emission rate per unit frequency, $dR^{(N)}_{\rm CL}/d\omega=(d\Gamma_{\rm CL}/d\omega)[I/e+\tau_{\rm meas} I^2\eta(\omega)/e^2]$. This result exhibits a quadratic dependence on the current $I=eN/\tau_{\rm meas}$, together with a coherent contribution that builds up linearly with the measurement time.

\clearpage

\twocolumngrid
\bibliographystyle{apsrev4-2}

\begin{thebibliography}{97}%
\makeatletter
\providecommand \@ifxundefined [1]{%
 \@ifx{#1\undefined}
}%
\providecommand \@ifnum [1]{%
 \ifnum #1\expandafter \@firstoftwo
 \else \expandafter \@secondoftwo
 \fi
}%
\providecommand \@ifx [1]{%
 \ifx #1\expandafter \@firstoftwo
 \else \expandafter \@secondoftwo
 \fi
}%
\providecommand \natexlab [1]{#1}%
\providecommand \enquote  [1]{``#1''}%
\providecommand \bibnamefont  [1]{#1}%
\providecommand \bibfnamefont [1]{#1}%
\providecommand \citenamefont [1]{#1}%
\providecommand \href@noop [0]{\@secondoftwo}%
\providecommand \href [0]{\begingroup \@sanitize@url \@href}%
\providecommand \@href[1]{\@@startlink{#1}\@@href}%
\providecommand \@@href[1]{\endgroup#1\@@endlink}%
\providecommand \@sanitize@url [0]{\catcode `\\12\catcode `\$12\catcode
  `\&12\catcode `\#12\catcode `\^12\catcode `\_12\catcode `\%12\relax}%
\providecommand \@@startlink[1]{}%
\providecommand \@@endlink[0]{}%
\providecommand \url  [0]{\begingroup\@sanitize@url \@url }%
\providecommand \@url [1]{\endgroup\@href {#1}{\urlprefix }}%
\providecommand \urlprefix  [0]{URL }%
\providecommand \Eprint [0]{\href }%
\providecommand \doibase [0]{https://doi.org/}%
\providecommand \selectlanguage [0]{\@gobble}%
\providecommand \bibinfo  [0]{\@secondoftwo}%
\providecommand \bibfield  [0]{\@secondoftwo}%
\providecommand \translation [1]{[#1]}%
\providecommand \BibitemOpen [0]{}%
\providecommand \bibitemStop [0]{}%
\providecommand \bibitemNoStop [0]{.\EOS\space}%
\providecommand \EOS [0]{\spacefactor3000\relax}%
\providecommand \BibitemShut  [1]{\csname bibitem#1\endcsname}%
\let\auto@bib@innerbib\@empty
\bibitem [{\citenamefont {Morigi}\ \emph {et~al.}(2025)\citenamefont {Morigi},
  \citenamefont {Bollinger}, \citenamefont {Drewsen}, \citenamefont
  {Podolsky},\ and\ \citenamefont {Shimshoni}}]{MBD25_arxiv}%
  \BibitemOpen
  \bibfield  {author} {\bibinfo {author} {\bibfnamefont {G.}~\bibnamefont
  {Morigi}}, \bibinfo {author} {\bibfnamefont {J.}~\bibnamefont {Bollinger}},
  \bibinfo {author} {\bibfnamefont {M.}~\bibnamefont {Drewsen}}, \bibinfo
  {author} {\bibfnamefont {D.}~\bibnamefont {Podolsky}},\ and\ \bibinfo
  {author} {\bibfnamefont {E.}~\bibnamefont {Shimshoni}},\ }\href
  {https://arxiv.orgabs/2508.07374} {\bibinfo {title} {Ion coulomb crystals: an
  exotic form of condensed matter}} (\bibinfo {year} {2025}),\ \Eprint
  {https://arxiv.org/abs/2508.07374} {arXiv:2508.07374 [physics.atom-ph]}
  \BibitemShut {NoStop}%
\bibitem [{\citenamefont {Pagano}\ \emph {et~al.}(2019)\citenamefont {Pagano},
  \citenamefont {Hess}, \citenamefont {Kaplan}, \citenamefont {Tan},
  \citenamefont {Richerme}, \citenamefont {Becker}, \citenamefont
  {Kyprianidis}, \citenamefont {Zhang}, \citenamefont {Birckelbaw},
  \citenamefont {Hernandez}, \citenamefont {Wu},\ and\ \citenamefont
  {Monroe}}]{PHK19}%
  \BibitemOpen
  \bibfield  {author} {\bibinfo {author} {\bibfnamefont {G.}~\bibnamefont
  {Pagano}}, \bibinfo {author} {\bibfnamefont {P.~W.}\ \bibnamefont {Hess}},
  \bibinfo {author} {\bibfnamefont {H.~B.}\ \bibnamefont {Kaplan}}, \bibinfo
  {author} {\bibfnamefont {W.~L.}\ \bibnamefont {Tan}}, \bibinfo {author}
  {\bibfnamefont {P.}~\bibnamefont {Richerme}}, \bibinfo {author}
  {\bibfnamefont {P.}~\bibnamefont {Becker}}, \bibinfo {author} {\bibfnamefont
  {A.}~\bibnamefont {Kyprianidis}}, \bibinfo {author} {\bibfnamefont
  {J.}~\bibnamefont {Zhang}}, \bibinfo {author} {\bibfnamefont
  {E.}~\bibnamefont {Birckelbaw}}, \bibinfo {author} {\bibfnamefont {M.~R.}\
  \bibnamefont {Hernandez}}, \bibinfo {author} {\bibfnamefont {Y.}~\bibnamefont
  {Wu}},\ and\ \bibinfo {author} {\bibfnamefont {C.}~\bibnamefont {Monroe}},\
  }\href {https://doi.org/10.1088/2058-9565/aae0fe} {\bibfield  {journal}
  {\bibinfo  {journal} {Quantum Sci. Technol.}\ }\textbf {\bibinfo {volume}
  {4}},\ \bibinfo {pages} {014004} (\bibinfo {year} {2019})}\BibitemShut
  {NoStop}%
\bibitem [{\citenamefont {Segal}\ and\ \citenamefont {Thompson}(1992)}]{ST92}%
  \BibitemOpen
  \bibfield  {author} {\bibinfo {author} {\bibfnamefont {D.}~\bibnamefont
  {Segal}}\ and\ \bibinfo {author} {\bibfnamefont {R.}~\bibnamefont
  {Thompson}},\ }\href {https://doi.org/https://doi.org/10.1038/357280a0}
  {\bibfield  {journal} {\bibinfo  {journal} {Nature}\ }\textbf {\bibinfo
  {volume} {357}},\ \bibinfo {pages} {280} (\bibinfo {year}
  {1992})}\BibitemShut {NoStop}%
\bibitem [{\citenamefont {Birkl}\ \emph {et~al.}(1992)\citenamefont {Birkl},
  \citenamefont {Kassner},\ and\ \citenamefont {Walther}}]{BKW92}%
  \BibitemOpen
  \bibfield  {author} {\bibinfo {author} {\bibfnamefont {G.}~\bibnamefont
  {Birkl}}, \bibinfo {author} {\bibfnamefont {S.}~\bibnamefont {Kassner}},\
  and\ \bibinfo {author} {\bibfnamefont {H.}~\bibnamefont {Walther}},\ }\href
  {https://doi.org/https://doi.org/10.1038/357310a0} {\bibfield  {journal}
  {\bibinfo  {journal} {Nature}\ }\textbf {\bibinfo {volume} {357}},\ \bibinfo
  {pages} {310} (\bibinfo {year} {1992})}\BibitemShut {NoStop}%
\bibitem [{\citenamefont {Sch\"{a}tz}\ \emph {et~al.}(2001)\citenamefont
  {Sch\"{a}tz}, \citenamefont {Schramm},\ and\ \citenamefont {Habs}}]{SSH01_2}%
  \BibitemOpen
  \bibfield  {author} {\bibinfo {author} {\bibfnamefont {T.}~\bibnamefont
  {Sch\"{a}tz}}, \bibinfo {author} {\bibfnamefont {U.}~\bibnamefont
  {Schramm}},\ and\ \bibinfo {author} {\bibfnamefont {D.}~\bibnamefont
  {Habs}},\ }\href {https://doi.org/https://doi.org/10.1038/35089045}
  {\bibfield  {journal} {\bibinfo  {journal} {Nature}\ }\textbf {\bibinfo
  {volume} {412}},\ \bibinfo {pages} {717} (\bibinfo {year}
  {2001})}\BibitemShut {NoStop}%
\bibitem [{\citenamefont {Schramm}\ and\ \citenamefont {Habs}(2004)}]{SH04}%
  \BibitemOpen
  \bibfield  {author} {\bibinfo {author} {\bibfnamefont {U.}~\bibnamefont
  {Schramm}}\ and\ \bibinfo {author} {\bibfnamefont {D.}~\bibnamefont {Habs}},\
  }\href {https://doi.org/10.1016/j.ppnp.2004.03.002} {\bibfield  {journal}
  {\bibinfo  {journal} {Progress in Particle and Nuclear Physics}\ }\textbf
  {\bibinfo {volume} {53}},\ \bibinfo {pages} {583} (\bibinfo {year}
  {2004})}\BibitemShut {NoStop}%
\bibitem [{\citenamefont {Hansen}(1973)}]{H1973_2}%
  \BibitemOpen
  \bibfield  {author} {\bibinfo {author} {\bibfnamefont {J.~P.}\ \bibnamefont
  {Hansen}},\ }\href {https://doi.org/https://doi.org/10.1103/PhysRevA.8.3096}
  {\bibfield  {journal} {\bibinfo  {journal} {Phys.\ Rev.\ A}\ }\textbf
  {\bibinfo {volume} {8}},\ \bibinfo {pages} {3096} (\bibinfo {year}
  {1973})}\BibitemShut {NoStop}%
\bibitem [{\citenamefont {Schiffer}(1996)}]{S96_2}%
  \BibitemOpen
  \bibfield  {author} {\bibinfo {author} {\bibfnamefont {J.~P.}\ \bibnamefont
  {Schiffer}},\ }\href {https://doi.org/https://doi.org/10.1007/BF02317344}
  {\bibfield  {journal} {\bibinfo  {journal} {Hyperfine Interactions}\ }\textbf
  {\bibinfo {volume} {103}},\ \bibinfo {pages} {71} (\bibinfo {year}
  {1996})}\BibitemShut {NoStop}%
\bibitem [{\citenamefont {Kim}\ \emph {et~al.}(2022)\citenamefont {Kim},
  \citenamefont {Bang}, \citenamefont {Lim}, \citenamefont {Lee}, \citenamefont
  {Hyun}, \citenamefont {Lee}, \citenamefont {Lee}, \citenamefont {Denlinger},
  \citenamefont {Huh}, \citenamefont {Kim}, \citenamefont {Song}, \citenamefont
  {Seo}, \citenamefont {Thapa}, \citenamefont {Kim}, \citenamefont {Lee},
  \citenamefont {Kim},\ and\ \citenamefont {Kim}}]{KBL22}%
  \BibitemOpen
  \bibfield  {author} {\bibinfo {author} {\bibfnamefont {S.}~\bibnamefont
  {Kim}}, \bibinfo {author} {\bibfnamefont {J.}~\bibnamefont {Bang}}, \bibinfo
  {author} {\bibfnamefont {C.-Y.}\ \bibnamefont {Lim}}, \bibinfo {author}
  {\bibfnamefont {S.~Y.}\ \bibnamefont {Lee}}, \bibinfo {author} {\bibfnamefont
  {J.}~\bibnamefont {Hyun}}, \bibinfo {author} {\bibfnamefont {G.}~\bibnamefont
  {Lee}}, \bibinfo {author} {\bibfnamefont {Y.}~\bibnamefont {Lee}}, \bibinfo
  {author} {\bibfnamefont {J.~D.}\ \bibnamefont {Denlinger}}, \bibinfo {author}
  {\bibfnamefont {S.}~\bibnamefont {Huh}}, \bibinfo {author} {\bibfnamefont
  {C.}~\bibnamefont {Kim}}, \bibinfo {author} {\bibfnamefont {S.~Y.}\
  \bibnamefont {Song}}, \bibinfo {author} {\bibfnamefont {J.}~\bibnamefont
  {Seo}}, \bibinfo {author} {\bibfnamefont {D.}~\bibnamefont {Thapa}}, \bibinfo
  {author} {\bibfnamefont {S.-G.}\ \bibnamefont {Kim}}, \bibinfo {author}
  {\bibfnamefont {Y.~H.}\ \bibnamefont {Lee}}, \bibinfo {author} {\bibfnamefont
  {Y.}~\bibnamefont {Kim}},\ and\ \bibinfo {author} {\bibfnamefont {S.~W.}\
  \bibnamefont {Kim}},\ }\href {https://doi.org/10.1038/s41563-022-01353-8}
  {\bibfield  {journal} {\bibinfo  {journal} {Nat.\ Mater.}\ }\textbf {\bibinfo
  {volume} {21}},\ \bibinfo {pages} {1269–1274} (\bibinfo {year}
  {2022})}\BibitemShut {NoStop}%
\bibitem [{\citenamefont {Wigner}(1934)}]{W1934}%
  \BibitemOpen
  \bibfield  {author} {\bibinfo {author} {\bibfnamefont {E.}~\bibnamefont
  {Wigner}},\ }\href {https://doi.org/https://doi.org/10.1103/PhysRev.46.1002}
  {\bibfield  {journal} {\bibinfo  {journal} {Phys.\ Rev.}\ }\textbf {\bibinfo
  {volume} {46}},\ \bibinfo {pages} {1002} (\bibinfo {year}
  {1934})}\BibitemShut {NoStop}%
\bibitem [{\citenamefont {Deshpande}\ and\ \citenamefont
  {Bockrath}(2008)}]{DB08}%
  \BibitemOpen
  \bibfield  {author} {\bibinfo {author} {\bibfnamefont {V.~V.}\ \bibnamefont
  {Deshpande}}\ and\ \bibinfo {author} {\bibfnamefont {M.}~\bibnamefont
  {Bockrath}},\ }\href {https://doi.org/https://doi.org/10.1038/nphys895}
  {\bibfield  {journal} {\bibinfo  {journal} {Nat.\ Phys.}\ }\textbf {\bibinfo
  {volume} {4}},\ \bibinfo {pages} {314} (\bibinfo {year} {2008})}\BibitemShut
  {NoStop}%
\bibitem [{\citenamefont {Shapir}\ \emph {et~al.}(2019)\citenamefont {Shapir},
  \citenamefont {Hamo}, \citenamefont {Pecker}, \citenamefont {Moca},
  \citenamefont {Legeza}, \citenamefont {Zarand},\ and\ \citenamefont
  {Ilani}}]{SHP19_2}%
  \BibitemOpen
  \bibfield  {author} {\bibinfo {author} {\bibfnamefont {I.}~\bibnamefont
  {Shapir}}, \bibinfo {author} {\bibfnamefont {A.}~\bibnamefont {Hamo}},
  \bibinfo {author} {\bibfnamefont {S.}~\bibnamefont {Pecker}}, \bibinfo
  {author} {\bibfnamefont {C.~P.}\ \bibnamefont {Moca}}, \bibinfo {author}
  {\bibfnamefont {O.}~\bibnamefont {Legeza}}, \bibinfo {author} {\bibfnamefont
  {G.}~\bibnamefont {Zarand}},\ and\ \bibinfo {author} {\bibfnamefont
  {S.}~\bibnamefont {Ilani}},\ }\href
  {https://doi.org/https://doi.org/10.1126/science.aat0905} {\bibfield
  {journal} {\bibinfo  {journal} {Science}\ }\textbf {\bibinfo {volume}
  {364}},\ \bibinfo {pages} {870} (\bibinfo {year} {2019})}\BibitemShut
  {NoStop}%
\bibitem [{\citenamefont {Li}\ \emph {et~al.}(2024)\citenamefont {Li},
  \citenamefont {Xiang}, \citenamefont {Reddy}, \citenamefont {Devakul},
  \citenamefont {Sailus}, \citenamefont {Banerjee}, \citenamefont {Taniguchi},
  \citenamefont {Watanabe}, \citenamefont {Tongay}, \citenamefont {Zettl},
  \citenamefont {Fu}, \citenamefont {Crommie},\ and\ \citenamefont
  {Wang}}]{LXR24}%
  \BibitemOpen
  \bibfield  {author} {\bibinfo {author} {\bibfnamefont {H.}~\bibnamefont
  {Li}}, \bibinfo {author} {\bibfnamefont {Z.}~\bibnamefont {Xiang}}, \bibinfo
  {author} {\bibfnamefont {A.~P.}\ \bibnamefont {Reddy}}, \bibinfo {author}
  {\bibfnamefont {T.}~\bibnamefont {Devakul}}, \bibinfo {author} {\bibfnamefont
  {R.}~\bibnamefont {Sailus}}, \bibinfo {author} {\bibfnamefont
  {R.}~\bibnamefont {Banerjee}}, \bibinfo {author} {\bibfnamefont
  {T.}~\bibnamefont {Taniguchi}}, \bibinfo {author} {\bibfnamefont
  {K.}~\bibnamefont {Watanabe}}, \bibinfo {author} {\bibfnamefont
  {S.}~\bibnamefont {Tongay}}, \bibinfo {author} {\bibfnamefont
  {A.}~\bibnamefont {Zettl}}, \bibinfo {author} {\bibfnamefont
  {L.}~\bibnamefont {Fu}}, \bibinfo {author} {\bibfnamefont {M.~F.}\
  \bibnamefont {Crommie}},\ and\ \bibinfo {author} {\bibfnamefont
  {F.}~\bibnamefont {Wang}},\ }\href
  {https://doi.org/https://doi.org/10.1126/science.adk1348} {\bibfield
  {journal} {\bibinfo  {journal} {Science}\ }\textbf {\bibinfo {volume}
  {385}},\ \bibinfo {pages} {86} (\bibinfo {year} {2024})}\BibitemShut
  {NoStop}%
\bibitem [{\citenamefont {Seiler}\ \emph {et~al.}(2025)\citenamefont {Seiler},
  \citenamefont {Statz}, \citenamefont {Eckel}, \citenamefont {Weimer},
  \citenamefont {P\"{o}hls}, \citenamefont {Watanabe}, \citenamefont
  {Taniguchi}, \citenamefont {Zhang},\ and\ \citenamefont {Weitz}}]{SSE25}%
  \BibitemOpen
  \bibfield  {author} {\bibinfo {author} {\bibfnamefont {A.~M.}\ \bibnamefont
  {Seiler}}, \bibinfo {author} {\bibfnamefont {M.}~\bibnamefont {Statz}},
  \bibinfo {author} {\bibfnamefont {C.}~\bibnamefont {Eckel}}, \bibinfo
  {author} {\bibfnamefont {I.}~\bibnamefont {Weimer}}, \bibinfo {author}
  {\bibfnamefont {J.}~\bibnamefont {P\"{o}hls}}, \bibinfo {author}
  {\bibfnamefont {K.}~\bibnamefont {Watanabe}}, \bibinfo {author}
  {\bibfnamefont {T.}~\bibnamefont {Taniguchi}}, \bibinfo {author}
  {\bibfnamefont {F.}~\bibnamefont {Zhang}},\ and\ \bibinfo {author}
  {\bibfnamefont {R.~T.}\ \bibnamefont {Weitz}},\ }\href
  {https://doi.org/https://doi.org/10.1038/s41467-025-64587-5} {\bibfield
  {journal} {\bibinfo  {journal} {Nat.\ Commun.}\ }\textbf {\bibinfo {volume}
  {16}},\ \bibinfo {pages} {8921} (\bibinfo {year} {2025})}\BibitemShut
  {NoStop}%
\bibitem [{\citenamefont {Fletcher}\ \emph {et~al.}(2023)\citenamefont
  {Fletcher}, \citenamefont {Park}, \citenamefont {Ryu}, \citenamefont {See},
  \citenamefont {Griffiths}, \citenamefont {Jones}, \citenamefont {Farrer},
  \citenamefont {Ritchie}, \citenamefont {Sim},\ and\ \citenamefont
  {Kataoka}}]{FPR23}%
  \BibitemOpen
  \bibfield  {author} {\bibinfo {author} {\bibfnamefont {J.~D.}\ \bibnamefont
  {Fletcher}}, \bibinfo {author} {\bibfnamefont {W.}~\bibnamefont {Park}},
  \bibinfo {author} {\bibfnamefont {S.}~\bibnamefont {Ryu}}, \bibinfo {author}
  {\bibfnamefont {P.}~\bibnamefont {See}}, \bibinfo {author} {\bibfnamefont
  {J.~P.}\ \bibnamefont {Griffiths}}, \bibinfo {author} {\bibfnamefont
  {G.~A.~C.}\ \bibnamefont {Jones}}, \bibinfo {author} {\bibfnamefont
  {I.}~\bibnamefont {Farrer}}, \bibinfo {author} {\bibfnamefont {D.~A.}\
  \bibnamefont {Ritchie}}, \bibinfo {author} {\bibfnamefont {H.-S.}\
  \bibnamefont {Sim}},\ and\ \bibinfo {author} {\bibfnamefont {M.}~\bibnamefont
  {Kataoka}},\ }\href {https://doi.org/10.1038/s41565-023-01369-4} {\bibfield
  {journal} {\bibinfo  {journal} {Nature Nanotechnology}\ }\textbf {\bibinfo
  {volume} {18}},\ \bibinfo {pages} {727} (\bibinfo {year} {2023})}\BibitemShut
  {NoStop}%
\bibitem [{\citenamefont {Ubbelohde}\ \emph {et~al.}(2023)\citenamefont
  {Ubbelohde}, \citenamefont {Freise}, \citenamefont {Pavlovska}, \citenamefont
  {Silvestrov}, \citenamefont {Recher}, \citenamefont {Kokainis}, \citenamefont
  {Barinovs}, \citenamefont {Hohls}, \citenamefont {Weimann}, \citenamefont
  {Pierz},\ and\ \citenamefont {Kashcheyevs}}]{UFP23}%
  \BibitemOpen
  \bibfield  {author} {\bibinfo {author} {\bibfnamefont {N.}~\bibnamefont
  {Ubbelohde}}, \bibinfo {author} {\bibfnamefont {L.}~\bibnamefont {Freise}},
  \bibinfo {author} {\bibfnamefont {E.}~\bibnamefont {Pavlovska}}, \bibinfo
  {author} {\bibfnamefont {P.~G.}\ \bibnamefont {Silvestrov}}, \bibinfo
  {author} {\bibfnamefont {P.}~\bibnamefont {Recher}}, \bibinfo {author}
  {\bibfnamefont {M.}~\bibnamefont {Kokainis}}, \bibinfo {author}
  {\bibfnamefont {G.}~\bibnamefont {Barinovs}}, \bibinfo {author}
  {\bibfnamefont {F.}~\bibnamefont {Hohls}}, \bibinfo {author} {\bibfnamefont
  {T.}~\bibnamefont {Weimann}}, \bibinfo {author} {\bibfnamefont
  {K.}~\bibnamefont {Pierz}},\ and\ \bibinfo {author} {\bibfnamefont
  {V.}~\bibnamefont {Kashcheyevs}},\ }\href
  {https://doi.org/10.1038/s41565-023-01370-x} {\bibfield  {journal} {\bibinfo
  {journal} {Nature Nanotechnology}\ }\textbf {\bibinfo {volume} {18}},\
  \bibinfo {pages} {733} (\bibinfo {year} {2023})}\BibitemShut {NoStop}%
\bibitem [{\citenamefont {Wang}\ \emph {et~al.}(2023)\citenamefont {Wang},
  \citenamefont {Edlbauer}, \citenamefont {Richard}, \citenamefont {Ota},
  \citenamefont {Park}, \citenamefont {Shim}, \citenamefont {Ludwig},
  \citenamefont {Wieck}, \citenamefont {Sim}, \citenamefont {Urdampilleta},
  \citenamefont {Meunier}, \citenamefont {Kodera}, \citenamefont {Kaneko},
  \citenamefont {Sellier}, \citenamefont {Waintal}, \citenamefont {Takada},\
  and\ \citenamefont {B{\"a}uerle}}]{WER23}%
  \BibitemOpen
  \bibfield  {author} {\bibinfo {author} {\bibfnamefont {J.}~\bibnamefont
  {Wang}}, \bibinfo {author} {\bibfnamefont {H.}~\bibnamefont {Edlbauer}},
  \bibinfo {author} {\bibfnamefont {A.}~\bibnamefont {Richard}}, \bibinfo
  {author} {\bibfnamefont {S.}~\bibnamefont {Ota}}, \bibinfo {author}
  {\bibfnamefont {W.}~\bibnamefont {Park}}, \bibinfo {author} {\bibfnamefont
  {J.}~\bibnamefont {Shim}}, \bibinfo {author} {\bibfnamefont {A.}~\bibnamefont
  {Ludwig}}, \bibinfo {author} {\bibfnamefont {A.~D.}\ \bibnamefont {Wieck}},
  \bibinfo {author} {\bibfnamefont {H.-S.}\ \bibnamefont {Sim}}, \bibinfo
  {author} {\bibfnamefont {M.}~\bibnamefont {Urdampilleta}}, \bibinfo {author}
  {\bibfnamefont {T.}~\bibnamefont {Meunier}}, \bibinfo {author} {\bibfnamefont
  {T.}~\bibnamefont {Kodera}}, \bibinfo {author} {\bibfnamefont {N.-H.}\
  \bibnamefont {Kaneko}}, \bibinfo {author} {\bibfnamefont {H.}~\bibnamefont
  {Sellier}}, \bibinfo {author} {\bibfnamefont {X.}~\bibnamefont {Waintal}},
  \bibinfo {author} {\bibfnamefont {S.}~\bibnamefont {Takada}},\ and\ \bibinfo
  {author} {\bibfnamefont {C.}~\bibnamefont {B{\"a}uerle}},\ }\href
  {https://doi.org/10.1038/s41565-023-01368-5} {\bibfield  {journal} {\bibinfo
  {journal} {Nature Nanotechnology}\ }\textbf {\bibinfo {volume} {18}},\
  \bibinfo {pages} {721} (\bibinfo {year} {2023})}\BibitemShut {NoStop}%
\bibitem [{\citenamefont {Shaju}\ \emph {et~al.}(2025)\citenamefont {Shaju},
  \citenamefont {Pavlovska}, \citenamefont {Suba}, \citenamefont {Wang},
  \citenamefont {Ouacel}, \citenamefont {Vasselon}, \citenamefont {Aluffi},
  \citenamefont {Mazzella}, \citenamefont {Geffroy}, \citenamefont {Ludwig},
  \citenamefont {Wieck}, \citenamefont {Urdampilleta}, \citenamefont
  {B{\"a}uerle}, \citenamefont {Kashcheyevs},\ and\ \citenamefont
  {Sellier}}]{SPSW25}%
  \BibitemOpen
  \bibfield  {author} {\bibinfo {author} {\bibfnamefont {J.}~\bibnamefont
  {Shaju}}, \bibinfo {author} {\bibfnamefont {E.}~\bibnamefont {Pavlovska}},
  \bibinfo {author} {\bibfnamefont {R.}~\bibnamefont {Suba}}, \bibinfo {author}
  {\bibfnamefont {J.}~\bibnamefont {Wang}}, \bibinfo {author} {\bibfnamefont
  {S.}~\bibnamefont {Ouacel}}, \bibinfo {author} {\bibfnamefont
  {T.}~\bibnamefont {Vasselon}}, \bibinfo {author} {\bibfnamefont
  {M.}~\bibnamefont {Aluffi}}, \bibinfo {author} {\bibfnamefont
  {L.}~\bibnamefont {Mazzella}}, \bibinfo {author} {\bibfnamefont
  {C.}~\bibnamefont {Geffroy}}, \bibinfo {author} {\bibfnamefont
  {A.}~\bibnamefont {Ludwig}}, \bibinfo {author} {\bibfnamefont {A.~D.}\
  \bibnamefont {Wieck}}, \bibinfo {author} {\bibfnamefont {M.}~\bibnamefont
  {Urdampilleta}}, \bibinfo {author} {\bibfnamefont {C.}~\bibnamefont
  {B{\"a}uerle}}, \bibinfo {author} {\bibfnamefont {V.}~\bibnamefont
  {Kashcheyevs}},\ and\ \bibinfo {author} {\bibfnamefont {H.}~\bibnamefont
  {Sellier}},\ }\href {https://doi.org/10.1038/s41586-025-09139-z} {\bibfield
  {journal} {\bibinfo  {journal} {Nature}\ }\textbf {\bibinfo {volume} {642}},\
  \bibinfo {pages} {928} (\bibinfo {year} {2025})}\BibitemShut {NoStop}%
\bibitem [{\citenamefont {Williams}\ and\ \citenamefont {Carter}(2009)}]{WC09}%
  \BibitemOpen
  \bibfield  {author} {\bibinfo {author} {\bibfnamefont {D.~B.}\ \bibnamefont
  {Williams}}\ and\ \bibinfo {author} {\bibfnamefont {C.~B.}\ \bibnamefont
  {Carter}},\ }\href {https://doi.org/10.1007/978-0-387-76501-3} {\emph
  {\bibinfo {title} {Transmission Electron Microscopy: A Textbook for Materials
  Science}}},\ \bibinfo {edition} {2nd}\ ed.\ (\bibinfo  {publisher}
  {Springer},\ \bibinfo {address} {New York, NY},\ \bibinfo {year}
  {2009})\BibitemShut {NoStop}%
\bibitem [{\citenamefont {Egerton}(2011)}]{E11}%
  \BibitemOpen
  \bibfield  {author} {\bibinfo {author} {\bibfnamefont {R.~F.}\ \bibnamefont
  {Egerton}},\ }\href {https://doi.org/10.1007/978-1-4419-9583-4} {\emph
  {\bibinfo {title} {Electron Energy-Loss Spectroscopy in the Electron
  Microscope}}},\ \bibinfo {edition} {3rd}\ ed.\ (\bibinfo  {publisher}
  {Springer},\ \bibinfo {address} {New York, NY},\ \bibinfo {year}
  {2011})\BibitemShut {NoStop}%
\bibitem [{\citenamefont {{Garc\'{\i}a de Abajo}}(2010)}]{paper149}%
  \BibitemOpen
  \bibfield  {author} {\bibinfo {author} {\bibfnamefont {F.~J.}\ \bibnamefont
  {{Garc\'{\i}a de Abajo}}},\ }\href
  {https://doi.org/10.1103/RevModPhys.82.209} {\bibfield  {journal} {\bibinfo
  {journal} {Rev.\ Mod.\ Phys.}\ }\textbf {\bibinfo {volume} {82}},\ \bibinfo
  {pages} {209} (\bibinfo {year} {2010})}\BibitemShut {NoStop}%
\bibitem [{\citenamefont {Boersch}(1954)}]{B1954_2}%
  \BibitemOpen
  \bibfield  {author} {\bibinfo {author} {\bibfnamefont {H.}~\bibnamefont
  {Boersch}},\ }\href@noop {} {\bibfield  {journal} {\bibinfo  {journal}
  {Zeitschrift ftir Physik}\ }\textbf {\bibinfo {volume} {139}},\ \bibinfo
  {pages} {115} (\bibinfo {year} {1954})}\BibitemShut {NoStop}%
\bibitem [{\citenamefont {Knauer}(1979)}]{K1979}%
  \BibitemOpen
  \bibfield  {author} {\bibinfo {author} {\bibfnamefont {W.}~\bibnamefont
  {Knauer}},\ }\href {https://doi.org/10.1116/1.570271} {\bibfield  {journal}
  {\bibinfo  {journal} {J. Vac. Sci. Technol.}\ }\textbf {\bibinfo {volume}
  {16}},\ \bibinfo {pages} {1676–1679} (\bibinfo {year} {1979})}\BibitemShut
  {NoStop}%
\bibitem [{\citenamefont {Kuwahara}\ \emph {et~al.}(2016)\citenamefont
  {Kuwahara}, \citenamefont {Nambo}, \citenamefont {Aoki}, \citenamefont
  {Sameshima}, \citenamefont {Jin}, \citenamefont {Ujihara}, \citenamefont
  {Asano}, \citenamefont {Saitoh}, \citenamefont {Takeda},\ and\ \citenamefont
  {Tanaka}}]{KNA16}%
  \BibitemOpen
  \bibfield  {author} {\bibinfo {author} {\bibfnamefont {M.}~\bibnamefont
  {Kuwahara}}, \bibinfo {author} {\bibfnamefont {Y.}~\bibnamefont {Nambo}},
  \bibinfo {author} {\bibfnamefont {K.}~\bibnamefont {Aoki}}, \bibinfo {author}
  {\bibfnamefont {K.}~\bibnamefont {Sameshima}}, \bibinfo {author}
  {\bibfnamefont {X.}~\bibnamefont {Jin}}, \bibinfo {author} {\bibfnamefont
  {T.}~\bibnamefont {Ujihara}}, \bibinfo {author} {\bibfnamefont
  {H.}~\bibnamefont {Asano}}, \bibinfo {author} {\bibfnamefont
  {K.}~\bibnamefont {Saitoh}}, \bibinfo {author} {\bibfnamefont
  {Y.}~\bibnamefont {Takeda}},\ and\ \bibinfo {author} {\bibfnamefont
  {N.}~\bibnamefont {Tanaka}},\ }\href {https://doi.org/10.1063/1.4955457}
  {\bibfield  {journal} {\bibinfo  {journal} {Appl.\ Phys.\ Lett.}\ }\textbf
  {\bibinfo {volume} {109}},\ \bibinfo {pages} {013108} (\bibinfo {year}
  {2016})}\BibitemShut {NoStop}%
\bibitem [{\citenamefont {Loeffler}(1969)}]{L1969}%
  \BibitemOpen
  \bibfield  {author} {\bibinfo {author} {\bibfnamefont {K.~H.}\ \bibnamefont
  {Loeffler}},\ }\href@noop {} {\bibfield  {journal} {\bibinfo  {journal}
  {Zeitschrift f{\"u}r Angewandte Physik}\ }\textbf {\bibinfo {volume} {27}},\
  \bibinfo {pages} {145} (\bibinfo {year} {1969})}\BibitemShut {NoStop}%
\bibitem [{\citenamefont {Jansen}(1988)}]{J1988_2}%
  \BibitemOpen
  \bibfield  {author} {\bibinfo {author} {\bibfnamefont {G.~H.}\ \bibnamefont
  {Jansen}},\ }\emph {\bibinfo {title} {Coulomb Interactions in Particle
  Beams}},\ \href@noop {} {Ph.D. thesis},\ \bibinfo  {school} {Delft University
  of Technology}, \bibinfo {address} {Delft, Netherlands} (\bibinfo {year}
  {1988})\BibitemShut {NoStop}%
\bibitem [{\citenamefont {Orloff}(2009)}]{O09}%
  \BibitemOpen
  \bibfield  {author} {\bibinfo {author} {\bibfnamefont {J.}~\bibnamefont
  {Orloff}},\ }\href@noop {} {\emph {\bibinfo {title} {Handbook of Charged
  Particle Optics}}},\ \bibinfo {edition} {2nd}\ ed.\ (\bibinfo  {publisher}
  {CRC Press},\ \bibinfo {address} {Boca Raton},\ \bibinfo {year}
  {2009})\BibitemShut {NoStop}%
\bibitem [{\citenamefont {Keramati}\ \emph {et~al.}(2021)\citenamefont
  {Keramati}, \citenamefont {Brunner}, \citenamefont {Gay},\ and\ \citenamefont
  {Batelaan}}]{KBG21}%
  \BibitemOpen
  \bibfield  {author} {\bibinfo {author} {\bibfnamefont {S.}~\bibnamefont
  {Keramati}}, \bibinfo {author} {\bibfnamefont {W.}~\bibnamefont {Brunner}},
  \bibinfo {author} {\bibfnamefont {T.~J.}\ \bibnamefont {Gay}},\ and\ \bibinfo
  {author} {\bibfnamefont {H.}~\bibnamefont {Batelaan}},\ }\href
  {https://doi.org/10.1103/PhysRevLett.127.180602} {\bibfield  {journal}
  {\bibinfo  {journal} {Physical Review Letters}\ }\textbf {\bibinfo {volume}
  {127}},\ \bibinfo {pages} {180602} (\bibinfo {year} {2021})}\BibitemShut
  {NoStop}%
\bibitem [{\citenamefont {Kuwahara}\ \emph {et~al.}(2021)\citenamefont
  {Kuwahara}, \citenamefont {Yoshida}, \citenamefont {Nagata}, \citenamefont
  {Nakakura}, \citenamefont {Furui}, \citenamefont {Ishida}, \citenamefont
  {Saitoh}, \citenamefont {Ujihara},\ and\ \citenamefont {Tanaka}}]{KYN21}%
  \BibitemOpen
  \bibfield  {author} {\bibinfo {author} {\bibfnamefont {M.}~\bibnamefont
  {Kuwahara}}, \bibinfo {author} {\bibfnamefont {Y.}~\bibnamefont {Yoshida}},
  \bibinfo {author} {\bibfnamefont {W.}~\bibnamefont {Nagata}}, \bibinfo
  {author} {\bibfnamefont {K.}~\bibnamefont {Nakakura}}, \bibinfo {author}
  {\bibfnamefont {M.}~\bibnamefont {Furui}}, \bibinfo {author} {\bibfnamefont
  {T.}~\bibnamefont {Ishida}}, \bibinfo {author} {\bibfnamefont
  {K.}~\bibnamefont {Saitoh}}, \bibinfo {author} {\bibfnamefont
  {T.}~\bibnamefont {Ujihara}},\ and\ \bibinfo {author} {\bibfnamefont
  {N.}~\bibnamefont {Tanaka}},\ }\href
  {https://doi.org/10.1103/PhysRevLett.126.125501} {\bibfield  {journal}
  {\bibinfo  {journal} {Physical Review Letters}\ }\textbf {\bibinfo {volume}
  {126}},\ \bibinfo {pages} {125501} (\bibinfo {year} {2021})}\BibitemShut
  {NoStop}%
\bibitem [{\citenamefont {Borrelli}\ \emph {et~al.}(2024)\citenamefont
  {Borrelli}, \citenamefont {{de Raadt}}, \citenamefont {{van der Geer}},
  \citenamefont {Mutsaers}, \citenamefont {{van Leeuwen}},\ and\ \citenamefont
  {Luiten}}]{BRG24}%
  \BibitemOpen
  \bibfield  {author} {\bibinfo {author} {\bibfnamefont {S.}~\bibnamefont
  {Borrelli}}, \bibinfo {author} {\bibfnamefont {T.~C.~H.}\ \bibnamefont {{de
  Raadt}}}, \bibinfo {author} {\bibfnamefont {S.~B.}\ \bibnamefont {{van der
  Geer}}}, \bibinfo {author} {\bibfnamefont {P.~H.~A.}\ \bibnamefont
  {Mutsaers}}, \bibinfo {author} {\bibfnamefont {K.~A.~H.}\ \bibnamefont {{van
  Leeuwen}}},\ and\ \bibinfo {author} {\bibfnamefont {O.~J.}\ \bibnamefont
  {Luiten}},\ }\href {https://doi.org/10.1103/PhysRevLett.132.115001}
  {\bibfield  {journal} {\bibinfo  {journal} {Physical Review Letters}\
  }\textbf {\bibinfo {volume} {132}},\ \bibinfo {pages} {115001} (\bibinfo
  {year} {2024})}\BibitemShut {NoStop}%
\bibitem [{\citenamefont {Haindl}\ \emph {et~al.}(2023)\citenamefont {Haindl},
  \citenamefont {Feist}, \citenamefont {Domr\"ose}, \citenamefont {M\"oller},
  \citenamefont {Gaida}, \citenamefont {Yalunin},\ and\ \citenamefont
  {Ropers}}]{HFD23}%
  \BibitemOpen
  \bibfield  {author} {\bibinfo {author} {\bibfnamefont {R.}~\bibnamefont
  {Haindl}}, \bibinfo {author} {\bibfnamefont {A.}~\bibnamefont {Feist}},
  \bibinfo {author} {\bibfnamefont {T.}~\bibnamefont {Domr\"ose}}, \bibinfo
  {author} {\bibfnamefont {M.}~\bibnamefont {M\"oller}}, \bibinfo {author}
  {\bibfnamefont {J.~H.}\ \bibnamefont {Gaida}}, \bibinfo {author}
  {\bibfnamefont {S.~V.}\ \bibnamefont {Yalunin}},\ and\ \bibinfo {author}
  {\bibfnamefont {C.}~\bibnamefont {Ropers}},\ }\href
  {https://doi.org/10.1038/s41567-023-02067-7} {\bibfield  {journal} {\bibinfo
  {journal} {Nat.\ Phys.}\ }\textbf {\bibinfo {volume} {19}},\ \bibinfo {pages}
  {1410} (\bibinfo {year} {2023})}\BibitemShut {NoStop}%
\bibitem [{\citenamefont {Kuttruff}\ \emph {et~al.}(2024)\citenamefont
  {Kuttruff}, \citenamefont {Nabben}, \citenamefont {Zimmermann}, \citenamefont
  {Ryabov},\ and\ \citenamefont {Baum}}]{KNZ24}%
  \BibitemOpen
  \bibfield  {author} {\bibinfo {author} {\bibfnamefont {J.}~\bibnamefont
  {Kuttruff}}, \bibinfo {author} {\bibfnamefont {D.}~\bibnamefont {Nabben}},
  \bibinfo {author} {\bibfnamefont {A.-C.}\ \bibnamefont {Zimmermann}},
  \bibinfo {author} {\bibfnamefont {A.}~\bibnamefont {Ryabov}},\ and\ \bibinfo
  {author} {\bibfnamefont {P.}~\bibnamefont {Baum}},\ }\href
  {https://doi.org/10.1126/sciadv.adl6543} {\bibfield  {journal} {\bibinfo
  {journal} {Science Advances}\ }\textbf {\bibinfo {volume} {10}},\ \bibinfo
  {pages} {eadl6543} (\bibinfo {year} {2024})}\BibitemShut {NoStop}%
\bibitem [{\citenamefont {Meier}\ \emph {et~al.}(2023)\citenamefont {Meier},
  \citenamefont {Heimerl},\ and\ \citenamefont {Hommelhoff}}]{MHH23}%
  \BibitemOpen
  \bibfield  {author} {\bibinfo {author} {\bibfnamefont {S.}~\bibnamefont
  {Meier}}, \bibinfo {author} {\bibfnamefont {J.}~\bibnamefont {Heimerl}},\
  and\ \bibinfo {author} {\bibfnamefont {P.}~\bibnamefont {Hommelhoff}},\
  }\href {https://doi.org/10.1038/s41567-023-02059-7} {\bibfield  {journal}
  {\bibinfo  {journal} {Nat.\ Phys.}\ }\textbf {\bibinfo {volume} {19}},\
  \bibinfo {pages} {1402} (\bibinfo {year} {2023})}\BibitemShut {NoStop}%
\bibitem [{\citenamefont {Hergert}\ \emph {et~al.}(2024)\citenamefont
  {Hergert}, \citenamefont {Lampe}, \citenamefont {W{\"o}ste},\ and\
  \citenamefont {Lienau}}]{HLW24}%
  \BibitemOpen
  \bibfield  {author} {\bibinfo {author} {\bibfnamefont {G.}~\bibnamefont
  {Hergert}}, \bibinfo {author} {\bibfnamefont {R.}~\bibnamefont {Lampe}},
  \bibinfo {author} {\bibfnamefont {A.}~\bibnamefont {W{\"o}ste}},\ and\
  \bibinfo {author} {\bibfnamefont {C.}~\bibnamefont {Lienau}},\ }\href
  {https://doi.org/10.1021/acs.nanolett.4c03240} {\bibfield  {journal}
  {\bibinfo  {journal} {Nano Letters}\ }\textbf {\bibinfo {volume} {24}},\
  \bibinfo {pages} {11067} (\bibinfo {year} {2024})}\BibitemShut {NoStop}%
\bibitem [{\citenamefont {Schr\"{o}der}\ \emph {et~al.}(2025)\citenamefont
  {Schr\"{o}der}, \citenamefont {Wendeln}, \citenamefont {Weber}, \citenamefont
  {Mukai}, \citenamefont {Kohno},\ and\ \citenamefont {Sch\"{a}fer}}]{SWW25}%
  \BibitemOpen
  \bibfield  {author} {\bibinfo {author} {\bibfnamefont {A.}~\bibnamefont
  {Schr\"{o}der}}, \bibinfo {author} {\bibfnamefont {A.}~\bibnamefont
  {Wendeln}}, \bibinfo {author} {\bibfnamefont {J.~T.}\ \bibnamefont {Weber}},
  \bibinfo {author} {\bibfnamefont {M.}~\bibnamefont {Mukai}}, \bibinfo
  {author} {\bibfnamefont {Y.}~\bibnamefont {Kohno}},\ and\ \bibinfo {author}
  {\bibfnamefont {S.}~\bibnamefont {Sch\"{a}fer}},\ }\href
  {https://doi.org/10.1016/j.ultramic.2025.114158} {\bibfield  {journal}
  {\bibinfo  {journal} {Ultramicroscopy}\ }\textbf {\bibinfo {volume} {275}},\
  \bibinfo {pages} {114158} (\bibinfo {year} {2025})}\BibitemShut {NoStop}%
\bibitem [{\citenamefont {Tziperman}\ \emph {et~al.}(2026)\citenamefont
  {Tziperman}, \citenamefont {Nabben}, \citenamefont {Ruimy}, \citenamefont
  {Holder}, \citenamefont {Nussinson}, \citenamefont {Fang}, \citenamefont
  {Gorlach}, \citenamefont {Kazenwadel}, \citenamefont {Karnieli},
  \citenamefont {Kaminer},\ and\ \citenamefont {Baum}}]{TNR26}%
  \BibitemOpen
  \bibfield  {author} {\bibinfo {author} {\bibfnamefont {O.}~\bibnamefont
  {Tziperman}}, \bibinfo {author} {\bibfnamefont {D.}~\bibnamefont {Nabben}},
  \bibinfo {author} {\bibfnamefont {R.}~\bibnamefont {Ruimy}}, \bibinfo
  {author} {\bibfnamefont {J.}~\bibnamefont {Holder}}, \bibinfo {author}
  {\bibfnamefont {E.}~\bibnamefont {Nussinson}}, \bibinfo {author}
  {\bibfnamefont {Y.}~\bibnamefont {Fang}}, \bibinfo {author} {\bibfnamefont
  {A.}~\bibnamefont {Gorlach}}, \bibinfo {author} {\bibfnamefont
  {D.}~\bibnamefont {Kazenwadel}}, \bibinfo {author} {\bibfnamefont
  {A.}~\bibnamefont {Karnieli}}, \bibinfo {author} {\bibfnamefont
  {I.}~\bibnamefont {Kaminer}},\ and\ \bibinfo {author} {\bibfnamefont
  {P.}~\bibnamefont {Baum}},\ }\href
  {https://doi.org/10.1038/s41567-026-03254-y} {\bibfield  {journal} {\bibinfo
  {journal} {Nature Physics}\ }\textbf {\bibinfo {volume} {22}},\ \bibinfo
  {pages} {763} (\bibinfo {year} {2026})}\BibitemShut {NoStop}%
\bibitem [{\citenamefont {Haindl}\ \emph {et~al.}(2025)\citenamefont {Haindl},
  \citenamefont {{Di Giulio}}, \citenamefont {Feist},\ and\ \citenamefont
  {Ropers}}]{HDF25}%
  \BibitemOpen
  \bibfield  {author} {\bibinfo {author} {\bibfnamefont {R.}~\bibnamefont
  {Haindl}}, \bibinfo {author} {\bibfnamefont {V.}~\bibnamefont {{Di Giulio}}},
  \bibinfo {author} {\bibfnamefont {A.}~\bibnamefont {Feist}},\ and\ \bibinfo
  {author} {\bibfnamefont {C.}~\bibnamefont {Ropers}},\ }\href
  {https://doi.org/10.1103/lylq-m63c} {\bibfield  {journal} {\bibinfo
  {journal} {Phys.\ Rev.\ Lett.}\ }\textbf {\bibinfo {volume} {135}},\ \bibinfo
  {pages} {165002} (\bibinfo {year} {2025})}\BibitemShut {NoStop}%
\bibitem [{\citenamefont {Gover}\ \emph {et~al.}(2012)\citenamefont {Gover},
  \citenamefont {Nause}, \citenamefont {Dyunin},\ and\ \citenamefont
  {Fedurin}}]{GND12}%
  \BibitemOpen
  \bibfield  {author} {\bibinfo {author} {\bibfnamefont {A.}~\bibnamefont
  {Gover}}, \bibinfo {author} {\bibfnamefont {A.}~\bibnamefont {Nause}},
  \bibinfo {author} {\bibfnamefont {E.}~\bibnamefont {Dyunin}},\ and\ \bibinfo
  {author} {\bibfnamefont {M.}~\bibnamefont {Fedurin}},\ }\href
  {https://doi.org/10.1038/nphys2443} {\bibfield  {journal} {\bibinfo
  {journal} {Nature Physics}\ }\textbf {\bibinfo {volume} {8}},\ \bibinfo
  {pages} {877} (\bibinfo {year} {2012})}\BibitemShut {NoStop}%
\bibitem [{\citenamefont {Feist}\ \emph {et~al.}(2025)\citenamefont {Feist},
  \citenamefont {Haindl},\ and\ \citenamefont {Ropers}}]{FHR25}%
  \BibitemOpen
  \bibfield  {author} {\bibinfo {author} {\bibfnamefont {A.}~\bibnamefont
  {Feist}}, \bibinfo {author} {\bibfnamefont {R.}~\bibnamefont {Haindl}},\ and\
  \bibinfo {author} {\bibfnamefont {C.}~\bibnamefont {Ropers}},\ }\href@noop {}
  {\bibinfo {title} {Electron beam apparatus and method for generating a pulsed
  electron beam and applications thereof}},\ \bibinfo {howpublished} {European
  Patent EP4526915B1} (\bibinfo {year} {2025})\BibitemShut {NoStop}%
\bibitem [{\citenamefont {Peter}\ \emph {et~al.}(1990)\citenamefont {Peter},
  \citenamefont {Garate}, \citenamefont {Main},\ and\ \citenamefont
  {Fisher}}]{PGM1990}%
  \BibitemOpen
  \bibfield  {author} {\bibinfo {author} {\bibfnamefont {W.}~\bibnamefont
  {Peter}}, \bibinfo {author} {\bibfnamefont {E.}~\bibnamefont {Garate}},
  \bibinfo {author} {\bibfnamefont {W.}~\bibnamefont {Main}},\ and\ \bibinfo
  {author} {\bibfnamefont {A.}~\bibnamefont {Fisher}},\ }\href
  {https://doi.org/10.1103/PhysRevLett.65.2989} {\bibfield  {journal} {\bibinfo
   {journal} {Physical Review Letters}\ }\textbf {\bibinfo {volume} {65}},\
  \bibinfo {pages} {2989} (\bibinfo {year} {1990})}\BibitemShut {NoStop}%
\bibitem [{\citenamefont {Ratner}\ \emph {et~al.}(2015)\citenamefont {Ratner},
  \citenamefont {Behrens}, \citenamefont {Ding}, \citenamefont {Huang},
  \citenamefont {Marinelli}, \citenamefont {Maxwell},\ and\ \citenamefont
  {Zhou}}]{RBD15}%
  \BibitemOpen
  \bibfield  {author} {\bibinfo {author} {\bibfnamefont {D.}~\bibnamefont
  {Ratner}}, \bibinfo {author} {\bibfnamefont {C.}~\bibnamefont {Behrens}},
  \bibinfo {author} {\bibfnamefont {Y.}~\bibnamefont {Ding}}, \bibinfo {author}
  {\bibfnamefont {Z.}~\bibnamefont {Huang}}, \bibinfo {author} {\bibfnamefont
  {A.}~\bibnamefont {Marinelli}}, \bibinfo {author} {\bibfnamefont
  {T.}~\bibnamefont {Maxwell}},\ and\ \bibinfo {author} {\bibfnamefont
  {F.}~\bibnamefont {Zhou}},\ }\href
  {https://doi.org/10.1103/PhysRevSTAB.18.030704} {\bibfield  {journal}
  {\bibinfo  {journal} {Physical Review Special Topics -- Accelerators and
  Beams}\ }\textbf {\bibinfo {volume} {18}},\ \bibinfo {pages} {030704}
  (\bibinfo {year} {2015})}\BibitemShut {NoStop}%
\bibitem [{\citenamefont {Adli}\ \emph {et~al.}(2018)\citenamefont {Adli} \emph
  {et~al.}}]{AAA18}%
  \BibitemOpen
  \bibfield  {author} {\bibinfo {author} {\bibfnamefont {E.}~\bibnamefont
  {Adli}} \emph {et~al.},\ }\href {https://doi.org/10.1038/s41586-018-0485-4}
  {\bibfield  {journal} {\bibinfo  {journal} {Nature}\ }\textbf {\bibinfo
  {volume} {561}},\ \bibinfo {pages} {363} (\bibinfo {year}
  {2018})}\BibitemShut {NoStop}%
\bibitem [{\citenamefont {Adli}\ \emph {et~al.}(2019)\citenamefont {Adli} \emph
  {et~al.}}]{AAA19}%
  \BibitemOpen
  \bibfield  {author} {\bibinfo {author} {\bibfnamefont {E.}~\bibnamefont
  {Adli}} \emph {et~al.},\ }\href
  {https://doi.org/10.1103/PhysRevLett.122.054802} {\bibfield  {journal}
  {\bibinfo  {journal} {Physical Review Letters}\ }\textbf {\bibinfo {volume}
  {122}},\ \bibinfo {pages} {054802} (\bibinfo {year} {2019})}\BibitemShut
  {NoStop}%
\bibitem [{\citenamefont {Gover}\ \emph {et~al.}(2019)\citenamefont {Gover},
  \citenamefont {Ianconescu}, \citenamefont {Friedman}, \citenamefont {Emma},
  \citenamefont {Sudar}, \citenamefont {Musumeci},\ and\ \citenamefont
  {Pellegrini}}]{GIF19}%
  \BibitemOpen
  \bibfield  {author} {\bibinfo {author} {\bibfnamefont {A.}~\bibnamefont
  {Gover}}, \bibinfo {author} {\bibfnamefont {R.}~\bibnamefont {Ianconescu}},
  \bibinfo {author} {\bibfnamefont {A.}~\bibnamefont {Friedman}}, \bibinfo
  {author} {\bibfnamefont {C.}~\bibnamefont {Emma}}, \bibinfo {author}
  {\bibfnamefont {N.}~\bibnamefont {Sudar}}, \bibinfo {author} {\bibfnamefont
  {P.}~\bibnamefont {Musumeci}},\ and\ \bibinfo {author} {\bibfnamefont
  {C.}~\bibnamefont {Pellegrini}},\ }\href
  {https://doi.org/10.1103/RevModPhys.91.035003} {\bibfield  {journal}
  {\bibinfo  {journal} {Rev.\ Mod.\ Phys.}\ }\textbf {\bibinfo {volume} {91}},\
  \bibinfo {pages} {035003} (\bibinfo {year} {2019})}\BibitemShut {NoStop}%
\bibitem [{\citenamefont {Saldin}\ \emph {et~al.}(2000)\citenamefont {Saldin},
  \citenamefont {Schneidmiller},\ and\ \citenamefont {Yurkov}}]{SSY1}%
  \BibitemOpen
  \bibfield  {author} {\bibinfo {author} {\bibfnamefont {E.}~\bibnamefont
  {Saldin}}, \bibinfo {author} {\bibfnamefont {E.}~\bibnamefont
  {Schneidmiller}},\ and\ \bibinfo {author} {\bibfnamefont {M.}~\bibnamefont
  {Yurkov}},\ }\href@noop {} {\emph {\bibinfo {title} {The Physics of Free
  Electron Lasers}}}\ (\bibinfo  {publisher} {Springer-Verlag Berlin
  Heidelberg},\ \bibinfo {address} {Berlin},\ \bibinfo {year}
  {2000})\BibitemShut {NoStop}%
\bibitem [{\citenamefont {Kruit}\ \emph {et~al.}(2004)\citenamefont {Kruit},
  \citenamefont {Steenbrink}, \citenamefont {Jager},\ and\ \citenamefont
  {Wieland}}]{KSJ04}%
  \BibitemOpen
  \bibfield  {author} {\bibinfo {author} {\bibfnamefont {P.}~\bibnamefont
  {Kruit}}, \bibinfo {author} {\bibfnamefont {S.}~\bibnamefont {Steenbrink}},
  \bibinfo {author} {\bibfnamefont {R.}~\bibnamefont {Jager}},\ and\ \bibinfo
  {author} {\bibfnamefont {M.}~\bibnamefont {Wieland}},\ }\href
  {https://doi.org/https://doi.org/10.1116/1.1821577} {\bibfield  {journal}
  {\bibinfo  {journal} {J. Vac. Sci. Technol. B}\ }\textbf {\bibinfo {volume}
  {22}},\ \bibinfo {pages} {2948} (\bibinfo {year} {2004})}\BibitemShut
  {NoStop}%
\bibitem [{\citenamefont {McMullan}\ \emph {et~al.}(2009)\citenamefont
  {McMullan}, \citenamefont {Clark}, \citenamefont {Turchetta},\ and\
  \citenamefont {Faruqi}}]{MCT09}%
  \BibitemOpen
  \bibfield  {author} {\bibinfo {author} {\bibfnamefont {G.}~\bibnamefont
  {McMullan}}, \bibinfo {author} {\bibfnamefont {A.}~\bibnamefont {Clark}},
  \bibinfo {author} {\bibfnamefont {R.}~\bibnamefont {Turchetta}},\ and\
  \bibinfo {author} {\bibfnamefont {A.~R.}\ \bibnamefont {Faruqi}},\ }\href
  {https://doi.org/https://doi.org/10.1016/j.ultramic.2009.07.004} {\bibfield
  {journal} {\bibinfo  {journal} {Ultramicroscopy}\ }\textbf {\bibinfo {volume}
  {109}},\ \bibinfo {pages} {1411} (\bibinfo {year} {2009})}\BibitemShut
  {NoStop}%
\bibitem [{\citenamefont {Hart}\ \emph {et~al.}(2017)\citenamefont {Hart},
  \citenamefont {Lang}, \citenamefont {Leff}, \citenamefont {Longo},
  \citenamefont {Trevor},\ and\ \citenamefont {Taheri}}]{HLL17}%
  \BibitemOpen
  \bibfield  {author} {\bibinfo {author} {\bibfnamefont {J.~L.}\ \bibnamefont
  {Hart}}, \bibinfo {author} {\bibfnamefont {A.~C.}\ \bibnamefont {Lang}},
  \bibinfo {author} {\bibfnamefont {A.~C.}\ \bibnamefont {Leff}}, \bibinfo
  {author} {\bibfnamefont {P.}~\bibnamefont {Longo}}, \bibinfo {author}
  {\bibfnamefont {C.}~\bibnamefont {Trevor}},\ and\ \bibinfo {author}
  {\bibfnamefont {R.~D. T. M.~L.}\ \bibnamefont {Taheri}},\ }\href
  {https://doi.org/https://doi.org/10.1038/s41598-017-07709-4} {\bibfield
  {journal} {\bibinfo  {journal} {Scientific Reports}\ }\textbf {\bibinfo
  {volume} {7}},\ \bibinfo {pages} {8243} (\bibinfo {year} {2017})}\BibitemShut
  {NoStop}%
\bibitem [{\citenamefont {Sekia}\ \emph {et~al.}(2018)\citenamefont {Sekia},
  \citenamefont {Ikuharaa},\ and\ \citenamefont {Shibata}}]{SIS18}%
  \BibitemOpen
  \bibfield  {author} {\bibinfo {author} {\bibfnamefont {T.}~\bibnamefont
  {Sekia}}, \bibinfo {author} {\bibfnamefont {Y.}~\bibnamefont {Ikuharaa}},\
  and\ \bibinfo {author} {\bibfnamefont {N.}~\bibnamefont {Shibata}},\ }\href
  {https://doi.org/https://doi.org/10.1016/j.ultramic.2018.06.014} {\bibfield
  {journal} {\bibinfo  {journal} {Ultramicroscopy}\ }\textbf {\bibinfo {volume}
  {193}},\ \bibinfo {pages} {118} (\bibinfo {year} {2018})}\BibitemShut
  {NoStop}%
\bibitem [{\citenamefont {Shi}\ \emph {et~al.}(2019)\citenamefont {Shi},
  \citenamefont {He}, \citenamefont {Peng}, \citenamefont {Xiao}, \citenamefont
  {Liu}, \citenamefont {Lin},\ and\ \citenamefont {Zhang}}]{SHP19}%
  \BibitemOpen
  \bibfield  {author} {\bibinfo {author} {\bibfnamefont {C.}~\bibnamefont
  {Shi}}, \bibinfo {author} {\bibfnamefont {X.}~\bibnamefont {He}}, \bibinfo
  {author} {\bibfnamefont {J.}~\bibnamefont {Peng}}, \bibinfo {author}
  {\bibfnamefont {G.}~\bibnamefont {Xiao}}, \bibinfo {author} {\bibfnamefont
  {F.}~\bibnamefont {Liu}}, \bibinfo {author} {\bibfnamefont {F.}~\bibnamefont
  {Lin}},\ and\ \bibinfo {author} {\bibfnamefont {H.}~\bibnamefont {Zhang}},\
  }\href@noop {} {\bibfield  {journal} {\bibinfo  {journal} {Opt.\ Laser\
  Technol.}\ }\textbf {\bibinfo {volume} {114}},\ \bibinfo {pages} {28}
  (\bibinfo {year} {2019})}\BibitemShut {NoStop}%
\bibitem [{\citenamefont {Feist}\ \emph {et~al.}(2017)\citenamefont {Feist},
  \citenamefont {Bach}, \citenamefont {N.~Rubiano~{da Silva}}, \citenamefont
  {M\"{o}ller}, \citenamefont {Priebe}, \citenamefont {Domr\"{o}se},
  \citenamefont {Gatzmann}, \citenamefont {Rost}, \citenamefont {Schauss},
  \citenamefont {Strauch}, \citenamefont {Bormann}, \citenamefont {Sivis},
  \citenamefont {Sch\"{a}fer},\ and\ \citenamefont {Ropers}}]{FBR17}%
  \BibitemOpen
  \bibfield  {author} {\bibinfo {author} {\bibfnamefont {A.}~\bibnamefont
  {Feist}}, \bibinfo {author} {\bibfnamefont {N.}~\bibnamefont {Bach}},
  \bibinfo {author} {\bibfnamefont {T.~D.}\ \bibnamefont {N.~Rubiano~{da
  Silva}}}, \bibinfo {author} {\bibfnamefont {M.}~\bibnamefont {M\"{o}ller}},
  \bibinfo {author} {\bibfnamefont {K.~E.}\ \bibnamefont {Priebe}}, \bibinfo
  {author} {\bibfnamefont {T.}~\bibnamefont {Domr\"{o}se}}, \bibinfo {author}
  {\bibfnamefont {J.~G.}\ \bibnamefont {Gatzmann}}, \bibinfo {author}
  {\bibfnamefont {S.}~\bibnamefont {Rost}}, \bibinfo {author} {\bibfnamefont
  {J.}~\bibnamefont {Schauss}}, \bibinfo {author} {\bibfnamefont
  {S.}~\bibnamefont {Strauch}}, \bibinfo {author} {\bibfnamefont
  {R.}~\bibnamefont {Bormann}}, \bibinfo {author} {\bibfnamefont
  {M.}~\bibnamefont {Sivis}}, \bibinfo {author} {\bibfnamefont
  {S.}~\bibnamefont {Sch\"{a}fer}},\ and\ \bibinfo {author} {\bibfnamefont
  {C.}~\bibnamefont {Ropers}},\ }\href@noop {} {\bibfield  {journal} {\bibinfo
  {journal} {Ultramicroscopy}\ }\textbf {\bibinfo {volume} {176}},\ \bibinfo
  {pages} {63} (\bibinfo {year} {2017})}\BibitemShut {NoStop}%
\bibitem [{\citenamefont {Cook}\ \emph {et~al.}(2010)\citenamefont {Cook},
  \citenamefont {Verduin}, \citenamefont {Hagen},\ and\ \citenamefont
  {Kruit}}]{CVH10}%
  \BibitemOpen
  \bibfield  {author} {\bibinfo {author} {\bibfnamefont {B.}~\bibnamefont
  {Cook}}, \bibinfo {author} {\bibfnamefont {T.}~\bibnamefont {Verduin}},
  \bibinfo {author} {\bibfnamefont {C.~W.}\ \bibnamefont {Hagen}},\ and\
  \bibinfo {author} {\bibfnamefont {P.}~\bibnamefont {Kruit}},\ }\href
  {https://doi.org/10.1116/1.3502642} {\bibfield  {journal} {\bibinfo
  {journal} {Journal of Vacuum Science \& Technology B}\ }\textbf {\bibinfo
  {volume} {28}},\ \bibinfo {pages} {C6C74} (\bibinfo {year}
  {2010})}\BibitemShut {NoStop}%
\bibitem [{\citenamefont {Dove}(1993)}]{D1993}%
  \BibitemOpen
  \bibfield  {author} {\bibinfo {author} {\bibfnamefont {M.~T.}\ \bibnamefont
  {Dove}},\ }\href@noop {} {\emph {\bibinfo {title} {Introduction to Lattice
  Dynamics}}}\ (\bibinfo  {publisher} {Cambridge University Press},\ \bibinfo
  {address} {Cambridge},\ \bibinfo {year} {1993})\BibitemShut {NoStop}%
\bibitem [{\citenamefont {Ashcroft}\ and\ \citenamefont
  {Mermin}(1976)}]{AM1976}%
  \BibitemOpen
  \bibfield  {author} {\bibinfo {author} {\bibfnamefont {N.~W.}\ \bibnamefont
  {Ashcroft}}\ and\ \bibinfo {author} {\bibfnamefont {N.~D.}\ \bibnamefont
  {Mermin}},\ }\href@noop {} {\emph {\bibinfo {title} {Solid State Physics}}}\
  (\bibinfo  {publisher} {Harcourt College Publishers},\ \bibinfo {address}
  {Philadelphia},\ \bibinfo {year} {1976})\BibitemShut {NoStop}%
\bibitem [{\citenamefont {Lindemann}(1910)}]{L1910}%
  \BibitemOpen
  \bibfield  {author} {\bibinfo {author} {\bibfnamefont {F.}~\bibnamefont
  {Lindemann}},\ }\href@noop {} {\bibfield  {journal} {\bibinfo  {journal}
  {Physikalische Zeitschrift}\ }\textbf {\bibinfo {volume} {11}},\ \bibinfo
  {pages} {609} (\bibinfo {year} {1910})}\BibitemShut {NoStop}%
\bibitem [{SI(2026)}]{SI}%
  \BibitemOpen
  \href@noop {} {} (\bibinfo {year} {2026}),\ \bibinfo {note} {see
  Supplementary Information at ... for additional information on
  ...}\BibitemShut {Stop}%
\bibitem [{\citenamefont {Wu}\ and\ \citenamefont {Neureuther}(2000)}]{WN00}%
  \BibitemOpen
  \bibfield  {author} {\bibinfo {author} {\bibfnamefont {B.}~\bibnamefont
  {Wu}}\ and\ \bibinfo {author} {\bibfnamefont {A.~R.}\ \bibnamefont
  {Neureuther}},\ }\href {https://doi.org/10.1116/1.1324645} {\bibfield
  {journal} {\bibinfo  {journal} {Journal of Vacuum Science \& Technology B}\
  }\textbf {\bibinfo {volume} {18}},\ \bibinfo {pages} {3004} (\bibinfo {year}
  {2000})}\BibitemShut {NoStop}%
\bibitem [{\citenamefont {Seurer}\ and\ \citenamefont {Toepffer}(1997)}]{ST97}%
  \BibitemOpen
  \bibfield  {author} {\bibinfo {author} {\bibfnamefont {M.}~\bibnamefont
  {Seurer}}\ and\ \bibinfo {author} {\bibfnamefont {C.}~\bibnamefont
  {Toepffer}},\ }\href {https://doi.org/10.1023/A:1012683301424} {\bibfield
  {journal} {\bibinfo  {journal} {Hyperfine Interactions}\ }\textbf {\bibinfo
  {volume} {108}},\ \bibinfo {pages} {333} (\bibinfo {year}
  {1997})}\BibitemShut {NoStop}%
\bibitem [{\citenamefont {Hasse}(1992)}]{H92_2}%
  \BibitemOpen
  \bibfield  {author} {\bibinfo {author} {\bibfnamefont {R.~W.}\ \bibnamefont
  {Hasse}},\ }\href {https://doi.org/10.1103/PhysRevA.46.5189} {\bibfield
  {journal} {\bibinfo  {journal} {Phys.\ Rev.\ A}\ }\textbf {\bibinfo {volume}
  {46}},\ \bibinfo {pages} {5189} (\bibinfo {year} {1992})}\BibitemShut
  {NoStop}%
\bibitem [{\citenamefont {Chen}\ and\ \citenamefont {Dubin}(1993)}]{CD93}%
  \BibitemOpen
  \bibfield  {author} {\bibinfo {author} {\bibfnamefont {S.-J.}\ \bibnamefont
  {Chen}}\ and\ \bibinfo {author} {\bibfnamefont {D.~H.~E.}\ \bibnamefont
  {Dubin}},\ }\href {https://doi.org/10.1103/PhysRevLett.71.2721} {\bibfield
  {journal} {\bibinfo  {journal} {Phys.\ Rev.\ Lett.}\ }\textbf {\bibinfo
  {volume} {71}},\ \bibinfo {pages} {2721} (\bibinfo {year}
  {1993})}\BibitemShut {NoStop}%
\bibitem [{\citenamefont {Hasse}\ and\ \citenamefont {Schiffer}(1990)}]{HS90}%
  \BibitemOpen
  \bibfield  {author} {\bibinfo {author} {\bibfnamefont {R.~W.}\ \bibnamefont
  {Hasse}}\ and\ \bibinfo {author} {\bibfnamefont {J.~P.}\ \bibnamefont
  {Schiffer}},\ }\href {https://doi.org/10.1016/0003-4916(90)90177-P}
  {\bibfield  {journal} {\bibinfo  {journal} {Ann. Phys.}\ }\textbf {\bibinfo
  {volume} {203}},\ \bibinfo {pages} {419} (\bibinfo {year}
  {1990})}\BibitemShut {NoStop}%
\bibitem [{\citenamefont {Coenen}\ and\ \citenamefont {Haegel}(2017)}]{CH17}%
  \BibitemOpen
  \bibfield  {author} {\bibinfo {author} {\bibfnamefont {T.}~\bibnamefont
  {Coenen}}\ and\ \bibinfo {author} {\bibfnamefont {N.~M.}\ \bibnamefont
  {Haegel}},\ }\href@noop {} {\bibfield  {journal} {\bibinfo  {journal} {Appl.\
  Phys.\ Rev.}\ }\textbf {\bibinfo {volume} {4}},\ \bibinfo {pages} {031103}
  (\bibinfo {year} {2017})}\BibitemShut {NoStop}%
\bibitem [{\citenamefont {Kociak}\ and\ \citenamefont {Zagonel}(2017)}]{KZ17}%
  \BibitemOpen
  \bibfield  {author} {\bibinfo {author} {\bibfnamefont {M.}~\bibnamefont
  {Kociak}}\ and\ \bibinfo {author} {\bibfnamefont {L.~F.}\ \bibnamefont
  {Zagonel}},\ }\href {https://doi.org/10.1016/j.ultramic.2017.03.014}
  {\bibfield  {journal} {\bibinfo  {journal} {Ultramicroscopy}\ }\textbf
  {\bibinfo {volume} {174}},\ \bibinfo {pages} {50} (\bibinfo {year}
  {2017})}\BibitemShut {NoStop}%
\bibitem [{\citenamefont {{Di Giulio}}\ \emph {et~al.}(2021)\citenamefont {{Di
  Giulio}}, \citenamefont {Kfir}, \citenamefont {Ropers},\ and\ \citenamefont
  {{Garc\'{\i}a de Abajo}}}]{paper373}%
  \BibitemOpen
  \bibfield  {author} {\bibinfo {author} {\bibfnamefont {V.}~\bibnamefont {{Di
  Giulio}}}, \bibinfo {author} {\bibfnamefont {O.}~\bibnamefont {Kfir}},
  \bibinfo {author} {\bibfnamefont {C.}~\bibnamefont {Ropers}},\ and\ \bibinfo
  {author} {\bibfnamefont {F.~J.}\ \bibnamefont {{Garc\'{\i}a de Abajo}}},\
  }\href {https://doi.org/10.1021/acsnano.1c00549} {\bibfield  {journal}
  {\bibinfo  {journal} {ACS\ Nano}\ }\textbf {\bibinfo {volume} {15}},\
  \bibinfo {pages} {7290} (\bibinfo {year} {2021})}\BibitemShut {NoStop}%
\bibitem [{\citenamefont {Schwarz}\ and\ \citenamefont {Hora}(1969)}]{SH1969}%
  \BibitemOpen
  \bibfield  {author} {\bibinfo {author} {\bibfnamefont {H.}~\bibnamefont
  {Schwarz}}\ and\ \bibinfo {author} {\bibfnamefont {H.}~\bibnamefont {Hora}},\
  }\href@noop {} {\bibfield  {journal} {\bibinfo  {journal} {Appl.\ Phys.\
  Lett.}\ }\textbf {\bibinfo {volume} {15}},\ \bibinfo {pages} {249} (\bibinfo
  {year} {1969})}\BibitemShut {NoStop}%
\bibitem [{\citenamefont {Favro}\ \emph {et~al.}(1971)\citenamefont {Favro},
  \citenamefont {Fradkin},\ and\ \citenamefont {Kuo}}]{FFK1971}%
  \BibitemOpen
  \bibfield  {author} {\bibinfo {author} {\bibfnamefont {L.~D.}\ \bibnamefont
  {Favro}}, \bibinfo {author} {\bibfnamefont {D.~M.}\ \bibnamefont {Fradkin}},\
  and\ \bibinfo {author} {\bibfnamefont {P.~K.}\ \bibnamefont {Kuo}},\
  }\href@noop {} {\bibfield  {journal} {\bibinfo  {journal} {Phys.\ Rev.\ D}\
  }\textbf {\bibinfo {volume} {35}},\ \bibinfo {pages} {2934} (\bibinfo {year}
  {1971})}\BibitemShut {NoStop}%
\bibitem [{\citenamefont {Andrews}\ \emph {et~al.}(2005)\citenamefont
  {Andrews}, \citenamefont {Brau},\ and\ \citenamefont {Jarvis}}]{ABJ05}%
  \BibitemOpen
  \bibfield  {author} {\bibinfo {author} {\bibfnamefont {H.~L.}\ \bibnamefont
  {Andrews}}, \bibinfo {author} {\bibfnamefont {C.~A.}\ \bibnamefont {Brau}},\
  and\ \bibinfo {author} {\bibfnamefont {J.~D.}\ \bibnamefont {Jarvis}},\
  }\href@noop {} {\bibfield  {journal} {\bibinfo  {journal} {Phys.\ Rev.\ ST\
  Accel.\ Beams}\ }\textbf {\bibinfo {volume} {8}},\ \bibinfo {pages} {110702}
  (\bibinfo {year} {2005})}\BibitemShut {NoStop}%
\bibitem [{\citenamefont {{Garc\'{\i}a de Abajo}}\ and\ \citenamefont {{Di
  Giulio}}(2021)}]{paper371}%
  \BibitemOpen
  \bibfield  {author} {\bibinfo {author} {\bibfnamefont {F.~J.}\ \bibnamefont
  {{Garc\'{\i}a de Abajo}}}\ and\ \bibinfo {author} {\bibfnamefont
  {V.}~\bibnamefont {{Di Giulio}}},\ }\href
  {https://doi.org/10.1021/acsphotonics.0c01950} {\bibfield  {journal}
  {\bibinfo  {journal} {ACS\ Photonics}\ }\textbf {\bibinfo {volume} {8}},\
  \bibinfo {pages} {945} (\bibinfo {year} {2021})}\BibitemShut {NoStop}%
\bibitem [{\citenamefont {{Di Giulio}}\ \emph {et~al.}(2025)\citenamefont {{Di
  Giulio}}, \citenamefont {Haindl},\ and\ \citenamefont {Ropers}}]{DHR25}%
  \BibitemOpen
  \bibfield  {author} {\bibinfo {author} {\bibfnamefont {V.}~\bibnamefont {{Di
  Giulio}}}, \bibinfo {author} {\bibfnamefont {R.}~\bibnamefont {Haindl}},\
  and\ \bibinfo {author} {\bibfnamefont {C.}~\bibnamefont {Ropers}},\ }\href
  {https://doi.org/10.1515/nanoph-2025-0040} {\bibfield  {journal} {\bibinfo
  {journal} {Nanophotonics}\ }\textbf {\bibinfo {volume} {14}},\ \bibinfo
  {pages} {1865} (\bibinfo {year} {2025})}\BibitemShut {NoStop}%
\bibitem [{\citenamefont {Matsumoto}\ \emph {et~al.}(2008)\citenamefont
  {Matsumoto}, \citenamefont {Ando}, \citenamefont {Inoue}, \citenamefont
  {Ohtake}, \citenamefont {Fukuchi},\ and\ \citenamefont {Hara}}]{MAI08}%
  \BibitemOpen
  \bibfield  {author} {\bibinfo {author} {\bibfnamefont {N.}~\bibnamefont
  {Matsumoto}}, \bibinfo {author} {\bibfnamefont {T.}~\bibnamefont {Ando}},
  \bibinfo {author} {\bibfnamefont {T.}~\bibnamefont {Inoue}}, \bibinfo
  {author} {\bibfnamefont {Y.}~\bibnamefont {Ohtake}}, \bibinfo {author}
  {\bibfnamefont {N.}~\bibnamefont {Fukuchi}},\ and\ \bibinfo {author}
  {\bibfnamefont {T.}~\bibnamefont {Hara}},\ }\href
  {https://doi.org/10.1364/JOSAA.25.001642} {\bibfield  {journal} {\bibinfo
  {journal} {Journal of the Optical Society of America A}\ }\textbf {\bibinfo
  {volume} {25}},\ \bibinfo {pages} {1642} (\bibinfo {year}
  {2008})}\BibitemShut {NoStop}%
\bibitem [{\citenamefont {Forbes}\ \emph {et~al.}(2021)\citenamefont {Forbes},
  \citenamefont {de~Oliveira},\ and\ \citenamefont {Dennis}}]{FOD21}%
  \BibitemOpen
  \bibfield  {author} {\bibinfo {author} {\bibfnamefont {A.}~\bibnamefont
  {Forbes}}, \bibinfo {author} {\bibfnamefont {M.}~\bibnamefont
  {de~Oliveira}},\ and\ \bibinfo {author} {\bibfnamefont {M.~R.}\ \bibnamefont
  {Dennis}},\ }\href {https://doi.org/10.1038/s41566-021-00780-4} {\bibfield
  {journal} {\bibinfo  {journal} {Nature Photonics}\ }\textbf {\bibinfo
  {volume} {15}},\ \bibinfo {pages} {253} (\bibinfo {year} {2021})}\BibitemShut
  {NoStop}%
\bibitem [{\citenamefont {Forbes}\ \emph {et~al.}(2016)\citenamefont {Forbes},
  \citenamefont {Dudley},\ and\ \citenamefont {McLaren}}]{FDM16}%
  \BibitemOpen
  \bibfield  {author} {\bibinfo {author} {\bibfnamefont {A.}~\bibnamefont
  {Forbes}}, \bibinfo {author} {\bibfnamefont {A.}~\bibnamefont {Dudley}},\
  and\ \bibinfo {author} {\bibfnamefont {M.}~\bibnamefont {McLaren}},\ }\href
  {https://doi.org/10.1364/AOP.8.000200} {\bibfield  {journal} {\bibinfo
  {journal} {Advances in Optics and Photonics}\ }\textbf {\bibinfo {volume}
  {8}},\ \bibinfo {pages} {200} (\bibinfo {year} {2016})}\BibitemShut {NoStop}%
\bibitem [{\citenamefont {Madan}\ \emph {et~al.}(2022)\citenamefont {Madan},
  \citenamefont {Leccese}, \citenamefont {Mazur}, \citenamefont {Barantani},
  \citenamefont {LaGrange}, \citenamefont {Sapozhnik}, \citenamefont {Tengdin},
  \citenamefont {Gargiulo}, \citenamefont {Rotunno}, \citenamefont {Olaya},
  \citenamefont {Kaminer}, \citenamefont {Grillo}, \citenamefont {{Garc\'{\i}a
  de Abajo}}, \citenamefont {Carbone},\ and\ \citenamefont
  {Vanacore}}]{paper397}%
  \BibitemOpen
  \bibfield  {author} {\bibinfo {author} {\bibfnamefont {I.}~\bibnamefont
  {Madan}}, \bibinfo {author} {\bibfnamefont {V.}~\bibnamefont {Leccese}},
  \bibinfo {author} {\bibfnamefont {A.}~\bibnamefont {Mazur}}, \bibinfo
  {author} {\bibfnamefont {F.}~\bibnamefont {Barantani}}, \bibinfo {author}
  {\bibfnamefont {T.}~\bibnamefont {LaGrange}}, \bibinfo {author}
  {\bibfnamefont {A.}~\bibnamefont {Sapozhnik}}, \bibinfo {author}
  {\bibfnamefont {P.~M.}\ \bibnamefont {Tengdin}}, \bibinfo {author}
  {\bibfnamefont {S.}~\bibnamefont {Gargiulo}}, \bibinfo {author}
  {\bibfnamefont {E.}~\bibnamefont {Rotunno}}, \bibinfo {author} {\bibfnamefont
  {J.-C.}\ \bibnamefont {Olaya}}, \bibinfo {author} {\bibfnamefont
  {I.}~\bibnamefont {Kaminer}}, \bibinfo {author} {\bibfnamefont
  {V.}~\bibnamefont {Grillo}}, \bibinfo {author} {\bibfnamefont {F.~J.}\
  \bibnamefont {{Garc\'{\i}a de Abajo}}}, \bibinfo {author} {\bibfnamefont
  {F.}~\bibnamefont {Carbone}},\ and\ \bibinfo {author} {\bibfnamefont {G.~M.}\
  \bibnamefont {Vanacore}},\ }\href
  {https://doi.org/10.1021/acsphotonics.2c00850} {\bibfield  {journal}
  {\bibinfo  {journal} {ACS\ Photonics}\ }\textbf {\bibinfo {volume} {9}},\
  \bibinfo {pages} {3215} (\bibinfo {year} {2022})}\BibitemShut {NoStop}%
\bibitem [{\citenamefont {Mihaila}\ \emph {et~al.}(2022)\citenamefont
  {Mihaila}, \citenamefont {Weber}, \citenamefont {Schneller}, \citenamefont
  {Grandits}, \citenamefont {Nimmrichter},\ and\ \citenamefont
  {Juffmann}}]{MWS22}%
  \BibitemOpen
  \bibfield  {author} {\bibinfo {author} {\bibfnamefont {M.~C.~C.}\
  \bibnamefont {Mihaila}}, \bibinfo {author} {\bibfnamefont {P.}~\bibnamefont
  {Weber}}, \bibinfo {author} {\bibfnamefont {M.}~\bibnamefont {Schneller}},
  \bibinfo {author} {\bibfnamefont {L.}~\bibnamefont {Grandits}}, \bibinfo
  {author} {\bibfnamefont {S.}~\bibnamefont {Nimmrichter}},\ and\ \bibinfo
  {author} {\bibfnamefont {T.}~\bibnamefont {Juffmann}},\ }\href
  {https://doi.org/10.1103/PhysRevX.12.031043} {\bibfield  {journal} {\bibinfo
  {journal} {Phys.\ Rev.\ X}\ }\textbf {\bibinfo {volume} {12}},\ \bibinfo
  {pages} {031043} (\bibinfo {year} {2022})}\BibitemShut {NoStop}%
\bibitem [{\citenamefont {Hoffrogge}\ \emph {et~al.}(2011)\citenamefont
  {Hoffrogge}, \citenamefont {Froehlich}, \citenamefont {Kasevich},\ and\
  \citenamefont {Hommelhoff}}]{HFK11}%
  \BibitemOpen
  \bibfield  {author} {\bibinfo {author} {\bibfnamefont {J.}~\bibnamefont
  {Hoffrogge}}, \bibinfo {author} {\bibfnamefont {R.}~\bibnamefont
  {Froehlich}}, \bibinfo {author} {\bibfnamefont {M.~A.}\ \bibnamefont
  {Kasevich}},\ and\ \bibinfo {author} {\bibfnamefont {P.}~\bibnamefont
  {Hommelhoff}},\ }\href {https://doi.org/10.1103/PhysRevLett.106.193001}
  {\bibfield  {journal} {\bibinfo  {journal} {Phys.\ Rev.\ Lett.}\ }\textbf
  {\bibinfo {volume} {106}},\ \bibinfo {pages} {193001} (\bibinfo {year}
  {2011})}\BibitemShut {NoStop}%
\bibitem [{\citenamefont {Huber}\ \emph {et~al.}(2022)\citenamefont {Huber},
  \citenamefont {Kern}, \citenamefont {Karnaushenko}, \citenamefont {Eisner},
  \citenamefont {Lepucki}, \citenamefont {Thampi}, \citenamefont
  {Mirhajivarzaneh}, \citenamefont {Becker}, \citenamefont {Kang},
  \citenamefont {Baunack}, \citenamefont {Büchner}, \citenamefont
  {Karnaushenko}, \citenamefont {Schmidt},\ and\ \citenamefont {Lubk}}]{HKK22}%
  \BibitemOpen
  \bibfield  {author} {\bibinfo {author} {\bibfnamefont {R.}~\bibnamefont
  {Huber}}, \bibinfo {author} {\bibfnamefont {F.}~\bibnamefont {Kern}},
  \bibinfo {author} {\bibfnamefont {D.~D.}\ \bibnamefont {Karnaushenko}},
  \bibinfo {author} {\bibfnamefont {E.}~\bibnamefont {Eisner}}, \bibinfo
  {author} {\bibfnamefont {P.}~\bibnamefont {Lepucki}}, \bibinfo {author}
  {\bibfnamefont {A.}~\bibnamefont {Thampi}}, \bibinfo {author} {\bibfnamefont
  {A.}~\bibnamefont {Mirhajivarzaneh}}, \bibinfo {author} {\bibfnamefont
  {C.}~\bibnamefont {Becker}}, \bibinfo {author} {\bibfnamefont
  {T.}~\bibnamefont {Kang}}, \bibinfo {author} {\bibfnamefont {S.}~\bibnamefont
  {Baunack}}, \bibinfo {author} {\bibfnamefont {B.}~\bibnamefont {Büchner}},
  \bibinfo {author} {\bibfnamefont {D.}~\bibnamefont {Karnaushenko}}, \bibinfo
  {author} {\bibfnamefont {O.~G.}\ \bibnamefont {Schmidt}},\ and\ \bibinfo
  {author} {\bibfnamefont {A.}~\bibnamefont {Lubk}},\ }\href
  {https://doi.org/10.1038/s41467-022-30703-y} {\bibfield  {journal} {\bibinfo
  {journal} {Nat.\ Commun.}\ }\textbf {\bibinfo {volume} {13}},\ \bibinfo
  {pages} {3220} (\bibinfo {year} {2022})}\BibitemShut {NoStop}%
\bibitem [{\citenamefont {Schmidt-Kaler}\ \emph {et~al.}(2026)\citenamefont
  {Schmidt-Kaler}, \citenamefont {Bode}, \citenamefont {Seidling},
  \citenamefont {Zimmermann}, \citenamefont {Walther},\ and\ \citenamefont
  {Hommelhoff}}]{SBS26}%
  \BibitemOpen
  \bibfield  {author} {\bibinfo {author} {\bibfnamefont {F.~D.~F.}\
  \bibnamefont {Schmidt-Kaler}}, \bibinfo {author} {\bibfnamefont {N.~R.}\
  \bibnamefont {Bode}}, \bibinfo {author} {\bibfnamefont {M.}~\bibnamefont
  {Seidling}}, \bibinfo {author} {\bibfnamefont {R.}~\bibnamefont
  {Zimmermann}}, \bibinfo {author} {\bibfnamefont {J.}~\bibnamefont
  {Walther}},\ and\ \bibinfo {author} {\bibfnamefont {P.}~\bibnamefont
  {Hommelhoff}},\ }\href {https://doi.org/10.1063/5.0279055} {\bibfield
  {journal} {\bibinfo  {journal} {Appl. Phys. Lett.}\ }\textbf {\bibinfo
  {volume} {128}},\ \bibinfo {pages} {031102} (\bibinfo {year}
  {2026})}\BibitemShut {NoStop}%
\bibitem [{\citenamefont {Swanson}\ and\ \citenamefont {Schwind}(2009)}]{SS09}%
  \BibitemOpen
  \bibfield  {author} {\bibinfo {author} {\bibfnamefont {L.~W.}\ \bibnamefont
  {Swanson}}\ and\ \bibinfo {author} {\bibfnamefont {G.~A.}\ \bibnamefont
  {Schwind}},\ }in\ \href {https://doi.org/10.1016/S1076-5670(09)59002-7}
  {\emph {\bibinfo {booktitle} {Advances in Imaging and Electron Physics}}},\
  Vol.\ \bibinfo {volume} {159}\ (\bibinfo  {publisher} {Elsevier},\ \bibinfo
  {year} {2009})\ pp.\ \bibinfo {pages} {63--100}\BibitemShut {NoStop}%
\bibitem [{\citenamefont {Houdellier}\ \emph {et~al.}(2018)\citenamefont
  {Houdellier}, \citenamefont {Caruso}, \citenamefont {Weber}, \citenamefont
  {Kociak},\ and\ \citenamefont {Arbouet}}]{HCW18_2}%
  \BibitemOpen
  \bibfield  {author} {\bibinfo {author} {\bibfnamefont {F.}~\bibnamefont
  {Houdellier}}, \bibinfo {author} {\bibfnamefont {G.~M.}\ \bibnamefont
  {Caruso}}, \bibinfo {author} {\bibfnamefont {S.}~\bibnamefont {Weber}},
  \bibinfo {author} {\bibfnamefont {M.}~\bibnamefont {Kociak}},\ and\ \bibinfo
  {author} {\bibfnamefont {A.}~\bibnamefont {Arbouet}},\ }\href
  {https://doi.org/10.1016/j.ultramic.2017.12.015} {\bibfield  {journal}
  {\bibinfo  {journal} {Ultramicroscopy}\ }\textbf {\bibinfo {volume} {186}},\
  \bibinfo {pages} {128} (\bibinfo {year} {2018})}\BibitemShut {NoStop}%
\bibitem [{\citenamefont {Zhang}\ \emph {et~al.}(2016)\citenamefont {Zhang},
  \citenamefont {Tang}, \citenamefont {Yuan}, \citenamefont {Yamauchi},
  \citenamefont {Suzuki}, \citenamefont {Shinya}, \citenamefont {Nakajima},\
  and\ \citenamefont {Qin}}]{ZTY16}%
  \BibitemOpen
  \bibfield  {author} {\bibinfo {author} {\bibfnamefont {H.}~\bibnamefont
  {Zhang}}, \bibinfo {author} {\bibfnamefont {J.}~\bibnamefont {Tang}},
  \bibinfo {author} {\bibfnamefont {J.}~\bibnamefont {Yuan}}, \bibinfo {author}
  {\bibfnamefont {Y.}~\bibnamefont {Yamauchi}}, \bibinfo {author}
  {\bibfnamefont {T.~T.}\ \bibnamefont {Suzuki}}, \bibinfo {author}
  {\bibfnamefont {N.}~\bibnamefont {Shinya}}, \bibinfo {author} {\bibfnamefont
  {K.}~\bibnamefont {Nakajima}},\ and\ \bibinfo {author} {\bibfnamefont
  {L.-C.}\ \bibnamefont {Qin}},\ }\href
  {https://doi.org/10.1038/nnano.2015.276} {\bibfield  {journal} {\bibinfo
  {journal} {Nature Nanotechnology}\ }\textbf {\bibinfo {volume} {11}},\
  \bibinfo {pages} {273} (\bibinfo {year} {2016})}\BibitemShut {NoStop}%
\bibitem [{\citenamefont {Gou}\ \emph {et~al.}(2025)\citenamefont {Gou},
  \citenamefont {Tang}, \citenamefont {Huang}, \citenamefont {Guan},
  \citenamefont {Zhan}, \citenamefont {Liang}, \citenamefont {Chen},
  \citenamefont {Shen},\ and\ \citenamefont {Deng}}]{GTH25}%
  \BibitemOpen
  \bibfield  {author} {\bibinfo {author} {\bibfnamefont {M.}~\bibnamefont
  {Gou}}, \bibinfo {author} {\bibfnamefont {S.}~\bibnamefont {Tang}}, \bibinfo
  {author} {\bibfnamefont {K.}~\bibnamefont {Huang}}, \bibinfo {author}
  {\bibfnamefont {H.}~\bibnamefont {Guan}}, \bibinfo {author} {\bibfnamefont
  {R.}~\bibnamefont {Zhan}}, \bibinfo {author} {\bibfnamefont {C.}~\bibnamefont
  {Liang}}, \bibinfo {author} {\bibfnamefont {C.}~\bibnamefont {Chen}},
  \bibinfo {author} {\bibfnamefont {Y.}~\bibnamefont {Shen}},\ and\ \bibinfo
  {author} {\bibfnamefont {S.}~\bibnamefont {Deng}},\ }\href
  {https://doi.org/10.1038/s43246-025-00860-w} {\bibfield  {journal} {\bibinfo
  {journal} {Communications Materials}\ }\textbf {\bibinfo {volume} {6}},\
  \bibinfo {pages} {133} (\bibinfo {year} {2025})}\BibitemShut {NoStop}%
\bibitem [{\citenamefont {McCulloch}\ \emph {et~al.}(2013)\citenamefont
  {McCulloch}, \citenamefont {Sheludko}, \citenamefont {Junker},\ and\
  \citenamefont {Scholten}}]{MSJ13}%
  \BibitemOpen
  \bibfield  {author} {\bibinfo {author} {\bibfnamefont {A.~J.}\ \bibnamefont
  {McCulloch}}, \bibinfo {author} {\bibfnamefont {D.~V.}\ \bibnamefont
  {Sheludko}}, \bibinfo {author} {\bibfnamefont {M.}~\bibnamefont {Junker}},\
  and\ \bibinfo {author} {\bibfnamefont {R.~E.}\ \bibnamefont {Scholten}},\
  }\href {https://doi.org/10.1038/ncomms2699} {\bibfield  {journal} {\bibinfo
  {journal} {Nature Communications}\ }\textbf {\bibinfo {volume} {4}},\
  \bibinfo {pages} {1692} (\bibinfo {year} {2013})}\BibitemShut {NoStop}%
\bibitem [{\citenamefont {Franssen}\ \emph {et~al.}(2019)\citenamefont
  {Franssen}, \citenamefont {{de Raadt}}, \citenamefont {{van Ninhuijs}},\ and\
  \citenamefont {Luiten}}]{FRN19}%
  \BibitemOpen
  \bibfield  {author} {\bibinfo {author} {\bibfnamefont {J.~G.~H.}\
  \bibnamefont {Franssen}}, \bibinfo {author} {\bibfnamefont {T.~C.~H.}\
  \bibnamefont {{de Raadt}}}, \bibinfo {author} {\bibfnamefont {M.~A.~W.}\
  \bibnamefont {{van Ninhuijs}}},\ and\ \bibinfo {author} {\bibfnamefont
  {O.~J.}\ \bibnamefont {Luiten}},\ }\href
  {https://doi.org/10.1103/PhysRevAccelBeams.22.023401} {\bibfield  {journal}
  {\bibinfo  {journal} {Physical Review Accelerators and Beams}\ }\textbf
  {\bibinfo {volume} {22}},\ \bibinfo {pages} {023401} (\bibinfo {year}
  {2019})}\BibitemShut {NoStop}%
\bibitem [{\citenamefont {Kubo}\ \emph {et~al.}(2005)\citenamefont {Kubo},
  \citenamefont {Mtingwa},\ and\ \citenamefont {Wolski}}]{KMW05}%
  \BibitemOpen
  \bibfield  {author} {\bibinfo {author} {\bibfnamefont {K.}~\bibnamefont
  {Kubo}}, \bibinfo {author} {\bibfnamefont {S.~K.}\ \bibnamefont {Mtingwa}},\
  and\ \bibinfo {author} {\bibfnamefont {A.}~\bibnamefont {Wolski}},\ }\href
  {https://doi.org/10.1103/PhysRevSTAB.8.081001} {\bibfield  {journal}
  {\bibinfo  {journal} {Physical Review Special Topics - Accelerators and
  Beams}\ }\textbf {\bibinfo {volume} {8}},\ \bibinfo {pages} {081001}
  (\bibinfo {year} {2005})}\BibitemShut {NoStop}%
\bibitem [{\citenamefont {Jarvis}\ \emph {et~al.}(2022)\citenamefont {Jarvis},
  \citenamefont {Lebedev}, \citenamefont {Romanov}, \citenamefont
  {Broemmelsiek}, \citenamefont {Carlson}, \citenamefont {Chattopadhyay},
  \citenamefont {Dick}, \citenamefont {Edstrom}, \citenamefont {Lobach},
  \citenamefont {Nagaitsev}, \citenamefont {Piekarz}, \citenamefont {Piot},
  \citenamefont {Ruan}, \citenamefont {Santucci}, \citenamefont {Stancari},\
  and\ \citenamefont {Valishev}}]{JLR22}%
  \BibitemOpen
  \bibfield  {author} {\bibinfo {author} {\bibfnamefont {J.}~\bibnamefont
  {Jarvis}}, \bibinfo {author} {\bibfnamefont {V.}~\bibnamefont {Lebedev}},
  \bibinfo {author} {\bibfnamefont {A.}~\bibnamefont {Romanov}}, \bibinfo
  {author} {\bibfnamefont {D.}~\bibnamefont {Broemmelsiek}}, \bibinfo {author}
  {\bibfnamefont {K.}~\bibnamefont {Carlson}}, \bibinfo {author} {\bibfnamefont
  {S.}~\bibnamefont {Chattopadhyay}}, \bibinfo {author} {\bibfnamefont
  {A.}~\bibnamefont {Dick}}, \bibinfo {author} {\bibfnamefont {D.}~\bibnamefont
  {Edstrom}}, \bibinfo {author} {\bibfnamefont {I.}~\bibnamefont {Lobach}},
  \bibinfo {author} {\bibfnamefont {S.}~\bibnamefont {Nagaitsev}}, \bibinfo
  {author} {\bibfnamefont {H.}~\bibnamefont {Piekarz}}, \bibinfo {author}
  {\bibfnamefont {P.}~\bibnamefont {Piot}}, \bibinfo {author} {\bibfnamefont
  {J.}~\bibnamefont {Ruan}}, \bibinfo {author} {\bibfnamefont {J.}~\bibnamefont
  {Santucci}}, \bibinfo {author} {\bibfnamefont {G.}~\bibnamefont {Stancari}},\
  and\ \bibinfo {author} {\bibfnamefont {A.}~\bibnamefont {Valishev}},\ }\href
  {https://doi.org/10.1038/s41586-022-04969-7} {\bibfield  {journal} {\bibinfo
  {journal} {Nature}\ }\textbf {\bibinfo {volume} {608}},\ \bibinfo {pages}
  {287} (\bibinfo {year} {2022})}\BibitemShut {NoStop}%
\bibitem [{\citenamefont {Marquardt}\ \emph {et~al.}(2007)\citenamefont
  {Marquardt}, \citenamefont {Chen}, \citenamefont {Clerk},\ and\ \citenamefont
  {Girvin}}]{MCC07}%
  \BibitemOpen
  \bibfield  {author} {\bibinfo {author} {\bibfnamefont {F.}~\bibnamefont
  {Marquardt}}, \bibinfo {author} {\bibfnamefont {J.~P.}\ \bibnamefont {Chen}},
  \bibinfo {author} {\bibfnamefont {A.~A.}\ \bibnamefont {Clerk}},\ and\
  \bibinfo {author} {\bibfnamefont {S.~M.}\ \bibnamefont {Girvin}},\ }\href
  {https://doi.org/10.1103/PhysRevLett.99.093902} {\bibfield  {journal}
  {\bibinfo  {journal} {Physical Review Letters}\ }\textbf {\bibinfo {volume}
  {99}},\ \bibinfo {pages} {093902} (\bibinfo {year} {2007})}\BibitemShut
  {NoStop}%
\bibitem [{\citenamefont {Vuleti{\'c}}\ \emph {et~al.}(2001)\citenamefont
  {Vuleti{\'c}}, \citenamefont {Chan},\ and\ \citenamefont {Black}}]{VCB01}%
  \BibitemOpen
  \bibfield  {author} {\bibinfo {author} {\bibfnamefont {V.}~\bibnamefont
  {Vuleti{\'c}}}, \bibinfo {author} {\bibfnamefont {H.~W.}\ \bibnamefont
  {Chan}},\ and\ \bibinfo {author} {\bibfnamefont {A.~T.}\ \bibnamefont
  {Black}},\ }\href {https://doi.org/10.1103/PhysRevA.64.033405} {\bibfield
  {journal} {\bibinfo  {journal} {Physical Review A}\ }\textbf {\bibinfo
  {volume} {64}},\ \bibinfo {pages} {033405} (\bibinfo {year}
  {2001})}\BibitemShut {NoStop}%
\bibitem [{\citenamefont {Behunin}\ \emph {et~al.}(2012)\citenamefont
  {Behunin}, \citenamefont {Zeng}, \citenamefont {Dalvit},\ and\ \citenamefont
  {Reynaud}}]{BZD12}%
  \BibitemOpen
  \bibfield  {author} {\bibinfo {author} {\bibfnamefont {R.~O.}\ \bibnamefont
  {Behunin}}, \bibinfo {author} {\bibfnamefont {Y.}~\bibnamefont {Zeng}},
  \bibinfo {author} {\bibfnamefont {D.~A.~R.}\ \bibnamefont {Dalvit}},\ and\
  \bibinfo {author} {\bibfnamefont {S.}~\bibnamefont {Reynaud}},\ }\href
  {https://doi.org/10.1103/PhysRevA.86.052509} {\bibfield  {journal} {\bibinfo
  {journal} {Phys.\ Rev.\ A}\ }\textbf {\bibinfo {volume} {86}},\ \bibinfo
  {pages} {052509} (\bibinfo {year} {2012})}\BibitemShut {NoStop}%
\bibitem [{\citenamefont {Paul}(1990)}]{P90}%
  \BibitemOpen
  \bibfield  {author} {\bibinfo {author} {\bibfnamefont {W.}~\bibnamefont
  {Paul}},\ }\href {https://doi.org/https://doi.org/10.1103/RevModPhys.62.531}
  {\bibfield  {journal} {\bibinfo  {journal} {Rev.\ Mod.\ Phys.}\ }\textbf
  {\bibinfo {volume} {62}},\ \bibinfo {pages} {531} (\bibinfo {year}
  {1990})}\BibitemShut {NoStop}%
\bibitem [{\citenamefont {Ghosh}(1995)}]{G1995}%
  \BibitemOpen
  \bibfield  {author} {\bibinfo {author} {\bibfnamefont {P.~K.}\ \bibnamefont
  {Ghosh}},\ }\href@noop {} {\emph {\bibinfo {title} {Ion Traps}}}\ (\bibinfo
  {publisher} {Clarendon Press},\ \bibinfo {address} {Oxford},\ \bibinfo {year}
  {1995})\BibitemShut {NoStop}%
\bibitem [{\citenamefont {Allen}\ \emph {et~al.}(1992)\citenamefont {Allen},
  \citenamefont {Beijersbergen}, \citenamefont {Spreeuw},\ and\ \citenamefont
  {Woerdman}}]{ABS92}%
  \BibitemOpen
  \bibfield  {author} {\bibinfo {author} {\bibfnamefont {L.}~\bibnamefont
  {Allen}}, \bibinfo {author} {\bibfnamefont {M.~W.}\ \bibnamefont
  {Beijersbergen}}, \bibinfo {author} {\bibfnamefont {R.~J.~C.}\ \bibnamefont
  {Spreeuw}},\ and\ \bibinfo {author} {\bibfnamefont {J.~P.}\ \bibnamefont
  {Woerdman}},\ }\href@noop {} {\bibfield  {journal} {\bibinfo  {journal}
  {Phys.\ Rev.\ A}\ }\textbf {\bibinfo {volume} {45}},\ \bibinfo {pages} {8185}
  (\bibinfo {year} {1992})}\BibitemShut {NoStop}%
\bibitem [{\citenamefont {Yang}\ \emph {et~al.}(2022)\citenamefont {Yang},
  \citenamefont {Li},\ and\ \citenamefont {Wang}}]{YLW22}%
  \BibitemOpen
  \bibfield  {author} {\bibinfo {author} {\bibfnamefont {Y.}~\bibnamefont
  {Yang}}, \bibinfo {author} {\bibfnamefont {Y.}~\bibnamefont {Li}},\ and\
  \bibinfo {author} {\bibfnamefont {C.}~\bibnamefont {Wang}},\ }\href
  {https://doi.org/10.1007/s12596-022-00857-5} {\bibfield  {journal} {\bibinfo
  {journal} {Journal of Optics}\ }\textbf {\bibinfo {volume} {51}},\ \bibinfo
  {pages} {910} (\bibinfo {year} {2022})}\BibitemShut {NoStop}%
\bibitem [{\citenamefont {Zimmermann}\ \emph {et~al.}(2021)\citenamefont
  {Zimmermann}, \citenamefont {Seidling},\ and\ \citenamefont
  {Hommelhoff}}]{ZSH21}%
  \BibitemOpen
  \bibfield  {author} {\bibinfo {author} {\bibfnamefont {R.}~\bibnamefont
  {Zimmermann}}, \bibinfo {author} {\bibfnamefont {M.}~\bibnamefont
  {Seidling}},\ and\ \bibinfo {author} {\bibfnamefont {P.}~\bibnamefont
  {Hommelhoff}},\ }\href
  {https://doi.org/https://doi.org/10.1038/s41467-020-20592-4} {\bibfield
  {journal} {\bibinfo  {journal} {Nat.\ Commun.}\ }\textbf {\bibinfo {volume}
  {12}},\ \bibinfo {pages} {390} (\bibinfo {year} {2021})}\BibitemShut
  {NoStop}%
\bibitem [{\citenamefont {Brush}\ \emph {et~al.}(1966)\citenamefont {Brush},
  \citenamefont {Sahlin},\ and\ \citenamefont {Teller}}]{BST1966}%
  \BibitemOpen
  \bibfield  {author} {\bibinfo {author} {\bibfnamefont {S.~G.}\ \bibnamefont
  {Brush}}, \bibinfo {author} {\bibfnamefont {H.~L.}\ \bibnamefont {Sahlin}},\
  and\ \bibinfo {author} {\bibfnamefont {E.}~\bibnamefont {Teller}},\ }\href
  {https://doi.org/https://doi.org/10.1063/1.1727895} {\bibfield  {journal}
  {\bibinfo  {journal} {J.\ Chem.\ Phys.}\ }\textbf {\bibinfo {volume} {45}},\
  \bibinfo {pages} {2102} (\bibinfo {year} {1966})}\BibitemShut {NoStop}%
\bibitem [{\citenamefont {Piacente}\ \emph {et~al.}(2004)\citenamefont
  {Piacente}, \citenamefont {Schweigert}, \citenamefont {Betouras},\ and\
  \citenamefont {Peeters}}]{PSP04}%
  \BibitemOpen
  \bibfield  {author} {\bibinfo {author} {\bibfnamefont {G.}~\bibnamefont
  {Piacente}}, \bibinfo {author} {\bibfnamefont {I.~V.}\ \bibnamefont
  {Schweigert}}, \bibinfo {author} {\bibfnamefont {J.~J.}\ \bibnamefont
  {Betouras}},\ and\ \bibinfo {author} {\bibfnamefont {F.~M.}\ \bibnamefont
  {Peeters}},\ }\href {https://doi.org/10.1103/PhysRevB.69.045324} {\bibfield
  {journal} {\bibinfo  {journal} {Phys.\ Rev.\ B}\ }\textbf {\bibinfo {volume}
  {69}},\ \bibinfo {pages} {045324} (\bibinfo {year} {2004})}\BibitemShut
  {NoStop}%
\bibitem [{\citenamefont {Abramowitz}\ and\ \citenamefont
  {Stegun}(1972)}]{AS1972}%
  \BibitemOpen
  \bibfield  {author} {\bibinfo {author} {\bibfnamefont {M.}~\bibnamefont
  {Abramowitz}}\ and\ \bibinfo {author} {\bibfnamefont {I.~A.}\ \bibnamefont
  {Stegun}},\ }\href@noop {} {\emph {\bibinfo {title} {Handbook of Mathematical
  Functions}}}\ (\bibinfo  {publisher} {Dover},\ \bibinfo {address} {New
  York},\ \bibinfo {year} {1972})\BibitemShut {NoStop}%
\bibitem [{\citenamefont {Croxton}(1974)}]{C1974_2}%
  \BibitemOpen
  \bibfield  {author} {\bibinfo {author} {\bibfnamefont {C.~A.}\ \bibnamefont
  {Croxton}},\ }\href {https://books.google.es/books?id=2HFpwgEACAAJ} {\emph
  {\bibinfo {title} {Liquid State Physics: A Statistical Mechanical
  Introduction}}},\ Cambridge monographs on physics\ (\bibinfo  {publisher}
  {Cambridge University Press},\ \bibinfo {address} {Cambridge},\ \bibinfo
  {year} {1974})\BibitemShut {NoStop}%
\end{thebibliography}

%

\clearpage
\onecolumngrid
\foreach \x in {1,...,5}{%
    \clearpage
    \includepdf[pages={\x}]{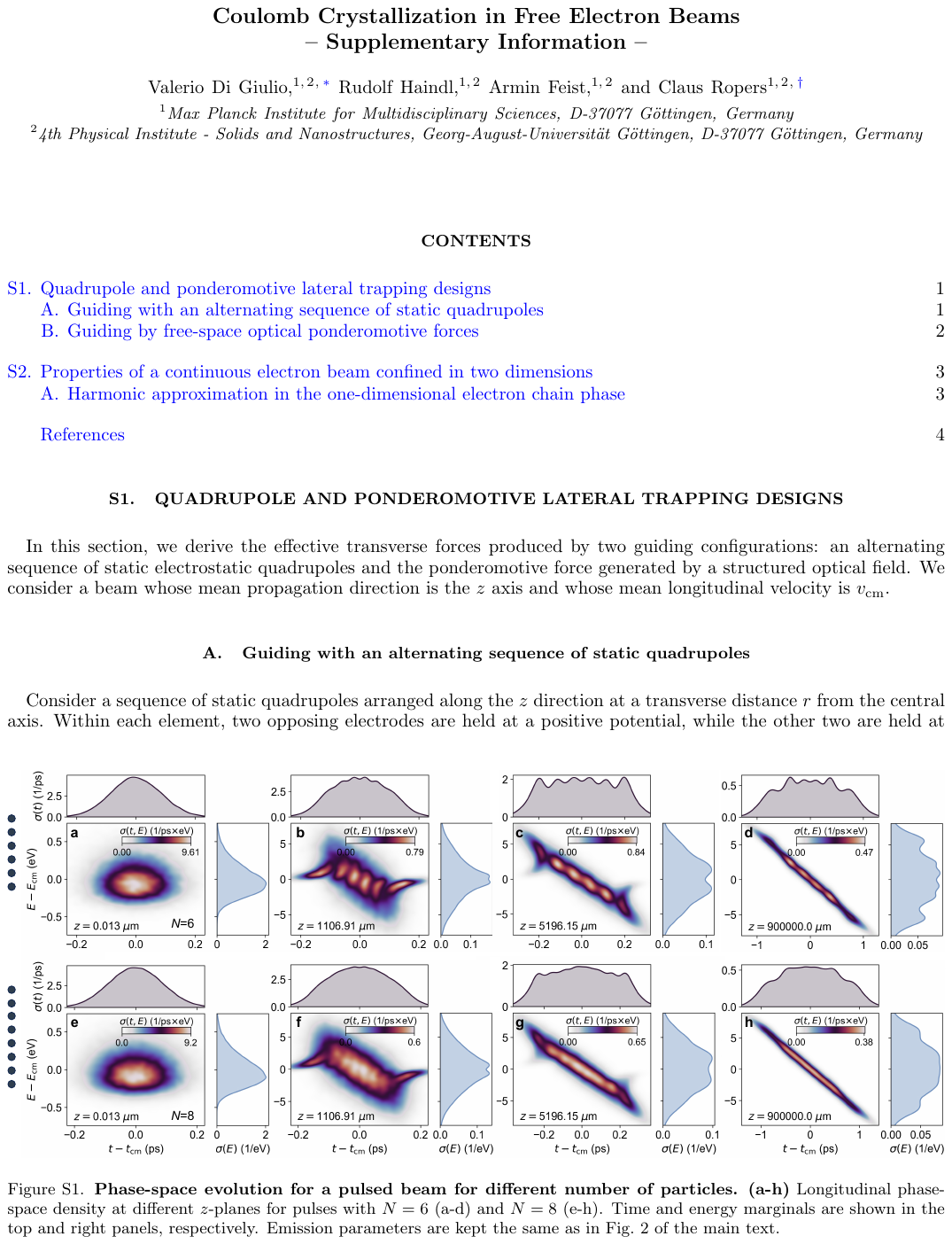}
}

\end{document}